# POLARISATION-MULTIPLEXING RING-CAVITY FIBRE LASER FOR DUAL-COMB GENERATION

Alberto Rodríguez Cuevas

Doctor of Philosophy

ASTON UNIVERSITY

June 2024

# Abstract

***Polarisation-Multiplexing Ring-Cavity Fibre Laser for Dual-Comb Generation***
Alberto Rodriguez Cuevas
Doctor of Philosophy
2024

This thesis investigates the development and characterization of a polarization-multiplexing ring-cavity fibre laser designed for dual-comb generation. It explores the underlying physics, practical implementation, and potential applications of this innovative laser system, particularly in LIDAR technology. Dual-frequency comb systems for metrology typically use two optical frequency combs synchronized via complex feedback loops, which can suffer from phase-locking issues and increase system cost and fragility. In contrast, single-cavity dual-comb systems generate two combs with slightly different repetition rates within the same cavity, ensuring mutual coherence and noise cancellation. However, these systems often generate unstable regimes and are generally demonstrated only in laboratories. The objective of this thesis is to design, build, characterize, and optimize a single-cavity polarization-multiplexed fibre laser capable of producing dual optical frequency combs with sufficient stability and precision for dual-comb LIDAR. This system aims to simplify the generation process while enhancing the practical applicability of dual-comb technology. Secondary objectives include studying the laser's intensity dynamics and collaborating with a company to address commercial needs and challenges in LIDAR technology, particularly under harsh environmental conditions. The results demonstrate the successful generation of dual optical frequency combs with minimal drift (1 Hz/hour) and high stability (over 250 hours of operation). The system shows potential for practical applications, achieving sub-millimetre precision in ambiguity ranges of 5m with a fundamental $f_{rep} = 39.25 MHz$ and a $\Delta f_{rep} = 869 Hz$. The findings highlight the robustness of the dual-comb regime for high-precision ranging applications while also revealing some precision limitations. Moreover, the thesis visualizes and analyses the build-up and propagation dynamics of the dual-comb in a single cavity, showcasing the two-stage build-up process and proving that the evolution of the energy of the initial spikes conditions the successful separation. Finally, this thesis offers mitigation strategies for commercial LIDAR devices operating under harsh environmental conditions.

*This thesis is dedicated to my family, whose unwavering support and encouragement made this journey possible.*

# Personal Acknowledgements

I would like to express my deepest gratitude to my family, whose unwavering support has been indispensable not only during the creation of this thesis but throughout my entire life. A special thanks to my parents, Raquel and Juan Antonio, and my brother, Juan Luis, for their love and encouragement. My extended family, including my grandparents, cousins, aunts, uncles, and others, has also provided a network of support for which I am incredibly thankful.

I am immensely grateful to my supervisor, Dr. Sergey Sergeyev, whose brilliance in science is matched by his kindness and generosity. His recognition and respect within the AiPT community are well-deserved and have greatly enriched my doctoral experience. I also extend my heartfelt thanks to my co-supervisor, Dr. Hani Kbashi for his guidance and support throughout this journey.

Special thanks go to Dr. Dmitrii Storialov, Dr. Igor Kudelin, Dr. Auro Perego, and Dr. Egor Maulovich for their invaluable scientific advice and for taking the time to explain complex concepts and provide thoughtful counsel.

I am fortunate to have worked alongside many brilliant minds in the AiPT. To all of you who have assisted me, whether in large ways or small, I extend my heartfelt gratitude.

To my friends outside of AiPT — Nelson, Ibon, Sara, Carmen, Marta, Julio, Dani, Sasi and many others — thank you for being there for me, for the laughter and respite you provided during this intense period. A special acknowledgment goes to Cassy, for being a part of this journey in ways words cannot fully express.

I am grateful to the Resident Tutor team, with whom I have shared three transformative years, gaining exceptional human experience.

Prior to this PhD, I had the opportunity to work with remarkable researchers at the GIF (Grupo de Ingenieria Fotónica), where I delved into the fascinating world of photonics.

This experience not only shaped my academic path but also led to significant contributions to the field through journal and conference papers.

Lastly, my sincere thanks to the European Union's Horizon 2020 research and innovation programme under the Marie Sklodowska-Curie project MEFISTA, grant agreement No 861152. This scholarship has made my PhD studies possible, and for that, I am eternally grateful.

# Collaborator Acknowledgements

I extend my sincere gratitude to Aurrigo Ltd. for hosting me during my six-month secondment. The opportunity to engage with a highly innovative company and contribute to its projects was both a fantastic experience and a valuable part of my professional development. I am confident in Aurrigo's bright future and grateful for the insights and experiences gained during my time there.

# List of Publications

**Journal Publications Arising from this Thesis:**

**Conference Publications Arising from this Thesis:**

**Additional Publications Completed Prior to this Thesis:**

# Contents

# List of Figures

# List of Tables

# List of Acronyms

**OFC** Optical Frequency Comb.

**TOF** Time of Flight.

**LOS** Linear Optical Sampling.

**WDM** Wavelength Division Multiplexing.

**SESAM** Semiconductor Saturable Absorber Mirror.

**DCS** Dual-Comb Spectroscopy.

**FSR** Free Spectral Range.

**NIST** National Institute of Standards and Technology.

**LO** Local Oscillator.

**FWHM** Full Width at Half Maximum.

**OSA** Optical Spectrum Analyser.

**PC** Polarisation Controllers.

**OSC** Oscilloscope.

**CEO** Carrier-Envelope Offset.

**DFT** Dispersive Fourier Transform.

**GVD** Group Velocity Dispersion.

**RF** Radio Frequency.

**SNR** Signal-to-Noise Ratio.

**FWM** Four-wave mixing.

# Chapter 1

# Introduction - Optical Frequency Comb - Literature Review

## 1.1 Introduction.

Optical Frequency Combs (OFCs) are precisely structured laser light sources that generate a series of equally spaced frequency lines, similar to the teeth of a comb. These frequency lines form an extremely stable and accurately calibrated ruler, allowing for precise measurement and manipulation of light. By generating a series of equally spaced frequency lines, OFCs can be used to measure the absolute frequency of light from various sources, such as atomic transitions and molecular vibrations. In the time domain, OFCs are extremely short pulses of light that are repeated at a fixed interval. The repetition rate of the pulses is equal to the spacing between the frequency lines of the comb, and the pulse duration determines the spectral width of the comb. The spacing between the frequency lines is precisely defined by the repetition rate, making the frequency comb an ideal tool for high-precision frequency and time measurements. OFCs have revolutionized a variety of fields and have become an essential tool for high-precision measurements [17]. They offer several advantages over traditional frequency measurement techniques, including unparalleled accuracy, stability, and versatility. In addition, OFCs provide a way to directly link the optical and radio frequency domains, allowing for precise comparisons of different types of frequencies. Their use has resulted in numerous breakthroughs and advancements in fields such as precision spectroscopy, optical atomic clocks, and optical communication, among others. As a result, OFCs have become critical components in a wide range of applications, from basic scientific research to industrial and commercial technologies. This chapter provides a historical overview of the development of OFCs. It covers early developments and experiments in frequency combs, as well as key innovations and breakthroughs that have led to their current widespread use. The document also examines the modern applications of OFCs, including their use in precision spectroscopy, astronomy, telecommunications, ranging, microwave generation, and optical clocks among others. Additionally, the document discusses current research and development in the field of OFCs, including generation methods and designs.

## 1.2 The concept of optical frequency comb and frequency comb equation.

One of the key breakthroughs that set the bases of the Fundamental principle of OFCs and converted the field into a new type was the self-referencing technology. The optical-frequency-comb formula that defines every tooth of the comb is the next one (1.1).

$$f_n = f_{\text{CEO}} + n * f_{\text{rep}} \tag{1.1}$$

The frequency of each tooth depends on an offset frequency and an integer number, n, that represents one tooth in a discrete succession of a number of the teeth. By multiplying both sides of the equation by two 1.2 we can deduce the equation 1.3.

$$2f_n = 2f_{CEO} + 2n \times f_{rep} \tag{1.2}$$

When the spectral width of OFC reaches one octave, the $n^{th}$ order comb is frequency-multiplied, and the frequency of the new light wave after frequency multiplication can be expressed as

$$f_{2n} = f_{CEO} + 2n \times f_{rep} \tag{1.3}$$

Usually, in fully stabilized OFCs, the cavity length of the laser is controlled through a piezoelectric device connected to a feedback loop. This allows locking the repetition rate frequency ($f_{rep}$), or in other words, stabilizing one of the degrees of freedom of the OFC. At the same time, a variable pump power modulator is introduced, using another or the same feedback loop, to fix the second degree of freedom of the OFC, ($f_{CEO}$). In this way, the OFC is fully stabilized. The feedback loop is typically generated starting with ultra-stable optical references or, in other cases, using self-referencing techniques such as the most common one, f-2f, which is shown in Figure 1.1.

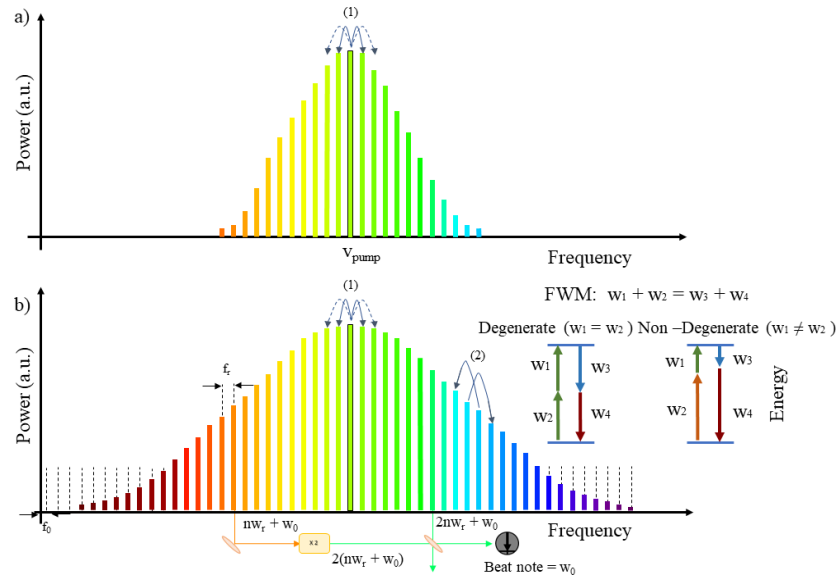

**Figure 1.1:** OFC Broadening a) Original Optical Frequency Comb, b) Broaden OFC of the Figure a). Adapted from Zhang, Hao, et al. "Coherent optical frequency combs: From principles to applications." Journal of Electronic Science and Technology 20.2 (2022): 100157. [18]

## 1.3 Historical overview.

The history of Optical Frequency Combs (OFCs) is a succession of breakthrough discoveries that have evolved into today's advanced technology. It began with the discovery of the Continuous-Wave Helium-Neon (CW HeNe) gas laser and the development of optical heterodyne diagnostic techniques for frequency behaviour in the early 1960s [19]. These advancements paved the way for frequency-based control of laser frequency and optical frequency measurements, enabling rapid progress in harmonic frequency measurements and new spectroscopic capabilities with tunable, frequency-controlled lasers.

The concept of frequency combs emerged in the late 1990s [20, 21, 22], with early experiments using mode-locked lasers and frequency-doubling techniques. Theodor W. Hänsch and his students focused on stabilising the modes of an OFC, the carrier-envelope offset (CEO) phase, and utilising OFCs for accurate optical frequency measurements by directly converting optical frequencies to radio frequencies [17]. These early experiments highlighted the potential of frequency combs for high-precision frequency measurements.

In the 1970s, frequency comb spectroscopy first utilised a mode-locked laser's regular train of picosecond pulses for Doppler-free two-photon excitation spectroscopy of simple atoms [23, 24, 25]. Despite the narrow spectral span, the role of the carrier-envelope offset frequency in the comb spectrum was understood.

A significant advancement occurred in 2000 with the publication of works that reinvigorated the field [26, 27, 28]. Before this, measuring optical frequencies required large, complex harmonic frequency chains. The developments in phase stabilisation and phase-locked femtosecond lasers introduced new tools for optical clock developments and precision spectroscopy [28]. Scott et al. [26] demonstrated precision measurements of optical frequencies compared to microwave counterparts, enabling direct microwave to visible frequency synthesis and transferring optical frequency stability to combs.

The 2000s saw the establishment of the self-referencing method for femtosecond laser synthesizers, enhancing comb-assisted precision spectroscopy and attracting significant research interest [29, 30, 31, 32, 33, 34, 35, 36]. This method reduced the difficulty of generating highly stable, widely tunable optical oscillators and determining their absolute frequencies relative to caesium microwave standards. Beams were matched and dispersed with a grating, and photodiode beating frequencies measured the difference between continuous wave lasers and adjacent frequency modes of the comb, allowing direct optical frequency measurements across the visible and near-infrared spectrum [26].

In the early 2000s, significant discoveries included the improvement in ultrashort pulse generation [37] and the stabilisation of the carrier-envelope phase of femtosecond mode-locked laser pulses using frequency-domain laser stabilisation [27]. This technique allowed phase locking without external optical input, enabling consistent phase pulses or phase variation, making the broad spectral comb of optical lines known frequencies, multiples of the pulse-repetition frequency plus a user-defined offset.

For their pioneering contributions to the development of self-referenced frequency combs, Theodor W. Hänsch (Max Planck Institute) and John L. Hall (University of Colorado Boulder) received the Nobel Prize in Physics in 2005. Their Nobel lectures recount the history of OFCs [38, 39].

Subsequent breakthroughs have further advanced this technology. Improvements in laser technologies, including ultrafast and fibre lasers, have significantly enhanced the stability and accuracy of frequency combs. Key developments include advances in frequency comb stabilisation and the introduction of microresonators. Frequency comb stabilisation, essential for maintaining long-term stability and accuracy, has evolved through new algorithms [40], advanced feedback systems [41], and the integration of frequency combs with ultra-stable references and self-referencing techniques [41]. These innovations have expanded the frequency range, spectral coverage, signal-to-noise ratio, and resolution of frequency combs, solidifying their role in modern precision spectroscopy.

These discoveries, in combination with early experiments in optical frequency metrology, opened the way for a multitude of diverse optical, atomic, molecular, and solid-state systems, including X-ray and attosecond pulse generation [42]. These early experiments and developments laid the foundation for the widespread use of OFCs in modern applications. In fact, over the past several decades, there have been several key breakthroughs and innovations in the field of OFCs that have led to their current widespread use in precision ranging [43], coherent control in field-dependent processes [44, 45], molecular fingerprinting [29], calibration of atomic spectrographs [32], tests of fundamental physics with atomic clocks [31], precision time/frequency transfer over fibre and free-space [33], trace gas sensing in the oil and gas industry [30], and arbitrary waveform measurements for optical communication [34].

Many of these applications make use of what is known as dual combs. A dual-comb system utilises two OFCs with slightly different repetition rates. These two combs can be generated in the same cavity or, as is the case in most applications, in two independent cavities. For most applications, they are locked together, meaning they are locked to the same optical references to ensure they maintain mutual coherence [35]. Since both combs have slightly different repetition rates, they can be beaten with each other on a photodiode and down-converted from the optical domain to the RF domain, allowing for their detection. This process reveals and quantifies the optical information that cannot be measured directly due to the ultra-high frequencies (optical frequencies), enabling measurement in the RF domain (MHz to GHz) frequencies. Mathematically, assuming two optical frequencies $f_1$ and $f_2$ we can define their electric fields as:

$$E_1(t) = E_0 \cos(2\pi f_1 t) \tag{1.4}$$

$$E_2(t) = E_0 \cos(2\pi f_2 t) \tag{1.5}$$

When these two optical signals interfere on a photodiode, the resulting signal $I(t)$ is proportional to the intensity of the combined electric fields:

$$I(t) \propto |E_1(t) + E_2(t)|^2 \tag{1.6}$$

Expanding this expression, we get:

$$I(t) \propto E_1(t)E_1^*(t) + E_2(t)E_2^*(t) + 2\text{Re}\{E_1(t)E_2^*(t)\} \tag{1.7}$$

The first two terms represent the intensities of the individual lasers, which are constant and do not contribute to the RF beat frequency. The third term represents the interference term, which contains the beat frequencies:

$$I(t) \propto 2\text{Re}\{E_1(t)E_2^*(t)\} \tag{1.8}$$

Substituting the expressions for $E_1(t)$ and $E_2(t)$:

$$I(t) \propto \cos(2\pi f_1 t)\cos(2\pi f_2 t) \tag{1.9}$$

Using the trigonometric identity:

$$\cos(A)\cos(B) = \frac{1}{2}[\cos(A - B) + \cos(A + B)] \tag{1.10}$$

we get:

$$I(t) \propto \frac{1}{2}[\cos(2\pi(f_1 - f_2)t) + \cos(2\pi(f_1 + f_2)t)] \tag{1.11}$$

The term $\cos(2\pi(f_1 - f_2)t)$ corresponds to the beat frequency, which lies in the RF domain, whereas $\cos(2\pi(f_1 + f_2)t)$ is twice the optical frequency and is filtered out by the photodetector.

Therefore, the down-converted RF frequency $f_{\text{RF}}$ is given by the difference between the two optical frequencies:

$$f_{\text{RF}} = |f_1 - f_2| \tag{1.12}$$

## 1.4 OFC generation techniques.

There are various ways to classify OFCs according to their generation methods. They can be classified based on the modulation parameter used, either as frequency-modulated combs or amplitude-modulated combs. Another classification is based on their synchronization: in-phase synchronization (related to amplitude modulation with pulses in the time domain) or anti-phase synchronization (related to frequency modulation without pulses)[46]. Additionally, OFCs can be classified by the physical platform used for their generation. The most common classification includes three main methods: using mode-locked lasers, using electro-optical modulators, and using micro-resonators. Due to their relevance, we explain them here.

### 1.4.1 OFC based on mode-locked lasers.

The mode-locked laser was invented [19] in 1964. Soon after the invention of the laser by Theodore H. Maiman in 1960 at Hughes Research Laboratories using a synthetic ruby crystal to produce pulsed red laser light at a wavelength of 694 nanometers [47], and soon after the contributions made by Ali Javan, William R. Bennett Jr., and Donald R. Herriott to the development of the gas laser at Bell Laboratories [48], achieving the first continuous-wave operation of the HeNe laser. However, it wasn't until the beginning of the new century when the frequency offset could be measured and controlled [49, 28, 50, 20, 26, 51] allowing for a fully stabilised OFC. Controlling the frequency offset wasn't possible for a long time since there was no way to coherently expand the optical spectrum of the comb to the point where an octave-spanning frequency comb is created and the comb can be stabilised using the technique of the Figure 1.1. As previously mentioned, once the comb is expanded to an octave spanning comb two of the comb teeth are extracted using a composite filtering scheme. Then using the equation 1.3, the frequency offset can be calculated and stabilised.

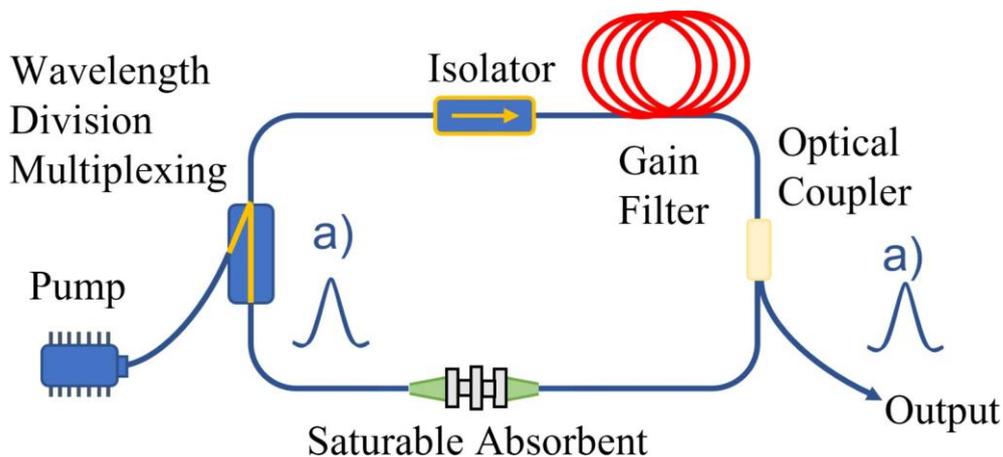

**Figure 1.2:** Mode-Locked OFC Scheme

Combs generated from mode-locked lasers are perhaps the most mature technology and the repetition rate that can be generated with this technology is in the order of 10MHz to 1 GHz.

Simple mode-locked lasers produce what's called free-running combs which are opposed to the fully stabilized combs in the fact that their $f_{rep}$ and $f_{CEO}$ are not fixed and varies over time. Indeed, there is an ongoing debate about what an OFC is, that confronts two different positions. Those who consider that only fully stabilized mode-locked lasers are OFC (orthodox position). And those who consider free-running combs as OFCs (unorthodox position). Taking this difference of opinions into account, the fact is that stabilized or not, when it comes to OFC laser cavities there are several types of cavities according to their geometry. Among them, we can find linear, circular, figure of eight, or figure of nine cavities. The linear cavities are usually fenced by SESAMs at each side having the other optical components in the middle while the circular cavities allow for a constant re-circulation of the light within them and are usually locked using saturable absorbers, such as in the example of the Figure 1.2. There are also artificial saturable absorbers such as nonlinear polarisation rotation (NPR) and nonlinear amplifying loop mirrors (NALM). They tend to have a much faster time response and require soliton pulse shaping to achieve a larger bandwidth. Once a mode-locked laser is stabilised using one of the different stabilisation mechanisms [41] the stability of its optical spectrum increases to a level where ultra-high precision metrology is possible.

### 1.4.2 OFC based on electro-optical modulators.

An electro-optical modulator, positioned between two highly reflective mirrors within an optical resonator, allows for the modulation of a microwave signal to impose a grid of sidebands on a continuous-wave laser. By placing the modulator inside the optical resonator and driving it with a microwave signal that matches the optical resonator mode spacing, we can generate resonantly enhanced sidebands, which in turn generate additional sidebands. This process eventually produces a succession of ultra-short pulses or a frequency comb [52], as illustrated in Figure 1.3. Using this type of laser, it is possible to extract a pulse that can be expanded in the optical domain using nanophotonics, ultimately converting the laser into a very broad, octave-spanning comb. These OFCs based on electro-optical modulators were conceived at an early stage of the OFC history, in fact, it was Dr. Kourogi who introduced the idea of inserting an electro-optical modulator inside a cavity and in that way creating a very efficient method of generating a frequency comb [21]. These were the first frequency combs used at the beginning of the century. Recently this idea of inserting an electro-optical modulator inside a cavity has been implemented in thin-film lithium niobate laser resulting in a much more integrated system [53]. This example is a clear way of integrating photonic circuits on a chip for OFC applications making use of the small dimensions of these lithium niobate micro-ring resonators.

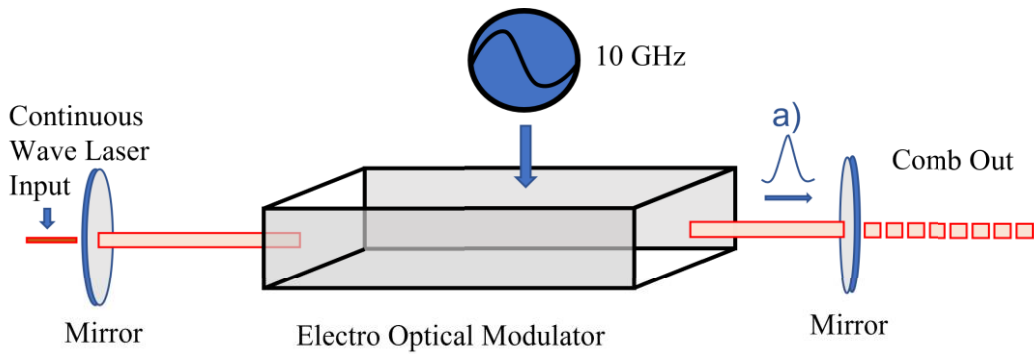

**Figure 1.3:** Electro-Optical Modulator OFC Scheme

### 1.4.3 OFC based on microresonators.

The third main way of OFC generation is the use of microresonators, Figure 1.4. This is the newest and perhaps the most exciting method nowadays to generate OFC. It is also the method where more effort is being put by the OFC community. The OFC generated in microresonators are known as microcombs [54, 55]. The way in which they work is by containing light into the whispering gallery modes of a small micro resonator and by four-wave mixing (FWM) multiplying the two original photon wavelengths into many more. If the wavelengths of the two original photons are the same it is called degenerate FWM, if they are different it is called non-degenerate FWM. See Figure 1.1.

The key factor of microresonators is the Q factor or quality factor that measures the resonator's ability to confine light and represents the efficiency of the resonator in terms of energy loss. Specifically, it is a dimensionless parameter that describes how under-damped an oscillator or resonator is, or equivalently, it characterises the resonator's bandwidth relative to its centre frequency. Higher Q factors indicate lower energy loss and thus more efficient confinement of light within the resonator. By varying the width of the resonator, the optical spectrum can be expanded or contracted to longer or shorter wavelengths. The prospect of integrating frequency combs on a chip is what make these devices so promising. The potential of reducing their dimensions and combining them with some other small-size devices in compact platforms together with the capability of cost-effective mass production is quite promising. Thus, since their discovery [54] they have been used in a wide range of applications [56] both in conventional areas and in emerging areas driven by the latest advances.

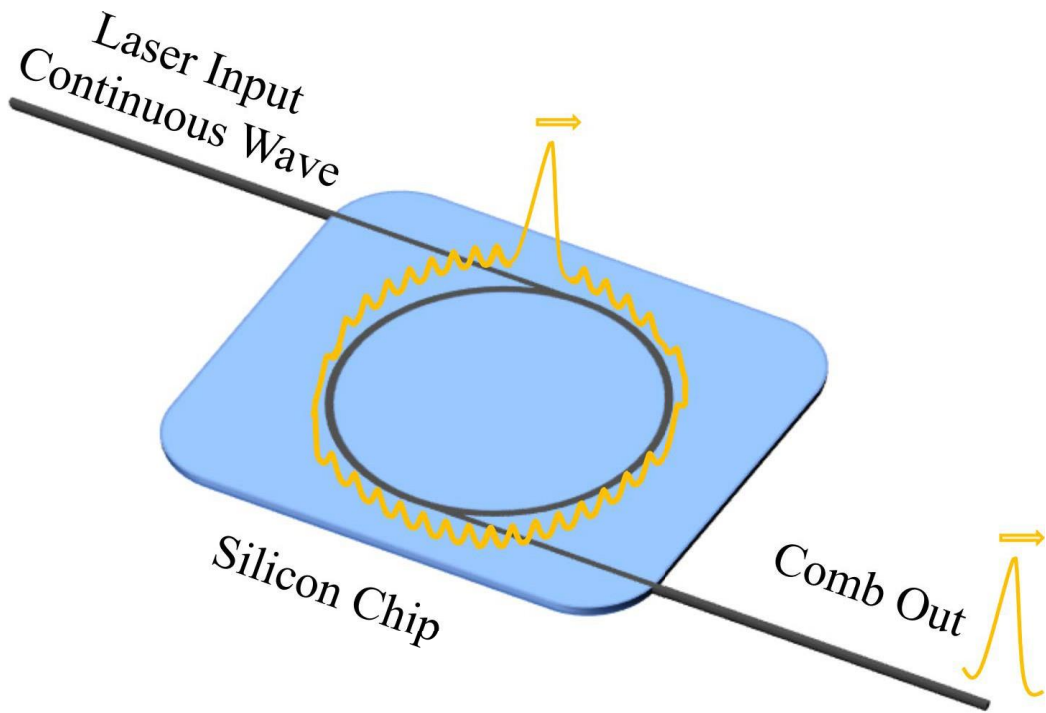

**Figure 1.4:** Microresonator OFC

## 1.5 Applications of OFCs.

The ultra-high precision of OFC to measure time and frequencies in combination with its versatility in terms of spectral wavelength, teeth separation, optical span, and repetition rate frequency, make OFC quite useful for several different metrology applications. In this section, I explain the main applications of OFCs.

### 1.5.1 Precision spectroscopy.

One of the most widely reckoned applications of OFCs is precision spectroscopy. Spectroscopy is the study of the interaction of light with matter and is used in a wide range of fields, including physics, chemistry, and biology. Precision spectroscopy requires high-precision frequency measurements, and frequency combs provide a way to directly link the optical and radio frequency domains, allowing for precise comparisons of different types of frequencies. Frequency combs have been used to perform high-resolution spectroscopy of a variety of species, including gases, liquids, and solids. For example, frequency combs have been used to detect trace amounts of toxic gases and pollutants, to identify the presence of specific compounds in biological samples, and to monitor the composition of industrial processes.

In addition, they have been used to perform high-precision measurements of atomic transitions, which are critical for the development of optical atomic clocks. Through their use in precision spectroscopy, frequency combs have enabled new discoveries and advancements in a variety of fields [35].

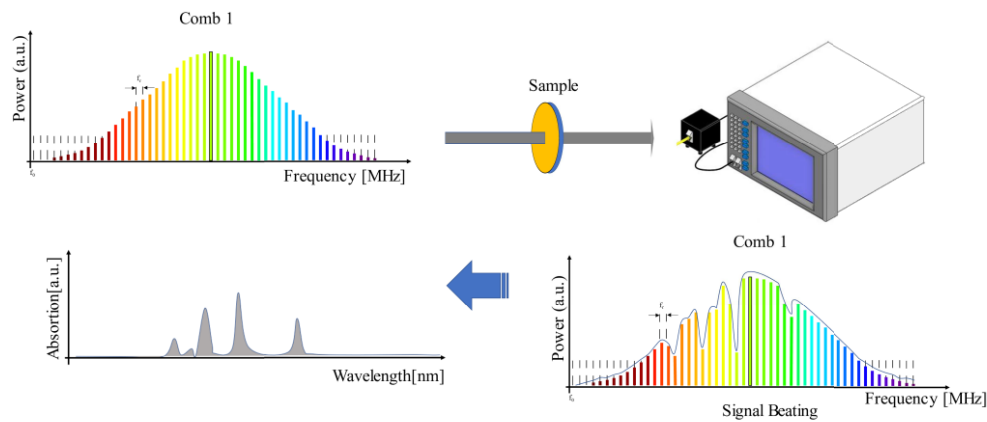

**Figure 1.5:** Direct OFC Spectroscopy. The Comb which is transmitted through the sample and detected by an OSA

The main advantage of OFC in spectroscopy applications is the fast resolution capacity that allows this system to measure spectrum with high update times as low as milliseconds or even lower, with the high spectral accuracy achieved when dual-comb spectroscopy is used. Traditional optical spectrum analysers require from a complex system to optically separate the optical frequencies using FBG (or alternative methodologies) and detect each frequency with an independent photodetector. This can also be done using direct OFC spectroscopy such as in Figure 1.5. However, by doing OFC direct spectroscopy we would lose all the benefits of the OFCs. Direct spectroscopy using an OSA is at the same time slow, complex, and limited in resolution. It is slow because the more narrower the frequency bandwidth wants to be separated and detected the more time need to pass to accumulate enough optical power in the detector. Since $\Delta T$ and $\Delta \nu$ are inversely proportional.

$$\Delta T \Delta \nu = 1 \tag{1.13}$$

It is complex because it requires a large array of photodetectors and precise calibration to direct every frequency to the right photodetector. In addition, it is limited in resolution because of the quality of the FBG used in the separation of every spectral component, limiting the optical resolution of the system. On the contrary, the heterodyne detection of dual-comb can directly map in a one-to-one scale the optical frequencies to RF frequencies that can be easily detected [35, 57, 41]. See Figure 1.6. This is fast since it only requires several beatings to average the signal. In fact, the minimum time to resolve the RF comb teeth, and therefore acquire a single spectrum, is simply $1/\Delta f_{\text{rep}}$. Finally, it has a high resolution given by the free spectral range of the combs used. The free spectral range is inversely proportional to the $f_{\text{rep}}$. And the smaller it is the higher the spectral resolution.

The spectral resolution is, however, inversely proportional to the signal-to-noise ratio (SNR). The fact that dual-comb spectroscopy (DCS) uses only one detector means that it captures millions of spectral points, introducing a "multiplexed" penalty because the fixed laser power must be distributed across multiple spectral elements or comb teeth. Thus Consequently, the signal-to-noise ratio (SNR) for each spectral element decreases proportionally with the square root of the number of elements, $1/\sqrt{M}$, leading to the overall SNR scaling with $1/M$. This fact highlights the necessity for prolonged coherent averaging time to maximize the SNR. The acquisition time ($\tau$) also influences the SNR leading to this final equation 1.14:

$$SNR = \sqrt{\tau} \, n_{comb}/M \tag{1.14}$$

Where $n_{comb}$ is the number of detected comb photons per second. The SNR is also conditioned by the Comb stability since unstable combs such as free-running combs prevent any sort of averaging and the more stable the pair of combs is the more averaging times can be used [35].

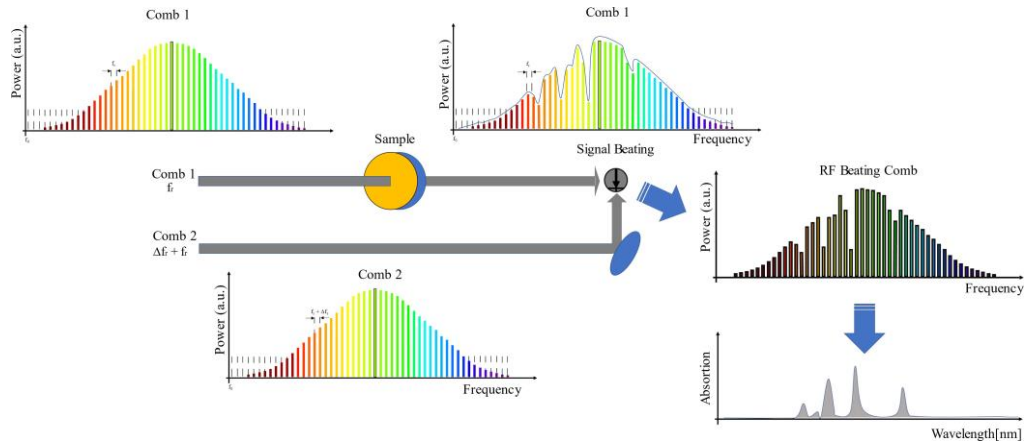

**Figure 1.6:** Dual Comb Spectroscopy. The Comb 1 with a repetition rate $f_{rep}$ is passed through the sample, the Comb 2 with a repetition rate $f_{rep} + \Delta f_{rep}$ is beaten with the Comb 1 in the photodiode. The resulting comb is an RF comb and the absorption spectrum can be obtained from this one.

Another important characteristic of the DCS is the spectral region in which it operates. Not all the spectral regions have the same value in terms of spectroscopic applications. Some of them have more intense interaction with the atoms, with the molecular bonds, or with the vibrational states. To date, most demonstrations of DCS have occurred in the approximately 1550 nm telecommunications band. This band is preferred due to the availability of relatively low-cost and durable erbium fibre frequency combs, fibre-optic components, and high-bandwidth, sensitive photoreceivers [35]. Electro-optic modulator frequency combs have been also widely proven in this spectral range. Nonetheless, spectral bands of 2 $\mu$m or more have more intense spectral interaction with most molecules, and the spectral signatures are acquired more clearly. Therefore, many efforts are being made in the development of OFCs with thulium-doped fibres and other gain media [58, 59].

Wavelengths greater than 3 $\mu$m, exhibit an even stronger spectral absorption cross sections. Therefore efforts to broader spectral coverage into the mid-IR (3−5 $\mu$m) and far-IR (6−13 $\mu$m) are currently happening [60]. Further in the spectral region, we find the THz which has a great potential for spectroscopy. As a matter of fact, terahertz radiation lies in the frequency range between microwaves and infrared light and has unique properties that make it useful for a wide range of applications. For example, terahertz radiation can penetrate through many materials that are opaque to visible light and infrared light, making it useful for imaging and sensing applications [60, 61]. Frequency combs have been used to generate and measure terahertz radiation, and have enabled new discoveries and advancements in fields such as material science, biology, and security [62].

### 1.5.2 Astronomy.

Closely related to the previous application and as a very useful niche of spectrometry application with OFCs, we find their use in the calibration of astronomical spectrometers. Spectrographs split light from celestial objects into its component wavelengths to study their spectra. Precise calibration ensures that measurements of wavelength are accurate. This accuracy is critical for determining various properties of celestial bodies, such as their chemical composition, temperature, velocity (via the Doppler shift), and distance. Indeed, the Doppler shift effect generated by the radial velocity of celestial bodies is perhaps the most critical evidence in the search for earth-like exoplanets, and the radial velocity Doppler shift precision needs to be as high as 2 cm/s [63]. These requirements evidence the need for an extremely stable wavelength reference for spectrograph calibration, and 'astrocombs', OFC tailored for astronomical spectrographs, play a key role in this. Astrocombs must have a spectral coverage in the range of the spectra that want to be measured. Moreover, they must match the comb mode spacing to the spectrograph resolving power [64, 65]. Provided this is achieved their use in astronomy is perhaps one of the applications where they have a more critical impact since the spectral resolution capacity of astrocombs is unrivaled.

### 1.5.3 Nonlinear optics and telecommunications.

Optical communications work by multiplexing several different optical frequencies into the same optical communication system, known as wavelength-division-multiplexed (WDM). Since OFCs provide a large number of equidistant frequencies, the idea of using OFC for optical communication systems has always been very attractive [66]. Since they have an intrinsically stable frequency spacing, their transmission performance is superior to what is possible with a free-running laser. In addition, the large number of comb teeth that can be generated with OFC could be a great advantage since it could multiplex way more channels using the same infrastructure. Nonetheless, current optical communication systems are highly effective and have been tested over decades. Their technology is mature, robust, and mass-manufactured, making it well-established and cost-effective. However, it has a significant drawback: high energy consumption. Using an optical frequency comb (OFC) to multiplex a large number of channels (comb lines) could help mitigate this issue, as many channels would be generated from the same OFC source, rather than having one source per channel, and measured using a second OFC source. This approach could potentially increase the amount of information sent through the same optical fiber by orders of magnitude, thereby reducing energy consumption in the emitter, receiver, and amplification stages. Currently, however, OFCs face a significant obstacle in optical communications due to the low power per comb line achieved. Consequently, the main efforts in OFC development for optical communications are focused on niche sectors, such as short-distance communications within data centres. Nonetheless, an increase in power per comb and more advanced noise cancellation techniques [67] are required for this application. Promising works have been done in this regard using microcombs to bring solitons generated in microresonators to optical communications [34].

### 1.5.4 Ranging.

The ultra-high resolution in the time domain of OFC can be transformed to ultra-high precision in measuring distances. In that way, dual-comb ranging was first demonstrated by Ian Coddington et. al., from NIST in 2009 [36]. The use of two OFCs combined on a photodiode allows the implementation of heterodyne detection. To do so, one of the combs is sent to a reference and a target simultaneously, getting two echoes back to the detector, these two echoes are beaten with the other comb that in this case acts as a local oscillator. Unlike traditional time of flight techniques in this case we can use an amplification factor which is $\frac{\Delta f_{rep}}{f_{rep}}$. These amplification factors are usually increasing the resolution by 4 to 5 orders of magnitude (since they are 10e-4 to 10-5). Considering the resolution of the oscilloscopes of 10-100 GHz we can estimate distance resolution in the order of nanometres. This magnification factor in combination with the relatively fast update rates (proportional to the $\Delta f_{rep}$) provides a rapid and precise optical ranging tool. Figure 1.7 represents a schematic visualization of the dual-comb working principle for ranging. More about the working principle of this technique and its variants are explained in Chapter 5.

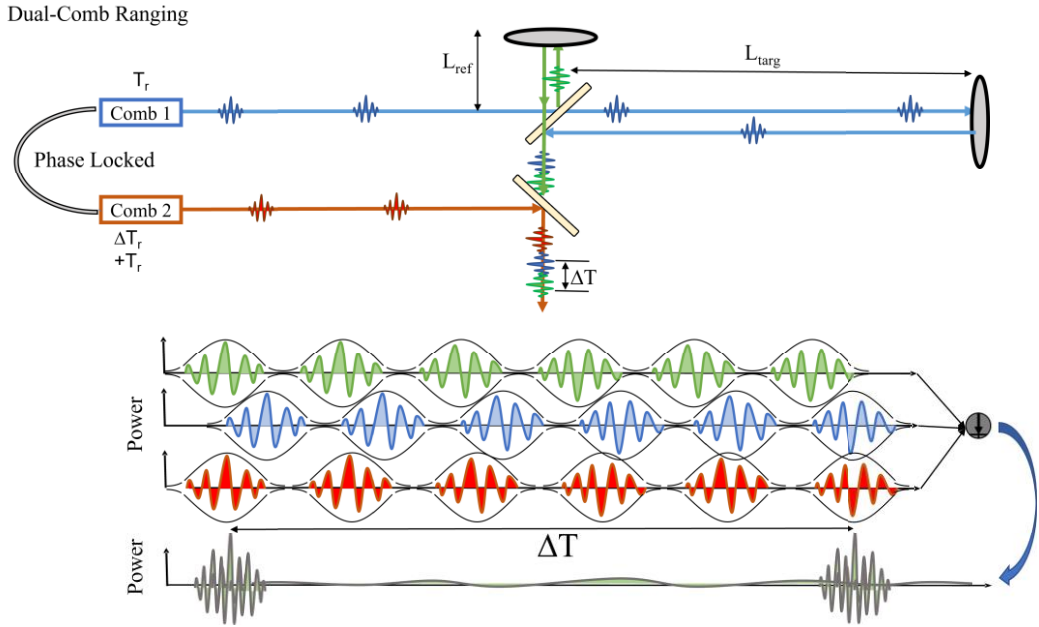

**Figure 1.7:** Dual OFC Ranging Scheme. The Comb 1 with a repetition rate $f_{rep}$ is sent to the target and reference while the Comb 2 with a repetition rate $f_{rep} + \Delta f_{rep}$ is beaten with the Comb 1 in the photodiode. The distance can be extracted by measuring the time between the spike generated by the target and the local oscillator and the spike generated by the reference and the local oscillator.

### 1.5.5 Ultra-stable microwave generation.

Another important application of OFCs is the synthesis of microwave frequencies out of optical frequencies. The key point of this application is the superior stability of optical references as compared to the microwave references, $10^{-13}$ at 1s for the microwave vs $10e^{-16}$ at 1s for the optical. Thus by locking the OFC to two of these ultra-stable optical references, we can stabilise all the comb modes or teeth to the same level of stability of the reference. Thus, using another ultra-stable optical source we can beat both frequencies and generate an ultra-stable microwave frequency [68]. Not only that but also we could potentially beat two ultra-stable frequency combs with slightly different repetition ranges and generate an ultra-stable microwave comb.

### 1.5.6 Precision timing - optical clocks.

Precision Timing is perhaps the most important characteristic in precision measurement applications of any kind. At the moment the definition of the second lies in the precision of the cesium atom. Currently, 1 second is defined as 9192631770 periods of radiation corresponding to the transition between the two hyperfine levels of the ground state of the

cesium-133 atom. There are several cesium clocks spread all over the world to keep reference measurements of the time. However, it is very likely that in the future this definition will be changed to a more precise definition in the THz domain. The reason for that is that at the moment we have been able to obtain higher-quality factors from measurements of narrow optical transitions. These optical transitions given in many atoms, Ca, Rb, Sr, Yb, Mg among others, can not be detected with ordinary microwave detectors, unlike Cesium-133 atomic transitions. But they can be detected using optical frequency combs [69].

The key parameter for ultraprecise time measurements is the fractional frequency and fractional frequency stability. The fractional frequency, $y(t)$ is defined as follows 1.15:

$$y(t) = \frac{f(t) - f_0}{f_0} = \frac{\Delta f}{f_0} \tag{1.15}$$

Where the relative deviation of the oscillation frequency, $f(t)$, after a time t is measured with respect to the original frequency, $f_0$. This concept, fractional frequency, is key when comparing different clocks without taking care of the nominal frequency of operation.

To measure the fractional frequency stability we use the Allan Variance and Allan Deviation. The Allan Variance, $\sigma_y^2(\tau)$ is the square root of the Allan deviation, $\sigma_y(\tau)$ and it is defined as follows 1.16:

$$\sigma_y^2(\tau) = \frac{1}{2(M-1)} \sum_{k=1}^{N-1} (\bar{y}_{k+1} - \bar{y}_k)^2 \tag{1.16}$$

Being M the number of time intervals and k the index for each consecutive interval.

Using the ultrahigh levels of stability provided by optical atomic clocks fundamental physic effects can be explored with much more precision. As an example, the effect of gravity on time has been measured in Japan [70] by placing two optical clocks, in the basement of a skyscraper, Skytree, and the other on top of it (450 m above) evidencing that the time passes faster on top of the building (lower gravity) than on the bottom (higher gravity). The difference in the atomic clocks ticking was 21Hz for their fundamental frequencies of 400 THz. As previously mentioned, terahertzs are not detectable with ordinary detectors. It is possible to measure them by using beating since the beat notes are in the MHz to GHz range and can be measured with standard detectors. Therefore, within the Optical clock architecture, the OFC is used as the frequency counter, which counts the optical transitions. The OFC is locked to certain atoms that have these optical transitions. An ultrastable laser is used as a local oscillator to generate the beating with the OFC. This is the architecture of an optical clock and how time can be measured with ultra-high precision.

Finally, another improvement derived from the increase in precision of time definition is the higher spatial resolution that can be achieved in GPS measurements [41].

### 1.5.7 Others.

The applications already explained are the main ones. Nonetheless, since they are sometimes organized in different ways, they all arise from their capacity to measure time or to measure frequency. Nonetheless, a couple of applications that are not related to the measure of time or frequency are applications for carrier-envelope phase control [71], and their use as frequency synthesizers [41, 40, 17].

## 1.6 Summary.

This introduction has laid the foundation for understanding the significance and multi-faceted applications of OFCs. By driving through the history of OFCs from the time prior to their invention to the current research the main discoveries in this field have been presented. The technical intricacies of their generation have been also explained here highlighting the 3 main generation mechanisms as well as the particularities of in-phase and anti-phase OFCs. The versatility of these combs is evidenced by their use in optical clocks, spectroscopy, ranging, or ultra-stable microwave generation among others. This versatility in their use underscores their potential to revolutionize various aspects of scientific research and industry. As we proceed, subsequent chapters delve deeper into the experimental setups, challenges, and innovations obtained during this thesis and in particular into the mechanisms and characteristics of single-cavity dual combs generated by polarisation multiplexing.

After introducing the OFCs in this chapter, 1. Later on, I present the results of the single-cavity, dual-comb laser in the next Chapter, 2. Its build-up and propagation dynamics are explained in Chapter 3 and the cavity optimisation in Chapter 4. Later on, I present the proofs-of-concept for single-cavity dual-comb ranging in Chapter 5. The following Chapter 6 discusses the real problems that a company, Aurrigo Ldt. faces when using ranging systems such as commercial LIDARs under harsh environmental conditions. And the solutions implemented within this company. Finally there in the last Chapter 7 I discuss the conclusions of this thesis and suggest potential future works that can be derived from it to continue this work.

# Chapter 2

# Stability, Tunability and Polarization Dynamics in a Dual-Comb Single-Cavity Polarization-Multiplexing Fiber Laser

---

*The work presented in this chapter has been adapted from the following publication:*
[2] Alberto Rodriguez Cuevas, Hani J Kbashi, Dmitrii Stoliarov, and Sergey Sergeyev. Polarization dynamics, stability and tunability of a dual-comb polarization-multiplexing ring-cavity fiber laser. *Results in Physics*, 46:106260, 2023.

---

## 2.1   Dual-comb generation in a single cavity fibre laser.

Dual frequency combs (DFC) consisting of two pulse trains with slightly different repetition rates have been used in several applications over the last two decades. The majority of the applications seen in the previous chapter use two combs such as in the case of LIDAR (Light Detection and Ranging) [36, 72], structural health monitoring [73, 74], gas sensing [35, 17, 57], optical communications [75], or calibration of instrumentation for astronomy [76] among others. To this date, in the large majority of cases, authors use dual optical frequency comb systems starting with two independent fibre lasers or solid-state lasers. These traditional implementations based on two synchronized ultra-fast lasers require addressing several technical challenges, such as the complexity of the phase-locking between both lasers. Therefore, some research efforts have been recently re-directed towards simpler systems based on a single laser.

Combs have two degrees of freedom: repetition rate $f_{rep}$ and carrier-envelope offset frequency $f_{CEO}$. Though individual $f_{rep}$ ($f_{CEO}$) of a free-running dual-comb in a single mode-locked laser drifts over time, the offset between the two combs $\Delta f_{rep}$ ($\Delta f_{CEO}$) tends to show superior long-term stability. This type of stability reflects intrinsic mutual coherence between the two pulse trains which is essential for dual-comb applications [77, 78]. Thus, a single cavity-based dual-comb is a cost-effective solution without complex $f_{rep}$ and $f_{CEO}$ stabilisation and phase-locked loop techniques [78]. The dual-comb single-cavity laser design approaches include wavelength multiplexing, polarisation multiplexing, circulation direction multiplexing, cavity space multiplexing, and extra-cavity fibre delay methods [78]. In most cases, dual-comb approaches are achieved using fiber lasers; however, other methodologies, such as semiconductor mode-locked integrated external-cavity surface-emitting lasers (MIX-ELS), are also employed [79]. In this section, I will briefly explain these methods, focusing primarily on the practical application of fiber lasers. The wavelength-multiplexing explores two trains of pulses with different central frequencies travelling simultaneously and in the same direction within the cavity [77, 78]. This is achieved by generating a critical balance between the gain, loss and filtering effects. Since the GVD is wavelength dependent the different central wavelengths of the pulses will generate a difference in propagation velocity and consequently a difference in $\Delta f_{rep}$.

$$\Delta f_{rep} = \frac{\lambda^2}{c \ D \ L \ \Delta\lambda} \qquad (2.1)$$

Being D the cavity dispersion, L the cavity Length and $\Delta\lambda$ the difference between the central wavelengths of both pulses. In polarisation multiplexing, two combs coexist in the same cavity circulating in the same direction but with two orthogonal states of polarisation (SOPs). Given the adjustment in the in-cavity birefringence, a difference in the refractive index of both polarization axes appears $n_x/= n_y$ leading to a difference in the repetition rate of both combs appears [80]. In the circulation-direction multiplexing, two trains of pulses travel in opposite directions of the same cavity. Given the non-symmetrical gain distribution in the gain media, the pulses end up interacting slightly differently with the cavity, generating combs with two different repetition rates [81], or two pulses with the same group velocity but different carrier frequencies. This second case would only be valid for certain applications. In some works there are combinations of more than one method such as in the work of Li, Bowen, et al. [82] where the bidirectional circulation of the combs is achieved with two different polarisations in combination with two polarisation beam splitters. In cavity-space multiplexing, a segment of the laser cavity is split into two segments with two different components, such as two different isolators or two different gain media for creating two different trains of pulses [83]. This method can also be combined with others such as with circulation-direction multiplexing [84]. Figure 2.1 shows a schematic representation of all of these methods. Finally, the so-called extra-cavity methods comprise a single frequency-comb laser source and, for example, the Michelson interferometer or other methods that produce two combs from the original output [78].

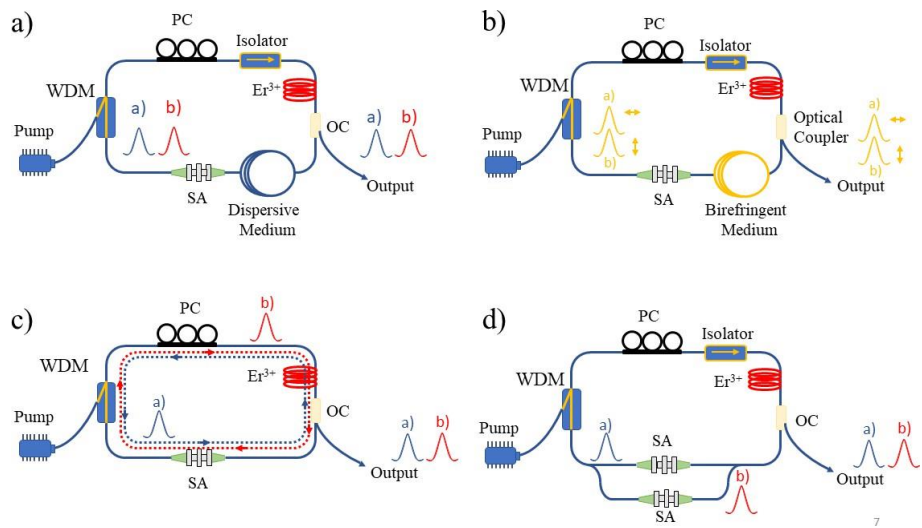

**Figure 2.1:** Methods of dual-comb laser generation in a single cavity system. (a) Wavelength-multiplexing; (b) Polarisation-multiplexing; (c) Circulation-direction-multiplexing. (d). Cavity-space-multiplexing.

To this date, various studies have been done where polarisation multiplexing has been used to generate two combs in microresonators [85, 86, 87, 88] or in single fibre lasers [77, 78, 89]. By introducing a piece of highly birefringence material in the medium, usually a segment of polarisation-maintaining (PM) fibre, the refractive index of the cavity is diverted into two axes, the slow and fast axis of the PM fibre $n_x \neq n_y$. Therefore, the trains of pulses with two orthogonal SOPs interact differently with the birefringent media in the cavity, for example, PM fibre, which results in different repetition rates. Unlike dual-wavelength lasers, the dual-polarisation comb takes the form of two pulse trains with an overlapping spectrum and supports intrinsic spectral coherence. To the best of our knowledge, there has not been any prior study evaluating the dynamics of the orthogonal SOPs in similar laser cavities. Nonetheless, the insight into SOPs' dynamics could extend the range of dual-comb laser applications to ellipsometry and environmental monitoring [90, 91]. To address this gap, in this chapter, we study experimentally the stability, tunability, and polarisation dynamics of this dual-comb laser based on the polarisation-multiplexing technique.

## 2.2 Setup design.

A schematic representation of the polarisation-multiplexed system based on a fibre ring laser is displayed in Figure 2.2. 0.85-meter-long Er-doped fibre (Liekki Er80-8/125) with group velocity dispersion (GVD) parameter of 0.016 $ps^2/km$ is pumped using a 980 nm laser diode through a 980/1550 nm WDM and an inline polarisation controller to control SOP of the pump. A one-meter-long PM-1550 fibre with GVD of $-0.030$ $ps^2/km$, a numerical aperture of 0.125 and core/cladding diameters of 8.5/125 $\mu m$ has also been used inside the cavity to generate polarisation multiplexing mode-locked regime, hence, to generate two trains of pulses in the same laser cavity. This polarisation multiplexing can only be achieved by adjusting the polarisation states inside the cavity with a polarisation controller. Thus, a 3-paddle polarisation controller comes after the PM fibre. A 51 dB dual-stage polarisation-independent optical isolator (Thorlabs IOT-H-1550A) is inserted in the laser cavity to ensure a single direction of propagation. A film-type homemade single-wall CNT absorber has been inserted between fibre connectors as a transmission-type mode-locker. An optical power characterisation of the CNT was done and can be found in the Appendix B. Finally, an output optical coupler redirects 10% of light outside the cavity. In total, the length of the cavity is 17.3 m, while the roundtrip time of a pulse is 83 ns (around repetition rate $f_{rep}$ = $12MHz$). Like the majority of Er-doped single-mode fibres, Er80-8/125 has a positive GVD, 0.016 $ps^2/km$, nonetheless, its influence in the cavity is neglectable. Consequently, we got an all-anomalous net cavity dispersion at 1560 nm with the parameter of - 0.021 $ps^2/km$.

For the polarisation-multiplexed comb separation, the polarisation controller and polarisation beam-splitter were connected to the laser output. The detection and measuring systems include a photodetector (InGaAsUDP-15-IR-2 FC) with a bandwidth of 17 GHz connected to a 2.5 GHz sampling oscilloscope (Tektronix DPO7254). A fast polarimeter (Novoptel PM1000-XL-FA-N20 D) with a sampling frequency of 100 MSa/s to measure the normalized Stokes parameters $S_1$, $S_2$, $S_3$, and degree of polarisation (DOP). An optical spectrum analyser (Yokogawa AQ6317B) with a maximum resolution of 20 pm, and a radio frequency spectrum analyser (FSV Rohde Schwarz) have also been used to measure the optical and electrical spectrums respectively.

The PM fibre and the in-cavity polarisation controller act as an optical filter inside the laser cavity by adjusting a high level of birefringence in the media and causing the formation of the two combs. Some authors have already demonstrated the ability of this configuration to generate tunable dual comb, they have used a different length of PM and dispersion compensated fibre segments and other components [80]. Additionally, a similar configuration was used to generate a phase-stable dual-wavelength regime [92]. Nonetheless, in this last work, they used a 10-meter-long high birefringence fibre and they did not use a saturable absorber.

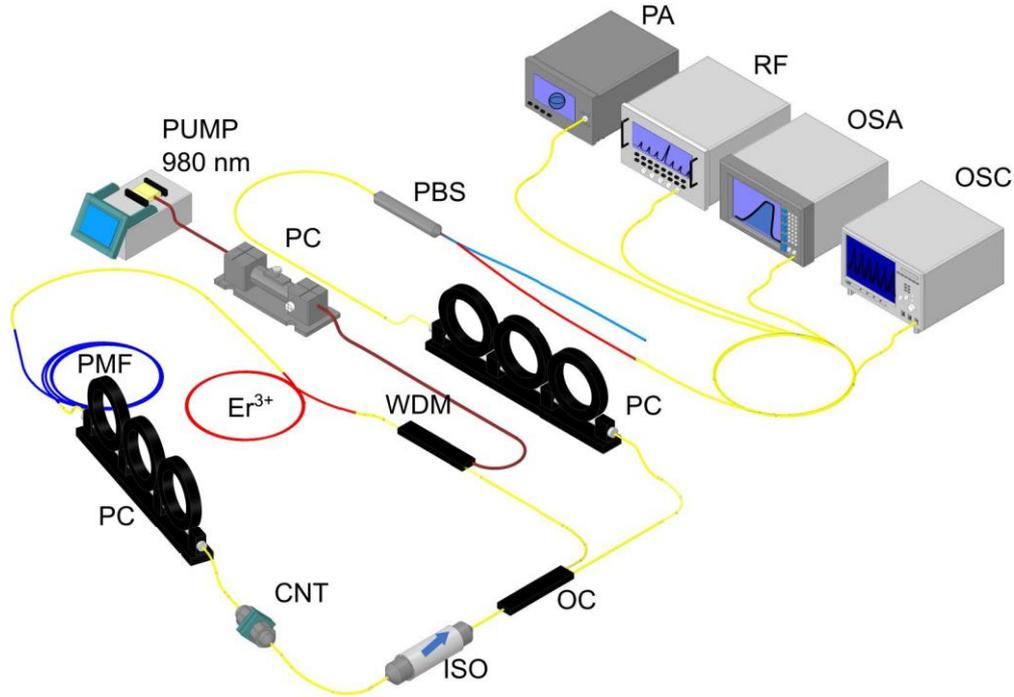

**Figure 2.2:** 3D representation of the laser cavity consisting of the following parts: isolator (ISO), optical coupler 90 / 10% (OC), wavelength-division multiplexing (WDM), polarisation controller (PC), saturable absorber based on a carbon nanotubes (CNT), segment of polarisation maintaining (PM) fibre, fibre segment of Erbium-doped fibre ($Er^{3+}$), and a continuous wave laser diode at 980nm (PUMP). Outside of the cavity, there is a polarisation controller (PC) and a polarisation beam splitter (PBS) in combination with the following measuring devices: polarimeter (PA), radiofrequency spectrum analyser (RF), optical spectrum analyser (OSA), and oscilloscope (OSC).

## 2.3  Results and discussion.

Using the setup described in the previous section, we demonstrated the emergence, tunability, long-term stability, and polarisation dynamics of the dual-comb system. As well as the optical power output of the laser characterized in Figure 2.3. The graph does not reach saturation power because the CNT can be quickly damaged if the peak power exceeds a certain value (observed at 1500 W) or if the pump power gets overheated. Therefore the trendline is adjusted, with a $R^2$ value equal to 0.9931, to the following equation Output Power = 0.0048 Pump Current − 0.6247.

The nominal pump currents supplied to the pump were around 220 mA which corresponds to an optical power output of around 0.5 mW. As a reference in Figure A.1 within Appendix A we present the amplified spontaneous emission vs the laser intensity. Measured inside the laser cavity when it is open and the multi-meter is connected right after the WDM.

This laser was assembled prioritizing the internal flexibility to be able to replace components, swap their positions or extend or contract segments of fibre optic. Since it was the initial dual-comb laser build within our facilities we did several tests and studied multiple configurations before ending up with this one. Therefore, some of the connections between components are based on standard FP/APC connectors instead of being spliced. That is the case for the connection between the Erbium fibre and the PM fibre; the PM fibre and the polarisation controller; the polarisation controller and the non-linear absorber; the non-linear absorber and the isolator; the isolator and the optical coupler; and finally the optical coupler and the WDM. All these connectors partially contribute to the reduction in the output power since they increase the losses within the cavity. Within the other alternatives that were tested, it is worth mentioning the Liot Filter configuration that has two polarisation controllers and a polariser between them. Nonetheless, this configuration failed to generate a dual-comb regime easily and instead, a dual-wavelength regime was detected sporadically in the system output but without two trains of pulses coexisting with different repetition rates. Indeed, that was never observed.

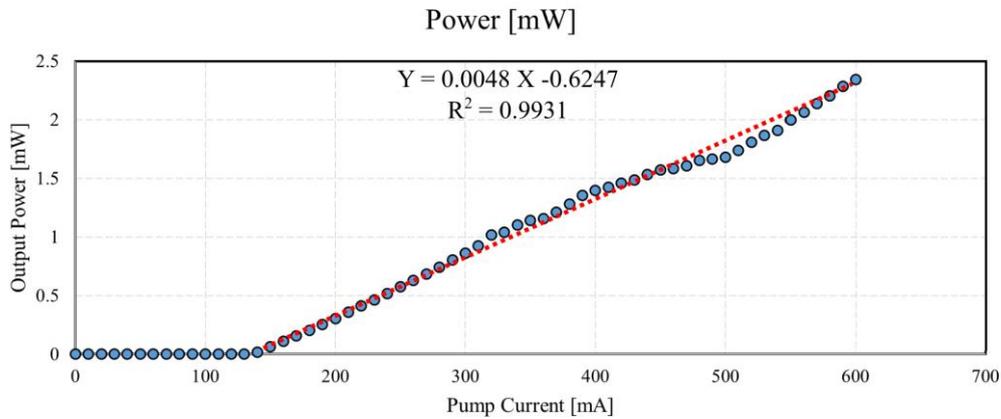

**Figure 2.3:** Pump current vs optical power output of the laser cavity.

The fundamental frequency of this laser is 11.95 MHz. Based on the fundamental frequency obtained for the standard mode-locked regime we can calculate the cavity length, $L_c$, using the following equation 2.2.

$$L_c = \frac{c}{nf} = \frac{299792458 \frac{m}{s}}{1.4475 * 11949738 \, Hz} = 17.332 \, m \tag{2.2}$$

Where c is the speed of light in vacuum, n is the refractive index of the propagation medium (laser cavity), and f is the fundamental frequency. Thus, the round-trip time can be calculated as the inverse of the fundamental frequency, f, equation 2.3.

$$T_{RT} = \frac{1}{f} = \frac{1}{11949738 Hz} = 83.68 \ ns \tag{2.3}$$

Once the cavity length is known, we can obtain the central frequencies of the slow, $f_1$, and fast comb, $f_2$, in the dual-comb regime and that way we calculate the refractive index of both slow and fast axes, as it is shown in the equations 2.4 and 2.5.

$$n_{slow} = \frac{c}{L * f_1} = 1.44728953 \tag{2.4}$$

$$n_{fast} = \frac{c}{L * f_2} = 1.44726531 \tag{2.5}$$

From these two values, we derived a refractive index difference of $\Delta_n = 2.422 * 10^{-5}$.

Finally, the difference in the propagation constant, $\Delta L_B$, the beat length, $L_B$, and the phase delay, $\Delta_\varphi$ can be derived from the previous values using the equations 2.6, 2.7 and 2.8 for each of them.

$$\Delta L_B = \frac{2\pi}{\lambda} \Delta_n = 146.363 \ m^{-1} \tag{2.6}$$

$$L_B = \frac{\lambda}{\Delta_n} = \frac{2\pi}{\Delta L_B} = \frac{1559.6 * 10^{-9}}{2.422 * 10^{-5}} = 0.0644 \ m \tag{2.7}$$

$$\Delta_\varphi = \Delta L_B \ L = 2536.35 \tag{2.8}$$

These values can change if the cavity is modified and even if the cavity isn't modified but the polarisation controller is repositioned the refractive index difference will change and therefore all the subsequent values.

We can also calculate the free spectral range (FSR) of the comb once we know the cavity length. The FSR is the frequency or wavelength interval between successive resonant modes [93]. In an optical frequency comb (OFC), it represents the separation between adjacent teeth of the optical spectrum. The equation for the FSR in terms of frequency is given by Equation 2.9, while the equation for the FSR in terms of wavelength is given by Equation 2.10, where $L$ is the length of the cavity and $n$ is the refractive index of the cavity.

$$FSR = \frac{V_{gr}}{2 \ L} = 5.99 KHz \tag{2.9}$$

$$FSR = \frac{\lambda^2}{2\,n\,L} = 0.048pm \qquad (2.10)$$

When it comes to tunability, Figure 2.4 a) presents the results of frequency difference tuning within a range from 110 to 250 Hz. With the correct polarisation state and curvature reduction of the PM fibre, the frequency difference can reach as high as 250 Hz or as low as 5 Hz. However, stable dual-comb regimes were only observed with frequency differences in the range of 110-250 Hz. If the frequency difference is lower, the regime tends to collapse after a few seconds or minutes. A similar phenomenon was observed for the frequency locking of coupled oscillators [94]. Additionally, the element composition of the system limits the upper range of the frequency difference to 250 Hz, regardless of the combination of intra-cavity polarisation and pump power. Multipulsing was often observed when the dual-comb regime was about to be obtained. The tunability of the frequency difference was achieved by precisely adjusting the polarisation controller within the cavity. Transitions of more than 10 Hz are not possible, but slight changes of up to 10 Hz can be obtained by carefully adjusting the intra-cavity polarisation controller while preserving the dual-comb regime. However, any intended changes in the frequency difference over 10 Hz cause the dual-comb regime to collapse abruptly, giving way to a different regime. Sometimes, the dual-comb regime reappears later with another frequency difference. Because the controllable birefringence in the cavity also modifies the optical spectral transmission [78], the spectra for different frequency differences change as shown in Figure 2.4 b).

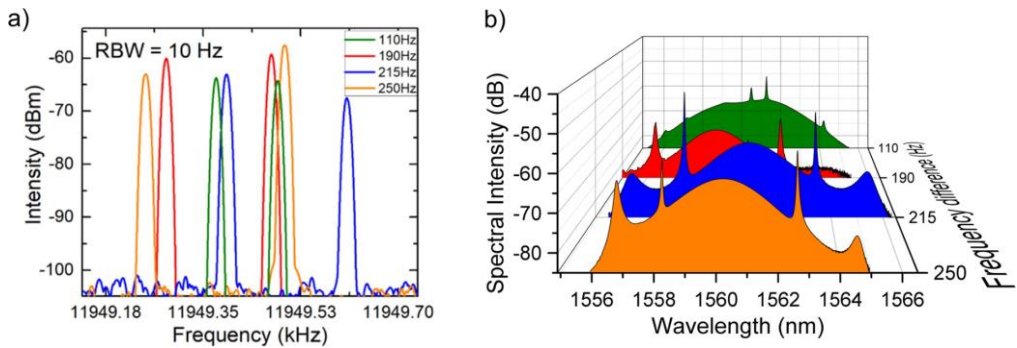

**Figure 2.4:** a) RF spectra tunability of the dual-comb frequency difference b) OSA spectra. The four plots correspond to the following frequency differences, starting from the front to the back 250Hz, 215 Hz, 190 Hz, 110 Hz.

Figure 2.5 presents the results for dual-comb temporal stability, namely for dual-comb operation with a repetition rate difference of 205 Hz. The dual-comb was monitored for 400 minutes in room temperature conditions. The regime tends to remain the same after that time, with a slight shift in repetition rate and central wavelength. Even though there is drift over time in the carrier-envelope offset frequency, the offset between the two combs shows stability with a drift of 6 Hz over 6 hours, going from 205 Hz to 199 Hz unevenly without any additional external stabilisation (Figure 2.5 a)). The resolution bandwidth (RBW) of the RF spectrum analyser was 1 Hz. As follows from Figure 2.5 a), the short-term stability over 1 min can be estimated as 15 mHz. The drift is more than twice less than the drift of 38 mHz/80 sec obtained in [80]. In terms of the stability of the optical spectrum, Figure 2.5 b) shows that its shape remains constant during the 6-hour test, with only a slow shift of 0.05 nm in the central wavelength. This change is correlated to the change in the frequency difference observed during the same test. The slow variation in both measures, combined with the lack of temperature stabilisation, suggests that slow temperature variations in the laboratory are causing this variation. Nonetheless, when the dual-comb regime is close to collapse, small perturbations can be observed in the optical spectrum analyser between consecutive measurements. This behaviour can serve as an early-warning signal, allowing for correction measurements to be triggered and the system to be stabilised before it reaches a critical state.

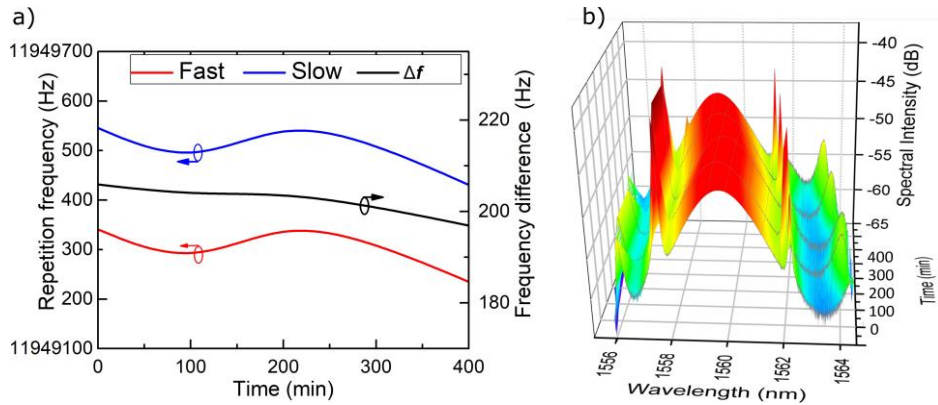

**Figure 2.5:** Stability of the dual-comb (a) frequencies of fast and slow axes and their difference; (b) optical spectra over time.

Recently, other polarisation-multiplexed single-cavity dual-comb laser systems have been demonstrated [80]. For example, Zhao et. al. reported the system generated repetition rate differences in the range of 228 to 773 Hz. Applications such as dual-comb LIDAR require pulse overlapping for several round trips so that a heterodyne detection can be performed. Therefore, smaller repetition rate differences are generally more adequate [91]. Our study expands the research of Zhao et. al, focusing on the polarisation dynamics of a dual-frequency comb generated from a single mode-locked Er-doped fibre laser with a lower repetition rate. A polarisation beam splitter (PBS1550SM-APC Thorlabs), in combination with a polarisation controller (PC), was employed at the laser output to separate the signal into two linear orthogonal states of polarisation. The PBS has an excitation ratio (splitter) $a \geq b$ 20 dB at an operating wavelength of 1550 ±40 nm. Figure 2.6 b) shows the RF spectra of the common signal from the output port before PBS. Two frequency components with a 205 Hz repetition rate difference were fixed at 11.949340 and 11.949545 MHz. From these frequencies, we obtain a refractive index difference of $2.573 \times 10^{-5}$, a beat length of 0.0605 m, a difference in propagation between both pulses of 103.925 $m^{-1}$ and a phase delay of 1700.52. Signal-to-noise ratios of more than 60 dB are observed for both RF peaks, which is evidence of good stability of the dual combs. These two peaks have FWHM bandwidths of 1.4 Hz and 1.5 Hz. Between the two combs, the weaker frequency peaks are beat notes (fbeat), equally spaced by the comb's frequency difference. A beat note has a FWHM bandwidth of 1.5 Hz. Figure 2.6 b) depicts the RF spectra of the two orthogonally polarized outputs at RBW at 1 Hz. The maximum difference in amplitude of around 15 dBm between the components was achieved by adjusting the PC before the PBS. In other words, two pulse trains are in a state of polarisation that is close to orthogonal. Figure 2.6 c) illustrates the common optical spectrum of polarisation-multiplexed mode-locking and spectra for each comb individually. In the slow axis, the centre wavelength is 1559.45 nm with $\Delta\lambda$ is 1.85 nm and in the fast axis, the centre wavelength is 1559.62 nm with $\Delta\lambda$ is 1.91 nm.

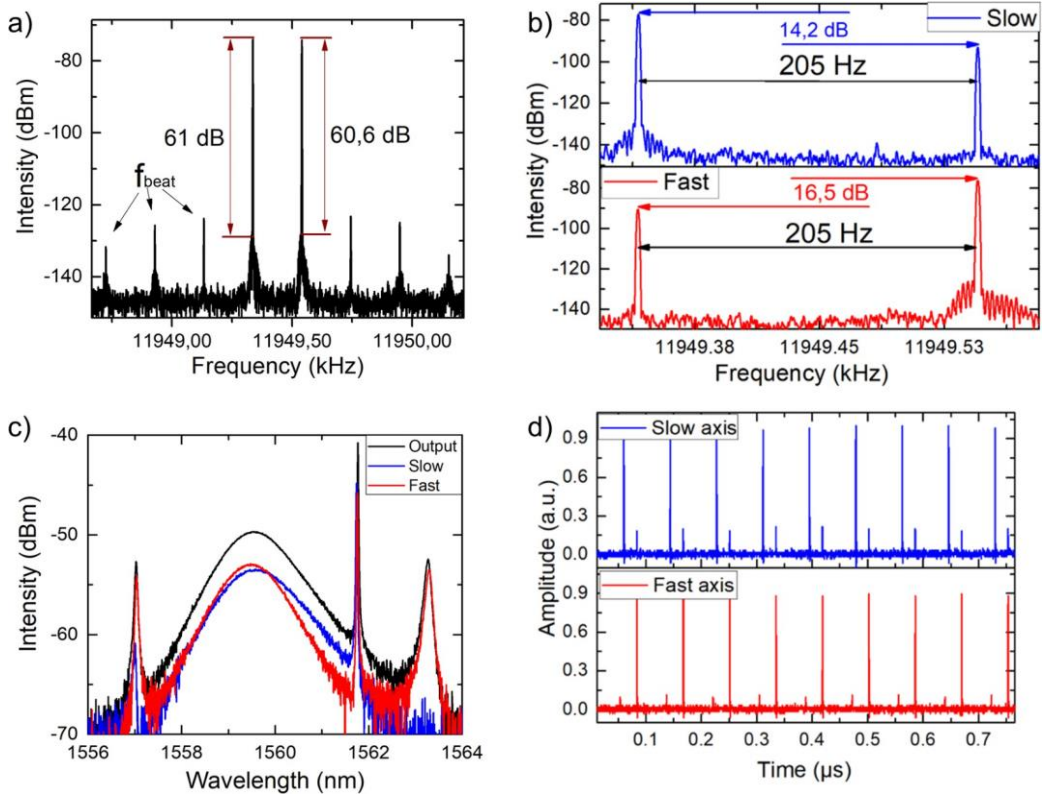

**Figure 2.6:** Laser cavity characterization. RF spectrum of the original signal with a frequency difference 205 Hz, (b) separated signal by PBS, in red colour the fast axis, in blue colour the slow axis, (c) optical spectra, (d) oscilloscope traces for each comb after PBS.

The optical spectra of the output, fast and slow axis, are overlapped with just a small swift of the fast axis towards the left (smaller frequencies), see Figure 2.6 c). This swift is considerably smaller than in the work of Zhao et. Al. [80]. There, the optical spectrums are overlapped but there is a substantial difference in the right and left edges of the spectrum between the fast and slow axes.

In terms of separation, even though the PBS can potentially separate the two combs to the level of the extinction ratio of 20 dB, the observed comb separation is lower than this extinction ratio. This fact can be caused by imperfections in the adjustment of the output polarisation controller to PBS. Autocorrelation traces (AC) for both pulse trains are also displayed in Figure 2.7 b). The pulse duration delivered by this laser was measured at its full width at half maximum (FWHM) for both axes using an autocorrelator (Femtochrome FR-103XL) and the results were 1.53 ps for the fast-moving pulse and 1.47 ps for the slow-moving pulse. As Figure 2.7 demonstrated both pulses have a shape close to Gaussian. Having in mind our final application, dual-comb LIDAR, and the need for a heterodyne detection we must take into account that there is a ratio between the pulse width and the repetition rate, that needs to be satisfied, defined by the following formula:

$$10 \ \Delta T \le \Delta \tau \tag{2.11}$$

That ratio must be over 10 times which in a way means that the pulses coexist while they overrun each other for at least 10 round trips and thus we can perform the heterodyne detection and extract the fringes information from them. Eventually increasing the overall resolution of the system. $\Delta T$ is the time that the fast comb takes to overrun the slow comb, described in the following equation.

$$\Delta T = \frac{\Delta f}{f_1^2} = \frac{205 Hz}{(1.1953 \ 10^7)^2 Hz^2} = 1.44 ps \tag{2.12}$$

Being $\Delta f$ the frequency difference and $f_1^2$ the square of the repetition rate of the fast comb. $\Delta \tau$ is the average pulse width of the pulses. As we can see the condition for LIDAR $10 \ \Delta T \le \ \Delta \tau$ is not fulfilled with this cavity, revealing the need to implement some improvements in the cavity.

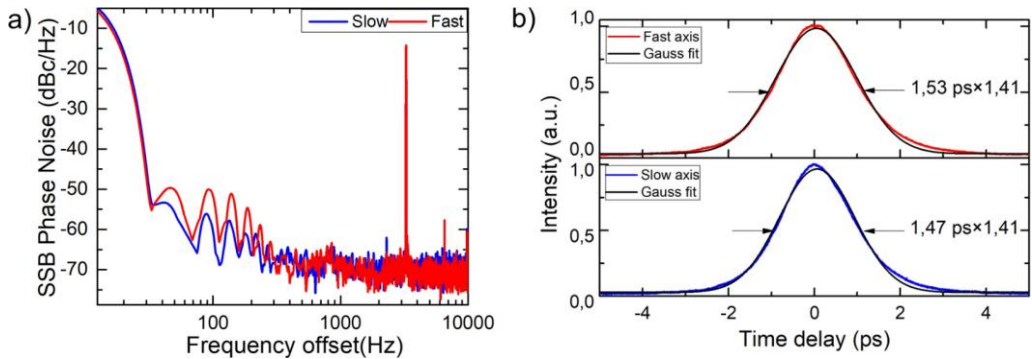

**Figure 2.7:** Laser cavity characterization. a) RF phase noise, b) autocorrelation traces.

The phase noise of a dual-comb system, Figure 2.7 a), is determined by the noise sources of the individual combs, such as the carrier-envelope phase noise, the repetition rate phase noise, and the residual phase noise [95]. The phase noise affects the quality and accuracy of the dual-comb measurements, as it introduces errors and uncertainties in the detection and processing of the signals. The values of phase noise required for different applications depend on the specific parameters and performance goals of the system. For example, for dual-comb spectroscopy, the phase noise should be low enough to avoid spectral broadening and distortion of the absorption features [96]. For dual-comb ranging, the phase noise should be low enough to achieve high precision and resolution in distance measurement.

One way to quantify the phase noise requirement is to use the signal-to-noise ratio (SNR) of the dual-comb signal. The SNR is defined as the ratio of the power of the signal to the power of the noise. A higher SNR means a lower phase noise and a better measurement quality. The SNR depends on various factors, such as the optical power, the repetition rate difference, the detection bandwidth, and the integration time. We can extract the SNR ratio from the RF spectrum analyser and consider the fundamental frequency of the most intense comb as the 0. Then we subtract the SNR comparing the value of dBm in the fundamental signal with the value of dBm in each and every point adjacent. These points are in Hz and the values of the phase noise can be measured in dBc/Hz. For high-resolution applications like molecular spectroscopy, phase noise levels in the range of -100 dBc/Hz to -140 dBc/Hz or lower are typically considered good. This ensures that narrow spectral lines can be resolved accurately. While for applications like distance or range measurements, the phase noise requirements may not be as stringent as in spectroscopy. Phase noise levels in the range of -70 dBc/Hz to -100 dBc/Hz may be sufficient, depending on the desired measurement precision. Our phase noise at an offset frequency of 1000 Hz is in the order of -70 dBc/Hz, so in the border of what is needed for ranging applications, see Figure 2.7 a).

The study of the polarisation dynamics of the fast and slow axis in comparison with the dual-comb signal has been performed with a fast polarimeter (Novoptel PM1000-XL-FA-N20-D) with a resolution of 10 ns. Polarisation data were acquired for every sample with different frequency differences. Each trace comprises samples of the Stokes parameters $S_0$, $S_1$, $S_2$, $S_3$, and degree of polarisation (DOP). Stock parameters can be defined as follows.

$$S_0 = |E_x|^2 + |E_y|^2 \tag{2.13}$$

$$S_1 = |E_x|^2 - |E_y|^2 \tag{2.14}$$

$$S_2 = 2\ |E_x|\ |E_y|\ cos(\Delta\varphi) \tag{2.15}$$

$$S_3 = 2\ |E_x|\ |E_y|\ sin(\Delta\varphi) \tag{2.16}$$

While the DOP is defined as:

$$DOP = \frac{\sqrt{((S_1))^2 + ((S_2))^2 + ((S_3))^2}}{((S_0))^2} \tag{2.17}$$

Where $\langle \dots \rangle$ means averaging over the 320 ns. Figure 2.8 (a-c) shows the normalized Stokes parameters

$$S_i = \sqrt{\frac{(\langle S_i \rangle)^2}{(\langle S_1 \rangle)^2 + (\langle S_2 \rangle)^2 + (\langle S_3 \rangle)^2}} \tag{2.18}$$

The output power $S_0$, DOP (a, b) and SOP evolution in the Poincaré sphere (Figure 2.8 c)) as a function of time. Given the different overlapping of the polarisation multiplexed combs as a function of time, the normalized Stokes parameters, the output power and DOP the maximum demonstrate the oscillatory behaviour in the time domain as shown in Figure 2.8 (a, b)) with an asymptotic value corresponding the absence of the overlapping. As a result of the oscillation of the Stokes parameters, the trajectory at the Poincare sphere takes the form of an arc as shown in Figure 2.8 c). The oscillatory behaviour of Stokes parameters can be used for the evolution of the polarimetric signatures of light scattered by the object in terms of the Mueller matrix [97]. Instead of generating four different input SOPs, analysis of Fourier components of the Stokes parameters oscillations for scattered light vs input Stokes parameters enables the extraction of the Mueller matrix elements. When the polarisation dynamics were measured after the comb separation using a PBS, the degree of polarisation of the fast and slow axes were 0.98 and 0.99 respectively for the optimal separation level, being stable points in the Poincaré sphere, as expected given the good extinction ratio obtained with the PBS.

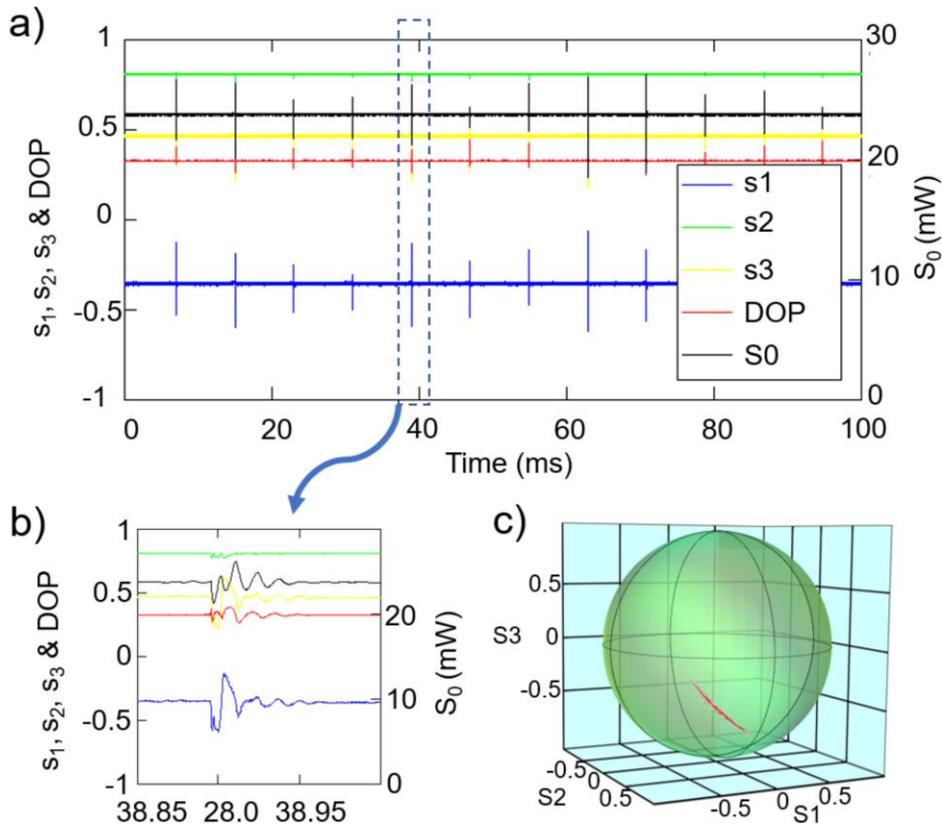

**Figure 2.8:** The normalized Stokes parameters $S_i$ (i=1,2,3), the output power So, DOP (a, b) and SOP evolution in the Poincar´e sphere (c) as a function of time.

Additionally, from the results obtained and from the Stock equations, and in particular from the equation 2.16 we can conclude that the phase difference is generally stable. If the phase difference, ($\Delta\varphi$), were otherwise not stable and varied randomly the average values of $\Delta\varphi$ would tend to 0 and therefore the total value of $S_3$ would tend to 0 as well. As we can see in Figure 2.8 a) and Figure 2.8 b) that is not the case.

## 2.4  Summary.

In conclusion, we designed and experimentally characterised a polarisation-multiplexed system capable of generating two optical frequency combs with a repetition rate difference in the range of a few hundred Hz. These dual-comb regimes exhibit relative stability, with a drift of 6 Hz over 6 hours (or 15 mHz in 60 seconds) without stabilisation. The use of a PBS can lead to an observed extinction ratio for the two combs of 16.5 dBm. The analysis of the polarisation dynamics and the combs' separation suggests that the system could be potentially used in dual-comb LIDAR and dual-comb polarimetric LIDAR.

However, the results from this laser system also highlight the need for improvement in the cavity design. These findings serve as a strong initial point for further development towards dual-comb LIDAR. Nonetheless, there are challenges in terms of stability, ease of generating the dual-comb regime, and the dual-comb separation, motivating the assembly of a new and improved laser cavity. This is addressed in Chapter 4 and Chapter 5 using the principles and lessons obtained in this chapter.

Moreover, to effectively use both combs for dual-comb ranging, they must satisfy another condition besides the separation (Aliasing Limit). The ratio between the fundamental repetition rate and the repetition difference between the two combs must be improved. Ideally, this can be achieved by reducing the length of the cavity and therefore increasing the $f_{rep}$ to a point where the aliasing condition is satisfied. Finally, to reduce frequency drift and increase stability, it is advisable to develop a housing for the laser to isolate it from external conditions. All of these improvements are implemented in the subsequent steps of this thesis, Chapters 4 and 5.

# Chapter 3

# Build-up and Propagation Dynamics of the Dual-Comb Regime in the Single-Cavity Fibre Laser

---

*The work presented in this chapter has been adapted from the following publication:*
[1] Alberto Rodriguez Cuevas, Igor Kudelin, Hani Kbashi, and Sergey Sergeyev. Single-shot dynamics of dual-comb generation in a polarization-multiplexing fiber laser. *Scientific Reports*, 13(1):19673, 2023.

---

## 3.1 Dispersive Fourier Transform as an analysis method.

The study of dual-comb formation and propagation within single laser cavities can yield valuable insights into understanding the dual-comb phenomenon and potentially improving its stability, generation, and control. Real-time observations of formation, commonly known as build-up dynamics, and propagation can be obtained from oscilloscope traces by properly timing and recoding the precise moments when these phenomena occur. The moments usually last over a few tens to thousands of round trips and are recorded as a continuous time-domain data package. They can be segmented into roundtrip intervals and organized into a two-dimensional matrix, creating a spatio-temporal representation of laser behaviour for each roundtrip [98]. These spatio-temporal representations are nothing but 2D maps from which it is possible to obtain complex temporal dynamics such as those seen in Raman lasers [99] and those seen in partially mode-locked lasers [98]. The spatiotemporal technique has also proven its effectiveness in observing phenomena such as laminar-to-turbulent transitions [100], the interaction of quasi-stationary localized structures [98], and rogue wave events [101]. More recently, it was used to study the initial dynamics of laser generation in mode-locked lasers [102, 103]. This method provides a straightforward way to capture fast-evolving dynamics in a single-shot measurement configuration. In a way this is the best diagnosis tool for complex dynamics which otherwise would be hidden and impossible to visualize and evaluate. Nonetheless, the level of information that we can extract from these 2D maps is directly related to the oscilloscope resolution or bandwidth. Modern oscilloscopes do not always have the capability to capture pulse dynamics over time with each roundtrip [98] as their spatio-temporal measurements are limited by their response time and by the response time of the photodetectors that are used with them. In fact, generally, the response time tends to be much longer than the duration of the ultrashort pulses being recorded. The following expression [104] estimates the temporal resolution of spatiotemporal measurements of an oscilloscope:

$$t_{\text{FWHM}} = \sqrt{t_{\text{FWHM PD}}^2 + t_{\text{FWHM DSO}}^2} \tag{3.1}$$

Here, $t_{F\,W\,HMPD}$ and $t_{F\,W\,HMDSO}$ represent the impulse response time of the photodetector and the digital storage oscilloscope, respectively. The response time is related to the rise time as $t_{F\,W\,HM} = 0.915 * t_R$. The response time, $t_R$, is the duration it takes for the signal to rise from a low value (typically 10% of the peak value) to a high value (typically 90% of the peak value). Currently, modern combinations of fast photodetectors and digital storage oscilloscopes have time resolutions in the range of tens or hundreds of picoseconds. One direct way to enhance the spatio-temporal resolution to yet another level is the use of the Dispersive Fourier Transform (DFT) method. DFT works on the principle of chromatic dispersion. When a pulse of light passes through a dispersive media, for example, standard telecommunication fibre, the pulse broadens due to the difference in propagation velocity of each wavelength given by the different interaction of each wavelength with the transmission media, dispersion. In normal dispersion media, the longest wavelengths will generally travel faster than shorter wavelengths, while in anomalous dispersion the opposite happens. In both cases, dispersion contributes to the pulse broadening. Eventually, each pulse spectrum is mapped to a temporal waveform that is analogue of the far-field (Fraunhofer) diffraction pattern in the spatial domain [105]. A common way of using DFT is passing a pulse through a dispersive element, like a long segment of fibre, and elongating the pulse's duration due to Group Velocity Dispersion (GVD). This process forms a spectral profile in the temporal domain (see Figure 3.1). Subsequently, a 2D map of optical spectral dynamics can be created, analogous to the spatio-temporal map. Therefore, when dealing with incoming pulses with complex temporal profiles that evolve rapidly in time the DFT technique efficiently captures these dynamic patterns.

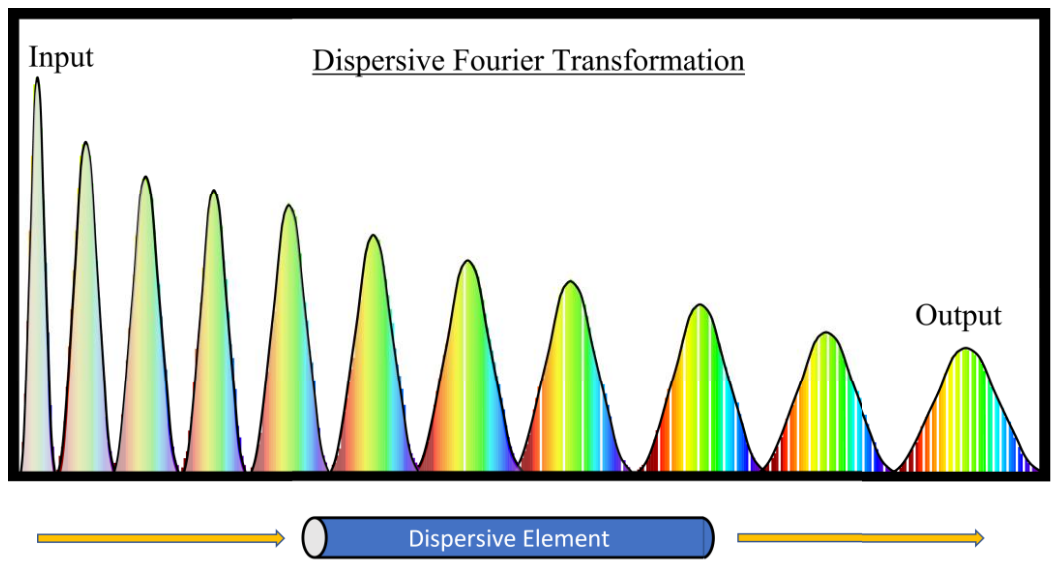

**Figure 3.1:** DFT, conceptual phenomenon representation. Spectroscopy technique that operates based on wavelength separation occurring as light propagates through a dispersive element with substantial group-velocity dispersion (GVD). This process maps the pulse spectrum to the temporal waveform of the pulse.

DFT enables single-shot recording of an optical spectrum directly from the oscilloscope trace. Unlike traditional tools for measuring optical spectra, such as optical spectrum analysers, which require lengthy integration times and offer data acquisition speeds on the order of a few Hz, the DFT technique allows for single-shot measurements at data rates in the tens and hundreds of MHz. Similar to the time-lens concept, DFT leverages the space-time duality and is analogous to paraxial diffraction. It is worth noting that even if DFT increases the resolution of the measurements, the temporal resolution of the DFT is limited by the sampling rate of the oscilloscope. Indeed, the spectral resolution of the DFT measurements, $\delta\lambda$, limited by the bandwidth of the used equipment, can be estimated as [106]:

$$\delta\lambda = \frac{1}{\Delta f_{OSC} \; D \; L} = \frac{1}{80 GHz \quad 0.0174 ns \; nm^{-1} \; km^{-1} \quad 11km} = 0.045nm \qquad (3.2)$$

where $\Delta f_{OSC}$ is the oscilloscope sampling frequency, D is the group velocity dispersion, $\lambda$ is the central wavelength, c is the speed of light, and L is the length of the dispersive media (for example, standard telecom fibre).

Being aware of the need for a better understating of the build-up and propagation phenomenon of dual-comb in a single cavity and being aware of the non-repetitive nature of the events that occur in ultra-short times when the dual-comb regimes are generated DFT was chosen as the best option to study these phenomenon. Previously, studies of similar soliton dynamics were done by several authors [107, 106, 108, 109, 110, 111, 112]. Some of them used DFT [113, 114, 115, 116, 105, 117, 118, 119, 120, 121, 122] since, as mention before, oscilloscope's GHz bandwidth can't provide real-time observation of the dual-comb build-up dynamics due to the fact that separation of two combs varies from hundreds of femtoseconds to tenths of nanoseconds. Some studies of the dual OFC build-up dynamics using DFT reveal different stages of dual-comb evolution, central wavelength shifts in opposite directions and vector-soliton collisions [113, 119]. Unlike the previous studies, herein, for the first time, we characterise the successful and unsuccessful dual-comb formation from spikes generated in the background noise, and the build-up dual-comb regime from a single-comb regime.

Collision dynamics of the solitons in the same dual-comb regime are also reported. Understanding how the dual-comb is generated inside the cavity can give us insights into optimal ways of generating dual comb with increased performance and optimise the relative long-term stability between each comb [117]. The importance of characterizing solitons is evidenced in other examples in the literature, where being able to fully characterise solitons or ultrashort pulses and having a reliable frequency conversion mechanism give way to an extensive number of applications: multi-photon microscopy, Raman spectroscopy, supercontinuum generation, frequency comb-based measurements or parametric-based two-photon interference among others [118].

## 3.2 DFT experimental setup.

The experimental setup is based on the one used in the previous chapter, Chapter 2. But in this case, few modifications have been introduced, see Figure 3.2. The total length of the cavity is 16 meters because one of the polarization controllers was replaced by another with the same characteristics but with a 1.3-meter reduction in connection fiber. Consequently, this led to an increase in the fundamental repetition rate to 12.657 MHz (79 ns). The repetition rate difference between each comb is around 255 Hz. The detection and measuring systems include a 50 GHz dual-window photodetector (XPDV2320R, II-IV) connected to a 32 GHz high-performance sampling oscilloscope (Agilent DSOX93204A Infiniium). Before the connection with the OSC, the experimental setup is connected to an 11 km-long standard telecom fibre with a group dispersion velocity of $17.4 ps\ nm^{-1}\ km^{-1}$. This 11-km 'link' allows the pulses that leave the dual-comb source to broaden and increase the spectral resolution of the measurements. The spectral resolution can be calculated by solving Equation 3.2, and with this setup, it is 0.045 nm. This means that every value of the OSC is separated by 0.045nm from the next one when evaluating the spectral shapes.

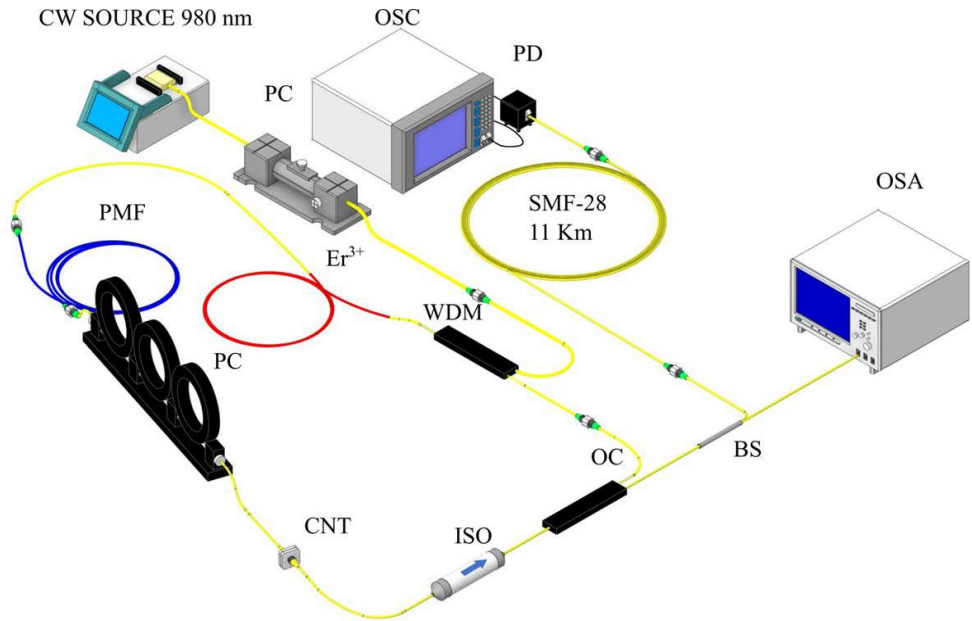

**Figure 3.2:** Representation of the laser cavity consisting of the following parts: isolator (ISO), optical coupler $90-10$ (OC), wavelength-division multiplexing (WDM), polarisation controller (PC), carbon nanotube (CNT), 1 m of polarisation maintaining fibre (PMF), 0.8 m of Erbium-doped fibre (Er3+), and a continuous wave laser pumping at 980 nm (CW SOURCE). In combination with a 1x2 beam splitter (BS) an optical spectrum analyser (OSA), and an 11km-long SMF-28 segment connected to a photodiode (PD) that is connected to an oscilloscope (OSC).

To obtain the necessary data, we meticulously adjusted the polarisation controller to establish a stable dual-comb generation regime. Subsequently, the trigger of the oscilloscope was set to a level where overlapped pulses would activate it. Then, the pump laser was switched off and on within a brief period of time. The newly generated combs triggered the oscilloscope, allowing us to record a 12.7 ms ($\sim$ 321500 round trips) trace both before and after the trigger activation. This provided valuable insights into the pulse onset dynamics, from initial noise to stable dual-comb generation. To investigate the build-up dynamics from a single comb to a dual comb, we employed a similar approach. Instead of switching the laser on and off, we strategically moved the polarisation controller to a position where the dual-comb generation occurs. Subsequently, we slightly adjusted the polarisation controller to transition back to a single comb regime. The trigger was then set, and the polarisation controller was rotated back to its original position. Through this meticulous method, we accurately recorded the dual-comb formation, as the trigger activated only when both pulses perfectly overlapped with each other.

In addition to the DFT intensity dynamics map, we can obtain the autocorrelation trace simply by doing the Fourier transform of the intensity dynamics map. This autocorrelation trace provides us with more information about the pulse dynamics. The span in the AC trace can be calculated using the following formula. Being $\lambda$ the centre wavelength of the spectrum, $\Delta\lambda$ the wavelength bandwidth, and $\Delta_\nu$ frequency bandwidth.

$$\Delta_\nu = \frac{c}{\lambda^2}\Delta\lambda = \frac{299792.458 \; nm/ps}{(1560nm)^2}0.8nm = 0.1 \; ps^{-1} \qquad (3.3)$$

Thus, a frequency bandwidth of $0.1 \; ps^{-1}$ means 1 cycle every 10 ps and so that is the total span of the autocorrelation trace.

## 3.3   Laser build-up and propagation dynamics.

Using DFT, the dual-comb build-up dynamics are recorded after the pump laser was switched on and this result is presented in Figure 3.3. Initially, in the system, there is just background noise, but after several thousand round trips, and due to the effect of the nonlinear absorber, some ultrashort spikes start to become dominant over the rest of the noise. Most of them collapse but some of them grow to a point where they experience spectral broadening. Eventually, they split into two independent and stable pulses. In this build-up process, we can observe that previous spikes get wider and split into two pulses but eventually collapse short after, see Figure 3. The observation in Figure 3 is not unique. Several energy spikes repeat the same process immediately after each other until one of them successfully separates into two. Once the two pulses start propagating along the cavity, the nonlinear absorber filters out the majority of the background energy noise and prevents the generation of more energy spikes or additional pulses.

The whole process of dual-comb generation from the point where the narrow spike appears took approximately 3500 round trips (276 µs) and it occurs in two steps. The time between each step took roughly 227 round trips or (17.93 µs). This two-step generation can be clearly identified in the pulse energy graph (Figure 2b), where the second energy increase is slightly higher and wider than the first one. After the second intensity peak, the overall intensity tends to get stable over time in correlation with the stability of propagation of both combs along the fibre ring cavity. The initial spike that triggers the formation of the dual-comb experiences a steady increase in energy. At a certain point, when this energy reaches approximately 20% of the final energy, the spike's energy undergoes an exponential increase. It surpasses the final energy by around one-third and subsequently decreases to a minimum level, which is approximately 25% lower than the reference value. This pattern repeats, with the spike's energy peaking at approximately 33% above the reference value before decreasing to a valley around 25% below it. Following this second valley, there is a gradual increase in energy until eventually reaching a stable regime, allowing the two solitons to propagate independently. The pulse energy refers to the sum of all the energy in the laser cavity during one round trip. This includes not only the energy in the laser pulse but also any noise or fluctuations that may be present in the cavity. To calculate it we divide the original 12.7 ms-long OSC trace into segments of 79.004968 ns (round trip time of the slow pulse). Then we sum the dB of each point inside these segments and plot it with the corresponding time. The obtention of the Auto-Correlation Function (ACF) consists of iterating through each row of the original DFT data and performing the inverse Fourier transform in each of them. The absolute values of each iteration are appended to the new matrix and represented in the figures.

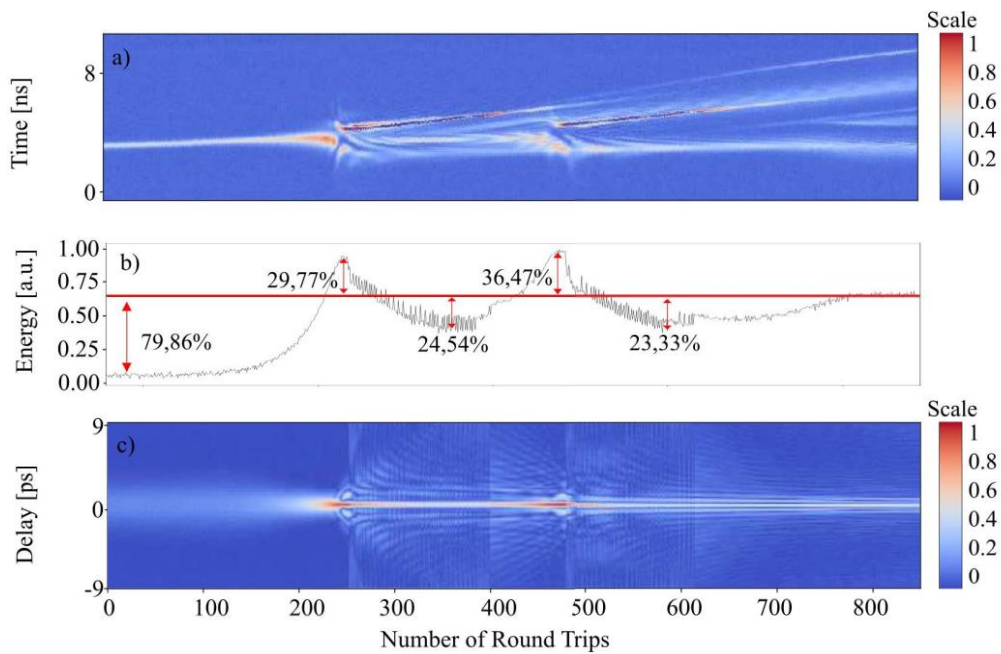

**Figure 3.3:** The build-up dynamic of the dual comb. A) Intensity dynamics obtained from the oscilloscope. B) The corresponding pulse energy. C) First-order autocorrelation trace.

In Figure 3.4, unlike in Figure 3.3, the separation of the spike into two pulses collapses and the dual-comb generation is unsatisfactory. Indeed, the energy evolution can provide us with additional information about that process. After the first energy increase the energy decays quickly and eventually collapses into the background noise. The collapse of this initial spike is immediately followed by the formation of the dual comb from a completely independent spike as it is shown in Figure 3.3, indeed the spike that collapsed and the spike that succeeded coexisted in competence just until one of them succeeded in becoming an actual dual comb traveling along the cavity. This is clearly seen in the Appendix C. Nonetheless, the pulse energy shows us that the first part of the build-up dynamics is roughly the same with an exponential increase in the spike's energy from 10% of the reference value to over 33% and then to a valley 25% below. However, once the energy of the spike gets to this point it fails to generate a second peak and its energy starts to decrease until the peak eventually vanishes. In short, there is an energy trigger that needs to be surpassed to achieve dual-comb generation. The significance of energy dynamics during build-up is similar to dynamics in the other dual-comb systems [108].

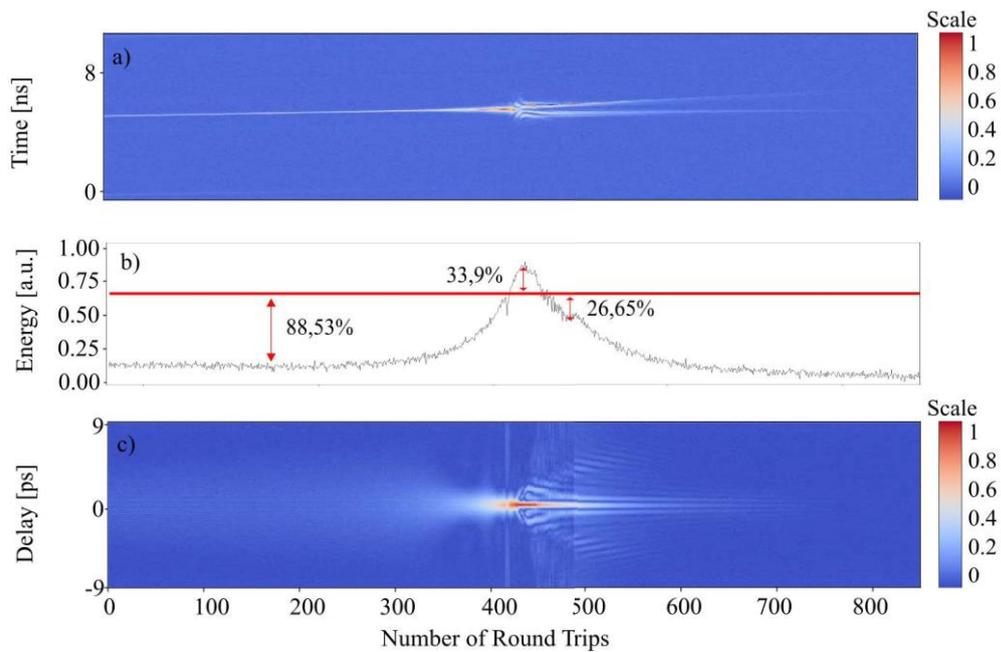

**Figure 3.4:** The build-up dynamic of the dual-comb for one of the spikes that collapse before the dual-comb formation. A) Intensity dynamics obtained from the oscilloscope. B) The corresponding pulse energy. C) First-order autocorrelation trace.

A successful dual comb build-up produces a frequency difference, measured in the RF spectrum analyser, of 255 Hz equivalent to a difference in pulse delay of 1407 ps per round trip. The round-trip time of the slow axes is 79.004968 ns while the round-trip time of the fast axes is 79.003561 ns.

When it comes to the collision of the solitons within the cavity once the dual-comb regime has been achieved the collision does not affect the shape of either soliton, as shown in Figure 3.5 a) and Figure 3.5 b). However, we have observed a case where a soliton molecule composed of multiple solitons can lose one of them after the interaction. Nonetheless, among the multiple observations of collision that have been captured with DFT there is not a single case where the intensity lines change the trajectory. Indicating that the collision does not alter the propagation velocity of the pulses in any way. Pulse collision information is highly relevant from an application perspective since some applications based on dual-comb such as dual-comb lidar require two independent combs that remain stable while they are colliding with each other [36]. This pulse behaviour can be observed in the autocorrelation traces of Figure 4b. Further observations of soliton molecules reveal a more unstable behaviour. Temporal evolution of the build-up dynamics and soliton collision are included in Appendix C.

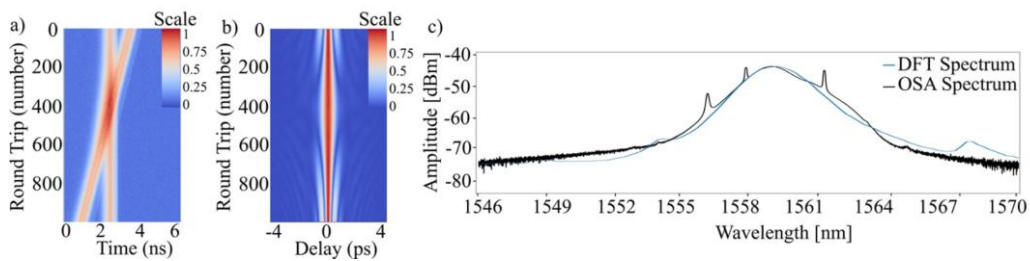

**Figure 3.5:** The interaction of the dual-comb at the moment of collision. A) DFT trace of the collision. B) Autocorrelator trace of the collision. C) Optical spectrum of the light pulses obtained with the OSA in black and with the DFT in blue.

In addition to the build-up and propagation dynamics, DFT can reveal important short-term spectral dynamics. These dynamics are important because soliton collisions could induce short-term instabilities which might affect the measurements [120]. Figure 3.5 c) offers a comparison of the optical spectrum of the pulses derived from the DFT and the optical spectrum captured with the OSA. When the optical spectrum is obtained with the DFT it must be averaged over a number of round trips, in our case 300. These round trips are acquired when the dual-comb regime is stable. The time-to-wavelength correspondence in the DFT intensity dynamics map is 1 ns equivalent to 5.214 nm. However, when applying these conversions to the figures that represent time vs DFT it must be taken into account that this conversion can only be applied directly when we have a section of the map with a single soliton, in the case there are more than one propagating at different velocities it can lead to a mistake since this time-to-wavelength conversion can only be applied to one of them.

Another important dynamic occurs when the dual-comb is generated from a single comb. In this last case, a new comb splits off from the original one, without altering the speed of propagation of the original one. Having a mode-locked regime with a single pulse per round trip we can adjust the polarisation controller inside the cavity to a position where two different trains of pulses start to circulate along the cavity. These pulses circulate at a slight difference in repetition frequency due to the polarisation-multiplexing nature of the fibre cavity. Figure 3.6 presents the build-up dynamics for this case. The AC traces obtained by calculating the Fourier transform of the intensity map show a pulse pattern that evolves in a way that the slow axis loses its multi-pulsing after some round trips while the fast axe develops a multi-pulsing pattern almost simultaneously.

Figure 3.6 a) and more closely Figure 3.6 b), show evidence of some differences with respect to the build-up dynamics shown in Figure 3.3. On the one hand the 1-to-2 transition is more chaotic than the build-up and the energy fluctuates for a longer period of time. On the other hand, there is a pulse energy transition from the slow to the fast axes after roughly 3000 round trips from its creation. The main reason for these two phenomena is the adjustment in the intracavity polarisation produced when the polarisation controller is moved during the process. While in the case of the build-up, the polarisation controller remains always in the same position. This difference can explain the succession of peaks and valleys presented in the energy level right before and after the separation into two combs and evidences the need for high precision electronically-driven polarisation controller. The original pulse (single-comb) becomes highly unstable in the round trips previous to its separation. Its pulse energy shows a noisy behaviour while the AC trace indicates complex multi-pulsing patterns. Once the separation has occurred and for the first 1000 round trips there are several consecutive peaks and valleys of energy. Another difference seen in Figure 3.6 c) and Figure 3.6 d) is the energy stability and pulse transition between each comb. The slow comb comes out of the separation with noisy behaviour while it transits to a more stable and less energetic state. Simultaneously, the fast comb experiences the opposite transition, increasing its overall energy and the autocorrelator trace shows an evolution from a single pulse towards a complex and unstable pulse distribution. This energy transfer from one comb to another can be explained by the change in the polarisation controller and therefore in the birefringence that affects the overall pulse propagation.

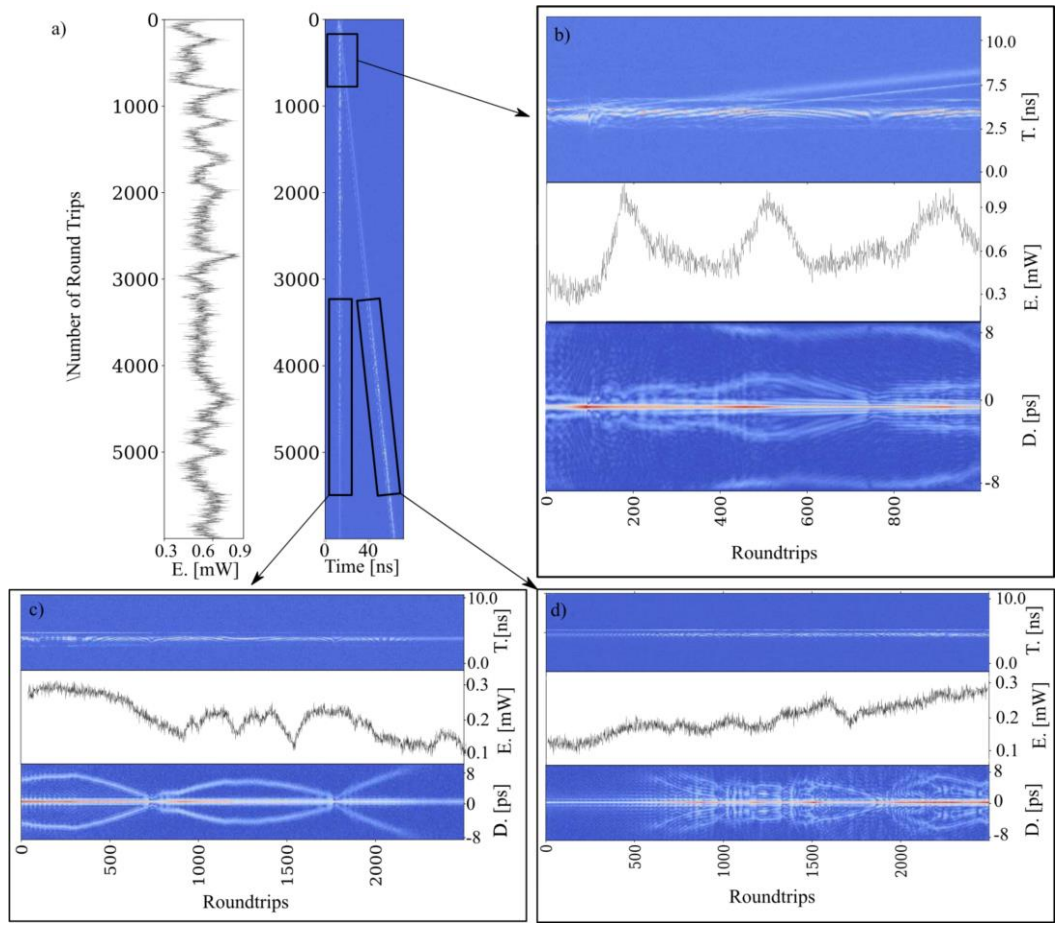

**Figure 3.6:** Dual-comb generation from a single comb regime. a) Dual-comb generation obtained with DFT and its corresponding pulse energy. b) Zoom-in of the actual dual-comb generation section, pulse energy, and autocorrelator trace. c) Zoom in on the slow axes section where there is an energy transition, its pulse energy, and its autocorrelator trace. d) Zoom in on the fast axes section where there is an energy transition, its pulse energy, and its autocorrelator trace.

Propagation and build-up dynamics are conditioned by the stability of the energy levels in every round trip. Internal variations in the energy level are present in both generation and collapse of the dual comb and therefore are intrinsically related to the stability of the regime. Therefore, an essential consideration is that any external perturbation that influences the energy level inside the cavity such as vibrations, or changes in the environmental conditions such as temperature or pump power instability can trigger the collapse of the dual-comb regime. To investigate this effect, we recorded several DFT maps every 10 min of a stable dual-comb regime that lasted for an hour, as well as in vivo observations. Over the first 56 min, the regime was stable, and the intensity dynamics of the DFT maps and OSA spectrum showed no differences, except for changes in the propagation speed resulting in differences in the round-trip time of 5-10 fs per hour, or 0.8 to 1.6 Hz in repetition rate frequency.

$$RT[ns] = 1/f_{rep1}[GHz^{-1}] \tag{3.4}$$

$$RT_1[ns] - RT_2[ns] = 1/f_{rep1}[GHz^{-1}] - 1/f_{rep2}[GHz^{-1}] \tag{3.5}$$

$$1/0.012931034[GHz^{-1}] - 1/0.0129310356[GHz^{-1}] = 0.0000095[ns] = 9.5[fs] \tag{3.6}$$

However, in the last 4 minutes, increased instability was observed, with clear variations in the light spectrum composition between successive samples obtained every 2 s. The more frequent and intense these variations were, the closer the dual-comb regime was to collapsing into a single-comb regime.

These instabilities in the spectrum composition can be seen as an early warning signal and can prevent the collapse of the regime by introducing corrective measures that extend the lifespan of the dual-comb regime. It should be noted that the authors of this research consistently observed the pattern of increased instability leading to a dual-comb collapse in previous works [2]. This observation opens the door to evaluating the influence of programmed adjustments in the system's power and internal polarisation in an effort to ease the dual-comb generation once the birefringence of the cavity has been adjusted to adequate levels.

In addition to all of these points, it is worth mentioning that an improvement in the tunability of the dual-comb can be developed by using the output autocorrelation trace resulting from DFT data combined with a machine learning-based evolution algorithm and electronically-driven polarisation controller such as in the following work [121].

In this chapter, it is important to note that additional data was generated during the course of this research. However, the data presented here and in the associated publication represents the highest quality and most relevant findings. This selective presentation ensures that the conclusions drawn are based on the most accurate and insightful data available.

## 3.4  Summary.

In this study, we have demonstrated the spectral and temporal dynamics of dual-comb formation within a fibre ring-cavity laser, from a mode-locked regime and from power off to the establishment of a stable dual-comb propagation along the cavity. Unlike the previous study of the dual-comb build-up dynamics [119, 113], our findings reveals the mechanism of successful and unsuccessful dual-comb formation from spikes generated in the background noise. The pulse energy of the initial short-lived mode-locked states plays a critical role in the build-up dynamics. If the energy level decreases under a certain level it tends to collapse, while if it remains over a certain value there is a successful separation into two orthogonally polarized independent pulses. Thus, these experimental observations pave the way to understanding the formation of a dual-comb regime in the same cavity in the context of the development of industrial-grade dual-comb laser sources.

# Chapter 4

# Laser Optimisation Towards Application

## 4.1    Optimisation requirements.

The stability results obtained in Chapter 2 highlighted the necessity to work on a more stable cavity able to preserve the dual-comb regime over hundreds of hours so it can be used in real applications. Moreover, this necessity also highlighted the fact that the entire system needed to be assembled within a portable housing so it can be used outside the lab. In addition to these two improvements, we needed to consider one of the original applications proposed for this dual-comb and that is ranging. To be able to accurately measure distances with high speed and high precision using dual-comb ranging we need to make sure the condition called aliasing limit is fulfilled. If that condition is not granted, then the dual-comb ranging will not provide its full potential [123]. That means it will not provide distances to the interferometric level [36]. That condition is presented in the following equation, 4.1, obtained from [123]:

$$\Delta f_{\text{rep}} < \frac{0.8\, f_{\text{rep}}^2}{2\, \Delta \nu} \tag{4.1}$$

Where $\Delta f_{\text{rep}}$ is the difference in repetition rate between the two combs, $f_{\text{rep}}$ is the fundamental repetition rate and $\Delta \nu$ is the common optical frequency bandwidth. If that condition is not passed but the condition of this second equation 4.2 is fulfilled the dual-comb ranging is still possible but with certain difficulties and applying other techniques such as dual-photon dual-comb, also obtained from [123], or some signal processing algorithm.

$$\Delta f_{\text{rep}} < \frac{4 f_{\text{rep}}^2}{2 \Delta \nu} \qquad (4.2)$$

However, if none of the two equations 4.1 or 4.2 are satisfied it won't be possible to use the system for dual-comb ranging. Since the pulses won't coexist together for enough time.

These conditions are referred to as the aliasing limit. They arise from the fact that the interference between two combs with repetition rates of approximately $f_{\text{rep}}$, sharing a common optical frequency bandwidth $\Delta \nu$, generates radio frequencies that map uniquely onto optical frequencies only when the repetition-rate difference $\Delta f_{\text{rep}}$ is below a maximum value of $\frac{f_{\text{rep}}^2}{2 \Delta \nu}$, As this limiting value is approached, the interferogram contains frequencies that approach DC, which modulate its shape and render time-of-flight analysis from envelope extraction unreliable [123].

In a similar way, the aliasing condition can be referenced in relation to the pulse width $\Delta \tau$ , as in the equation 2.12. This is, in fact, a better approach when the common optical bandwidth is relatively small such as in this case or when the interferometric measurements are not used. For the laser presented in Chapters 2 and 3 the proportion between $\Delta T$ and $\Delta \tau$ is $1.05 \Delta T \approx \Delta \tau$ (For a $\Delta f_{\text{rep}}$ of 200Hz, a $f_{\text{rep}}$ of 11.95MHz and a $\Delta \tau$ of 1.5 ps). This means that the ratio between repetition rate frequency and fundamental frequency needed to be ideally improved by at least 10 times for a constant pulse width.

In order to obtain a more favorable ratio between the repetition rate difference and the fundamental frequency I decided to reduce the cavity length as much as reasonably possible to increase the fundamental frequency of both combs, while at the same time reducing the PM fibre to compensate for the relative increase in the average birefringence of the cavity. This chapter describes the improvement in the laser system of chapter number 2 and 3.

## 4.2   Optimised setup design.

After an initial study where different cavity lengths were tested in combination with different lengths of the PM fibre, the resulting cavity consisted of the following components. The optimised laser cavity described in this chapter is presented in the 4.1 and follows the same structure as the one in Figure 2.2. It comprises a 5.25-meter-long cavity that is pumped by a 980 nm laser diode. The Wavelength Division Multiplexing (WDM) at the cavity input combines the continuous wave light of the laser at 980 nm with the pulses already circulating within the cavity at the 1560 nm wavelength range. Inside the cavity, there is a segment of 45 cm of $Er^{3+}$-doped fibre (ER110-4/125). A 0.5-meter PM-1550 fibre, possessing a GVD of $-0.030$ $ps^2/km$, facilitates the generation of a polarisation multiplexing mode-locked regime by adjusting the polarisation states within the cavity using an in-cavity polarisation controller positioned after the PM fibre. To ensure unidirectional propagation, a 51 dB dual-stage polarisation-independent optical isolator (Thorlabs IOT-H-1550A) is located in the laser cavity. Additionally, a film-type homemade single-wall carbon nanotube (CNT) absorber (inserted between two FC/APC connectors) serves as the main mode locking component. The cavity concludes with an output fibre coupler redirecting 25% of the light outside the cavity. When comparing the structure of this cavity with the cavity used in previous chapters 2 and 3, we can highlight four main changes. The cavity is shorter, 5.25 m compared to 17.3 m and 16 m in previous occasions, and so the fundamental repetition rate is now 39.25 MHz (25.47 ns round-trip time) instead of 11.95 MHz (83 ns round-trip time) and 12.657 MHz (79 ns round-trip time). Secondly, to reduce the cavity length, a shorter segment of Er-doped fibre, 0.45 m, is used but with a higher concentration, Er110 (GVD = 0.0135 $ps^2$/km), to compensate for the reduced gain. This results in a net anomalous dispersion GVD for the entire cavity of -0.0201 $ps^2/km$. Thirdly, the higher output percentage in the coupler, in combination with the shorter cavity length, generates a slightly smaller output power in the cavity, 0.27 mW, in comparison to 0.57 mW. Nonetheless, the circulating power inside the cavity is much smaller, according to our estimations, 5.3 times smaller (0.11 mW compared to 5.7 mW). This lower intra-cavity optical power helps with stability since the CNT can be fragile to high peak power.

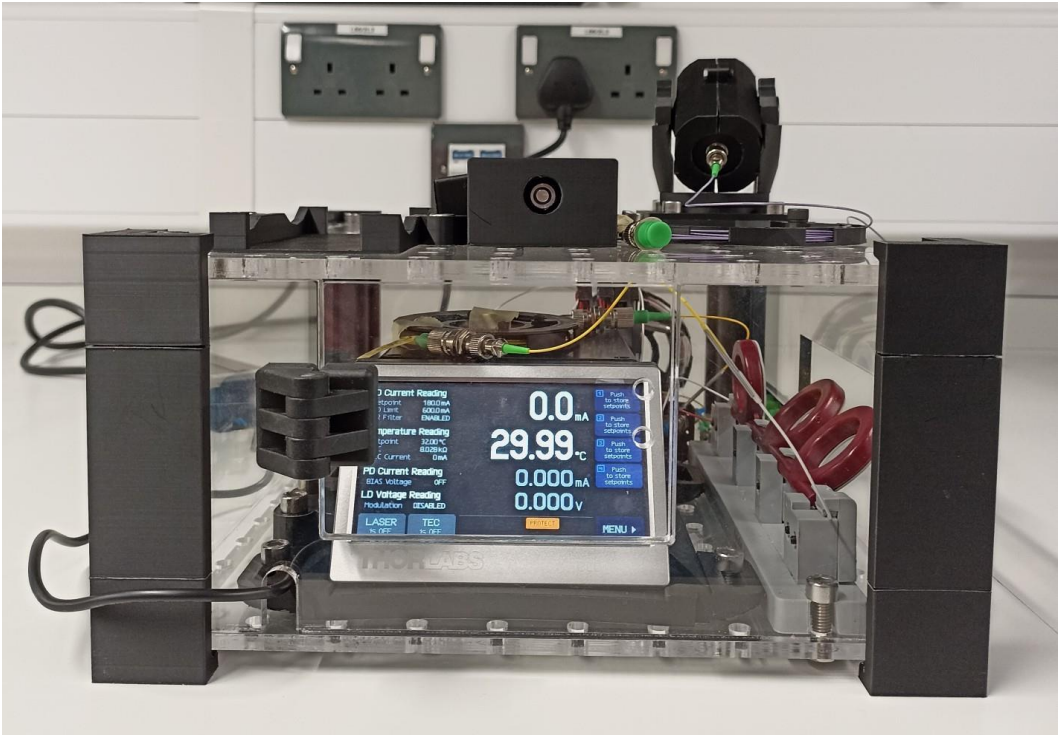

**Figure 4.1:** Dual-Comb Laser System assembled within a custom-made housing

In addition to the previous improvements in the cavity, the entire laser system was assembled within a custom-made housing laser cut in transparent Polycarbonate plastic. The complete housing design can be seen in the Appendix D. In addition to the surrounding plastic, many holders were 3D printed using Black ABS to improve the support of optical fibres and optical components of the laser.

## 4.3 Stability of the optimised cavity.

This combination of components resulted in the most stable and easy-to-obtain dual-comb regime of the ones tested. Moreover, since the beginning, it was clear that the dual-comb was easier to obtain than when using the previous cavity. In this section, we present the stability results of the dual-comb regime generated in the cavity of Figure 4.1. These results are classified into long-term stability and short-term stability.

### 4.3.1 Long-term stability.

In terms of the RF spectrum stability, we observe similar behavior to the previous cavity when evaluating the $\Delta f_{rep}$ stability over a given period of time. Nonetheless, in this case, it is evaluated after 125 hours. Initially, over 53.5 hours, the difference between the combs progressively increases at a rate of 0.6257 Hz per hour, elevating from 1108.725 Hz to 1142.2 Hz. Subsequently, a decreasing stage spanning 20 hours sees the frequency difference decrease to less than 1125.5 Hz. Notably, the highest drift deviation, lasting 4 hours, is reached at 2 Hz per hour. Modifications made to the laser setup compared to previous demonstrations 2, 3 have substantially improved stability by a factor of 17, extending it from 7 hours to 125 hours. This is a notable aspect of our laser compared to other works in the literature that tend to show similar results but for periods of time of 1 to 2 hours [80, 83, 124, 125].

Figures 4.2 and 4.3 illustrate the spectral and frequency stability of the laser cavity used in this work. Both vector solitons coexist for over 125 hours, although with notable fluctuations influenced by external temperature variations. The OSA stability results show an initial phase of instability within the first three measurements, 150 mins, followed by a stable phase and at the end, the OSA spectrum suffers the main instability accelerating towards the end. Usually, stability in OSA is the best indicator of the long-term stability of the dual-comb regime. The refined regime within the cavity not only becomes more accessible but also offers superior control.

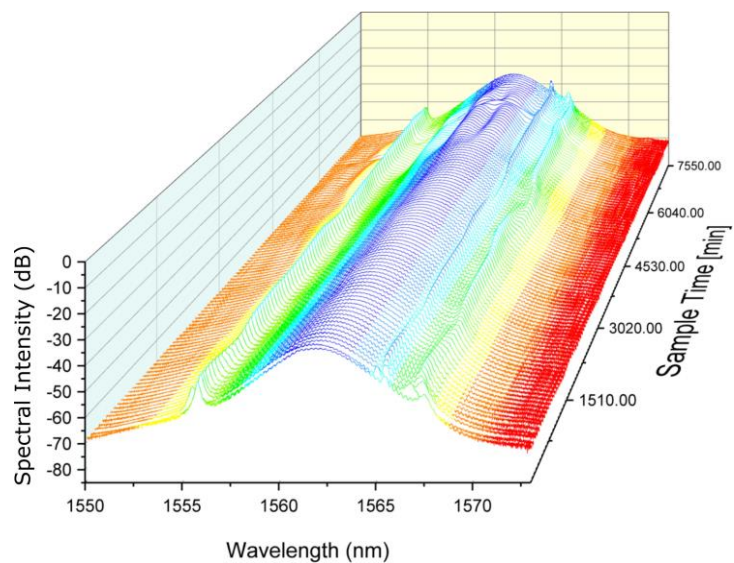

**Figure 4.2:** Dual comb stability over time. OSA Stability

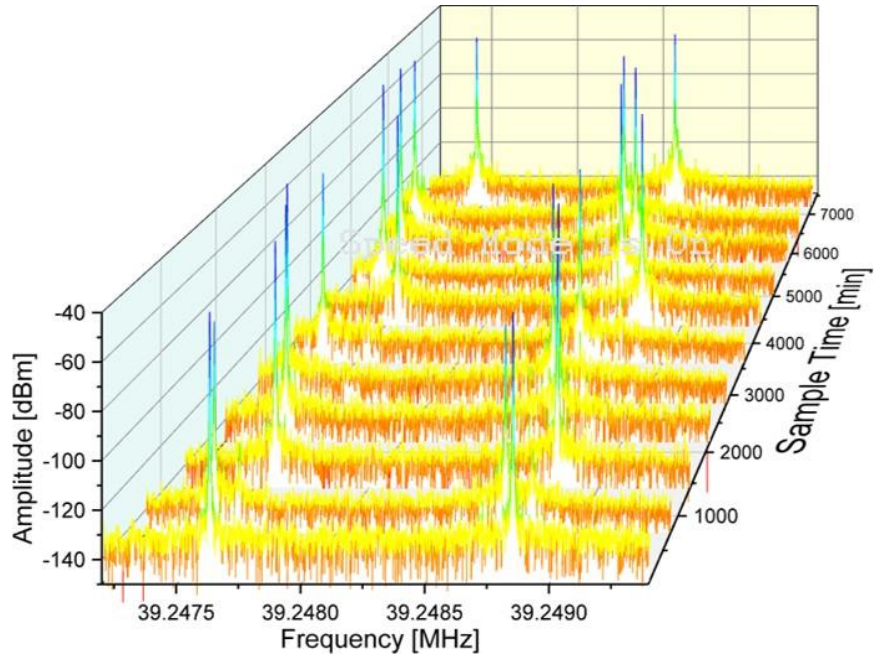

**Figure 4.3:** Dual comb stability over time. RF Stability

Figure 4.4 presents the results for dual-comb temporal stability, namely for dual-comb operation with a repetition rate difference of 1100 Hz. The dual-comb was monitored for 125 h at room temperature conditions. The regime tends to remain the same after that time, with a slight shift in repetition rate and central wavelength. Even though there is drift over time in the carrier-envelope offset frequency, the offset between the two combs shows stability with a drift of 1 Hz per hour, going from 1120 Hz to 1150 Hz unevenly without any additional external stabilisation. The resolution bandwidth (RBW) of the RF spectrum analyser was 1 Hz. Therefore, the short-term stability over 1 min can be estimated as 16 mHz.

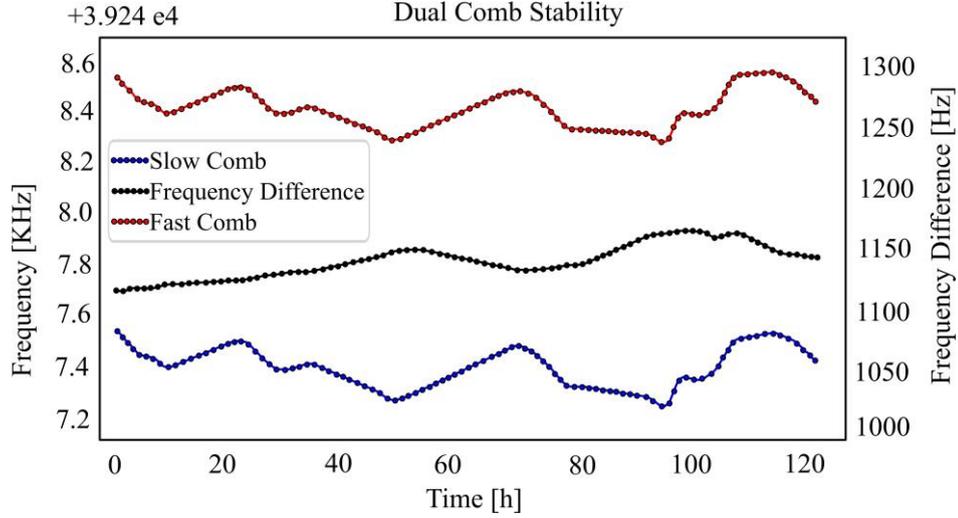

**Figure 4.4:** Dual comb $\Delta f_{rep}$ stability over time.

In the case of this new cavity, the aliasing limit is reduced down to 1.4 considering a $\Delta f_{rep}$ of 820Hz, a $f_{rep}$ of 39.25MHz and a $\Delta\tau$ of 1.28 ps. This means that the ratio between repetition rate frequency and fundamental frequency has been improved to the point where it is close to one-third of the pulse width, $2.4\Delta T \approx \Delta\tau$ . However, it is still not 10 times smaller, the limit that allows for interferometric measurements 2.12.

$$\Delta T = \frac{\Delta f_{rep}}{f_{rep}^2} = \frac{820Hz}{(3.92474 \ 10^7)^2 Hz^2} = 0.53 ps \tag{4.3}$$

Therefore, we decided to use DFT to expand the pulse so it fulfills the interferometric condition (pulse width of 21.39 ps and $40\Delta T \approx \Delta\tau$ ), more details will be explained in Chapter 5. I also used DFT in a similar way to the one done in our previous work [1] to obtain the intensity dynamics presented in Figure 4.5. These results show that the mode-locking of both combs is stable and pure from round trip to round trip since the lines are straight and keep the shape. It can also be appreciated that one of the lines moves away from the other, meaning that one of the combs travels slightly faster than the other within the round trip. In fact, the exact round trip time of the fast axis is 25.4800555666 ns while for the slow axis, it is 25.47954342786. It results in a round trip difference of 512.14 ps, (compared to the 1407 ps difference of the previous cavity). Unstable shapes or broadening and narrowing of the lines would have meant that the mode-locking is unstable, nonetheless, that doesn't appear in this regime, giving another proof of its stability. In this case of its short-term stability.

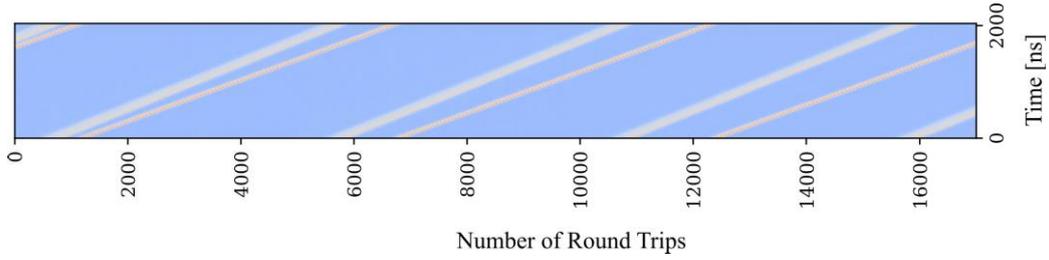

**Figure 4.5:** Intensity Dynamics of the dual-comb obtained using Dispersive Fourier Transform over a period of 17000 roundtrips.

Additionally, using Equations 2.4, 2.5, 2.6, 2.7, and 2.8 from Chapter 2, we can calculate the refractive indexes to be $n_{slow}$ = 1.454973 and $n_{fast}$ = 1.45494112 for the slow and fast axes, respectively. The refractive index difference is $\Delta n$ = 3.188 × 10⁻⁵. The difference in the propagation constant is $\Delta L_B$ = 128.49 m⁻¹, the beat length is $L_B$ = 0.049 m, and the phase delay is $\Delta\phi$ = 674.55. Using the Equations 2.9 and 2.10 we can calculate the FSR in frequency equal to 196.24 Hz and 0.159 pm in wavelength.

### 4.3.2 Short-term stability.

Figure 4.6 presents the RF phase noise measurements of the optimised laser cavity, highlighting significant improvements in the system's stability. Specifically, the phase noise has decreased from -70 dBc/Hz to -80 dBc/Hz at a 100 Hz offset frequency, indicating a quieter and more stable laser operation. These improvements are attributed to several modifications: reduction in total cavity length, decrease in gain media length while increasing doping levels, secure winding of optical fibres within 3D printed spools to prevent movement, and external isolation of all components within a custom housing, mitigating external perturbations. Similar single-cavity schematics to the one presented here tend to show phase noise around the same range. For example Kowalczyk et. al., [124] also presented -80 dBc/Hz at a 100 Hz offset frequency while in their case the phase noise decreases more notably when they expand the offset frequency and the effect of the beat notes disappears. In their case, their SNR is also around 60 dBm or quite similar to our case and the same happens with the $\Delta f_{rep}$ drift that remains in the order of 2Hz varying rapidly but remaining within this values over a 60 min sample.

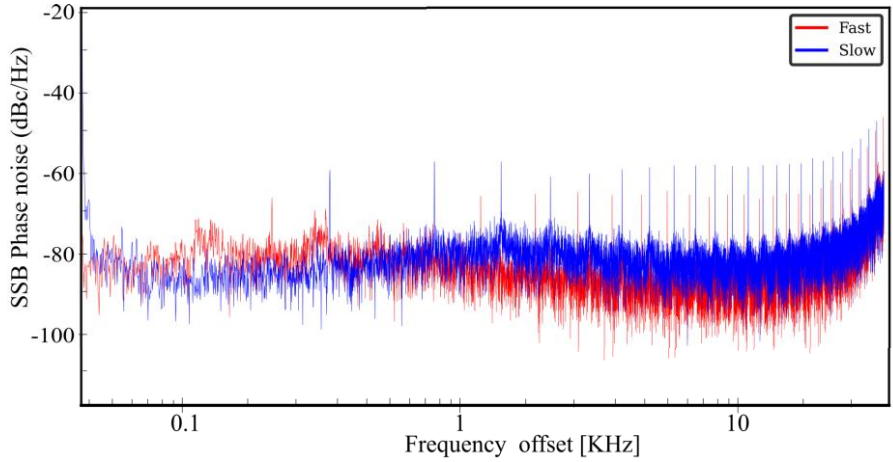

**Figure 4.6:** RF phase noise

Figure 4.7 illustrates the Allan variance and Allan deviation of the optimised laser system, which are key metrics used to evaluate the stability of frequency sources over time. The Allan variance and deviation provide insights into the frequency stability of the laser over various time intervals and indicate how the frequency of the laser changes over time. The data presented shows that the optimised laser system maintains a frequency output typical of unstabilized lasers (in the other of $10^{-3}Hz$, or free running combs, as compared to stabilized OFCs that generate a single frequency comb and are available with stabilities in the order of $10^{-13}Hz$ for a second average time. Nonetheless, as it has been shown before both combs drift together, and thus the mutual stability is superior to the individual of each comb. This stability is crucial for dual-comb applications. The Allan variance ($\sigma_y^2(\tau)$) is a measure of the frequency stability of a signal over a specified averaging time ($\tau$). It is defined as:

$$\sigma_y^2(\tau) = \frac{1}{2} \left\langle \left( \bar{y}_{n+1} - \bar{y}_n \right)_2 \right\rangle$$

(4.4)

where $\sigma_y^2(\tau)$ is the Allan variance for the averaging time $\tau$; $\bar{y}_n$ is the average fractional frequency over the $n$-th measurement period, and h·0 denotes the expectation value or the average over many measurements. The Allan deviation ($\sigma_y(\tau)$) is simply the square root of the Allan variance:

$$\sigma_y(\tau) = \sqrt{\sigma_y^2(\tau)}$$

(4.5)

where ($\tau$) is the time interval (averaging time) over which the frequency measurements are averaged. It can range from very short to very long periods, depending on the application and the desired stability analysis. $(\bar{y}_n)$ represents the normalized frequency deviation during the $n$-th averaging time $\tau$. It is the fractional frequency and is typically calculated as the difference between the measured frequency and the nominal frequency, divided by the nominal frequency. And $(\langle \cdot \rangle)$ denotes the statistical average over multiple measurement periods. It ensures that the Allan variance and deviation represent typical behaviour rather than fluctuations from a single measurement. While the Allan variance provides a measure of the spread or dispersion of frequency deviations over time, the Allan deviation offers a more intuitive understanding of this spread by presenting it in the same units as the original measurements. We can find more info about them in the 'IEEE Standard Definitions of Physical Quantities for Fundamental Frequency and Time Metrology–Random Instabilities [126].

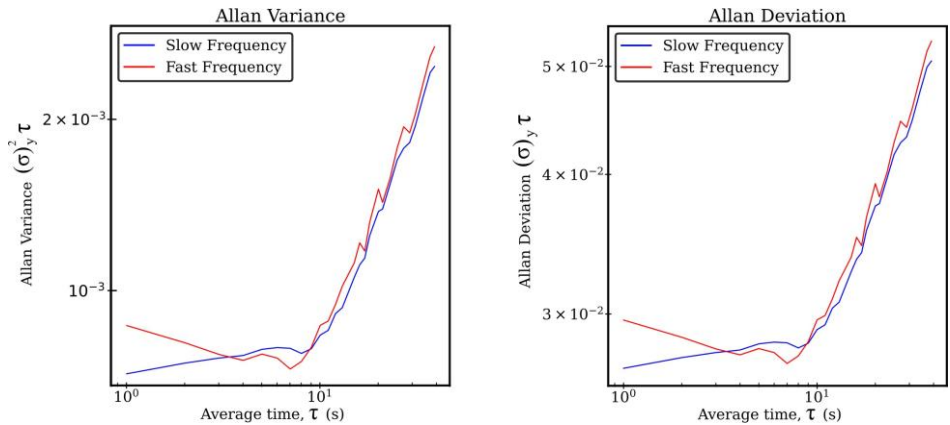

**Figure 4.7:** Allan Variance and Allan Deviation

## 4.4 Discussion and summary.

In addition to the results already presented, several observations and discussions are worth mentioning. A notable difficulty encountered in this system, as predictably happens in many other lasers with polarization controllers (PCs) inside the cavity [78, 80, 82], is the challenge of adjusting the polarization within the cavity accurately. It is not possible to measure the polarization within the cavity, and often, the same position of the three paddles in the PC does not produce the same regime. Thus, while it is possible to generate a dual-comb regime by adjusting the polarization in a specific way, replicating that regime using the same PC adjustment can fail. This randomness in the generation of mode-locking regimes and the inability to determine the polarization within the cavity makes it difficult to isolate variables and draw definitive conclusions about the influencing parameters and their effects.

However, several observations have been consistently noted throughout this thesis. First, the temperature of the pump (980 nm laser diode) significantly impacts stability; in our case, 32 degrees Celsius produces the most stable regimes. Second, achieving dual-comb regimes is easier right after turning on the pump, particularly within the first 5 minutes; beyond this period, it becomes more challenging. Third, the build-up dynamics of the dual-comb regime can be facilitated by a slight adjustment to one of the three paddles in the cavity. Once this regime is achieved and stable (as indicated by clear stability in both the OSA and RF signals), minor readjustments of the paddle positions do not collapse the regime, even when returning to the previous position. These behaviour patterns have been consistently observed during this work. However, no structured study has been conducted to analyse these behaviours, so these observations should be taken as preliminary findings.

Additionally, it was observed in this cavity, as well as in those from previous chapters that comb regimes with multipulsing are often generated. The $\Delta f_{rep}$ is usually around 820 Hz; however, $\Delta f_{rep}$ values in the order of 5 Hz are also occasionally observed. Nonetheless, these regimes tend to last only for a few seconds or minutes before evolving into a harmonic mode-locking regime.

In summary, this chapter presents the results of improving the laser cavity, representing a significant step in this doctoral thesis. It marks the transition from a laser capable of generating a dual-comb regime to one that can produce this regime in a stable and straightforward manner, suitable for multiple metrology applications. Moreover, the improvement in the aliasing limit due to a higher ratio between $f_{rep}$ and $\Delta f_{rep}$ significantly enhances precision in distance measurement and spectroscopy. Finally, the laser is now assembled within a custom-made housing measuring 15 x 25 x 35 cm³, making it a compact, robust, and portable device suitable for off-lab applications. This chapter highlights the engineering work that has transitioned the system from scientific interest to practical and potentially commercial uses.

# Chapter 5

# Ranging Using a Portable Single-Cavity Polarisation Multiplexing System

## 5.1 Dual-comb LIDAR ranging.

A brief introduction to ranging using OFCs is presented in Chapter 1. Here I expand that information paying special attention to the heterodyne detection technique and the current state of the art in terms of dual-comb ranging. The most widely acknowledged paper about ranging using dual-comb is the one by Coddington et al., [36]. In this paper, the authors use two fully stabilised optical frequency combs with a frep of 100.021 and 100.016 MHz for each comb. Before this complete and important work done at NIST there were previous works in which authors used different methods to measure distances. For example [43, 127, 128, 129, 130], however, this work presented at the same time a highly precise system and a technique to largely extend its ambiguity range.

Some of these initial papers in this field highlight a distinction between ranging and LIDAR. On the one hand, laser ranging involves determining the phase shift of a signal after it traverses a given distance. When using a continuous wave (CW) laser, the most widely used source, the ambiguity range is limited to half of the laser wavelength. Therefore, shorter-wavelength signals offer greater resolution, while longer-wavelength signals offer a greater ambiguity range. On the other hand, LIDAR measures distance using pulse-to-pulse or radio-frequency modulated waveforms. The ambiguity range is equivalent to the repetition rate of the modulation frequency multiplied by the speed of light in the air and divided by two (accounting for the signal traveling to the target and returning). LIDAR typically offers higher resolution compared to laser ranging.

Multi-wavelength interferometry is the first step towards merging both techniques since it uses several optical wavelengths to effectively generate a synthetic wavelength and increase the ambiguity range. However, this technique is vulnerable to systematic errors, and extending the ambiguity range beyond a millimeter requires slow scanning. Optical frequency combs, however, merge both techniques in a more advantageous way. The combination of many frequencies, equidistantly separated in the frequency domain, allows for the application of laser ranging. Simultaneously, these frequencies generate a succession of pulses in the time domain, enabling LIDAR application.

Using two combs with slightly different repetition rates allows for heterodyne detection, enabling distance measurements with interferometric precision and the ambiguity range of LIDAR. See Figure 5.1.

Dual-comb ranging can be calculated using the following equation:

$$d = \frac{v_g \tau^{rt}}{2} \cdot \frac{\Delta f_{\text{rep}}}{f_{\text{rep}}} \tag{5.1}$$

where $v_g$ is the group velocity, $\tau^{rt}$ is the time between reference and target echoes, and $\Delta f_{\text{rep}}/f_{\text{rep}}$ is the amplification factor.

In our case the $\Delta f_{\text{rep}}$ is around 800 Hz and the $f_{\text{rep}}$ is in the order of 40 MHz. So the amplification factor can be calculated as:

$$\frac{\Delta f_{\text{rep}}}{f_{\text{rep}}} \approx \frac{0.8 \text{ kHz}}{40 \text{ MHz}} \approx 2 \times 10^{-5} \tag{5.2}$$

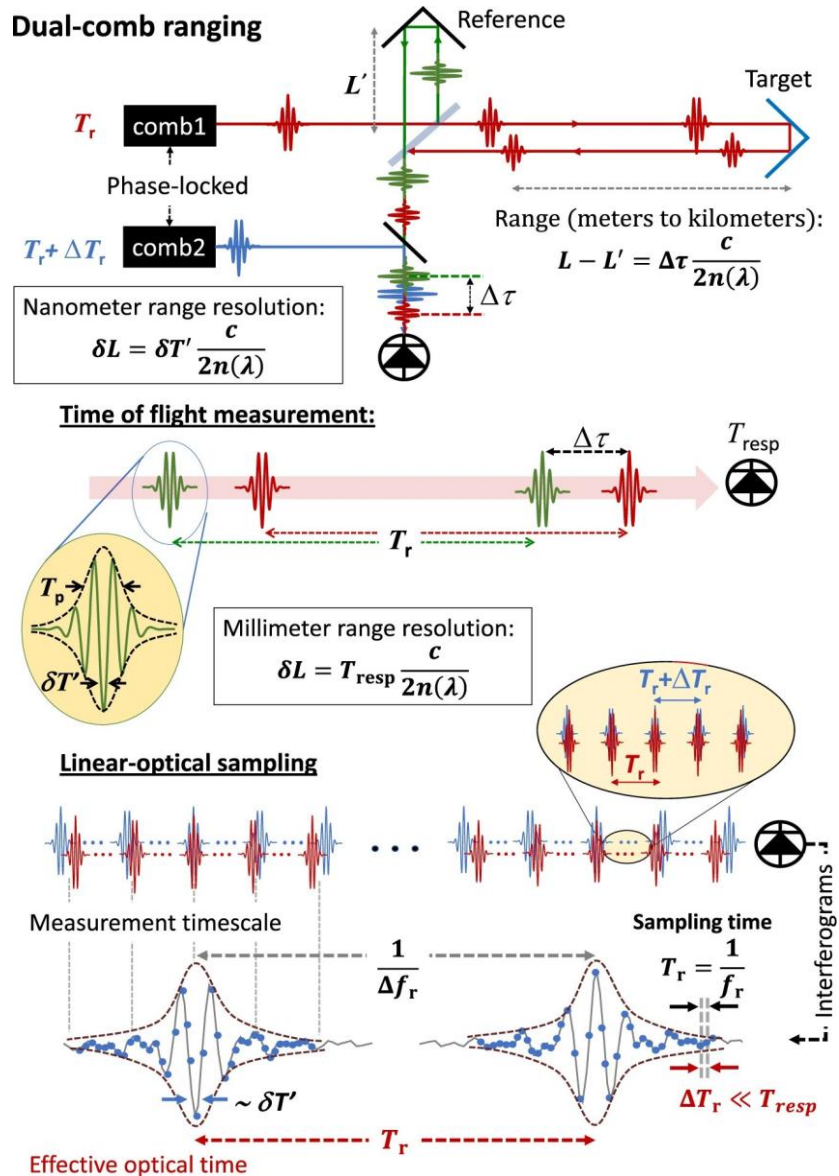

**Figure 5.1:** Dual-comb ranging and linear-optical sampling. In direct time of flight ranging with a single comb, the range is determined by measuring the delay between pulses travelling to a known reference (green pulses) and a target (red pulses). Ideally, the distance resolution, L, should be limited by the width of the optical pulses, Tp. However, when the optical pulses are detected directly, it is the much slower response time of the photodetector, TrespTp, that limits the distance/timing resolution. By employing a second comb, linear-optical sampling (LOS) is used to circumvent the photodetector response limit. LOS creates an interferogram between two OFCs with slightly offset repetition rates, fr. This optical cross-correlation between the blue and red pulses yields information about the relative optical pulse envelopes, Tp, and the optical carriers (interferometric fringes), T, accessible with standard microwave electronics and megahertz sampling rates. This interferometric LOS improves the coarse distance resolution of direct time-of-flight measurements from the millimeter- to nanometer scale. Linear-optical sampling also enables the measurement of high-resolution time/frequency information. In two-way optical time/frequency transfer, interferograms from LOS are collected at remote sites to compare, syntonize, and synchronize remote clocks, yielding time/frequency measurement accuracy and precision of parts $10^{-17}$. Source: Obtained from Fortier, Tara, and Esther Baumann. '20 years of developments in optical frequency comb technology and applications.' Communications Physics 2, no. 1 (2019): 153. [17]

By knowing the echo back from the target and the one back from the reference, and using this amplification factor, the distance, d, can be accurately determined. Just by using these formulas and averaging over a period of time usually a few milliseconds [36, 123, 131, 132, 133] we can obtain accuracies in the order of microseconds. The longer the average time the higher the accuracy, $T_{update}/T$ where T is the averaging period and $T_{update}$ is described by the following formula.

$$T_{update} = (T_r)^2/\Delta T_r \tag{5.3}$$

Where in our case $T_r \approx 25.5ns$ , and $\Delta T_r \approx 0.53ps$ (for $\Delta f_{rep} = 820Hz$ and $f_{rep} = 39.247MHz$), obtaining as a result:

$$T_{update} = \frac{25.5^2 \; ns^2}{0.00053 \; ns} = 1226887 \; ns \approx 1.22 \; ms. \tag{5.4}$$

$T_{update}$ can also be calculated as $1/\Delta f_{rep}$. $T_{update}$ condition is not only applied for TOF (Time of flight measurements) but also for interferometric measurements where we measure the optical carrier phase difference of both combs, and we can obtain measuring accuracies as precise as a few micrometers.

## 5.2   Setup.

The power output of the new optimized cavity is quite similar to that of the previous cavity. In this case, the trendline is adjusted, with a $R^2$ value equal to 0.9989, to the following equation Output Power = 0.0055 Pump Current − 0.8237.

Comparing Figures 2.3 and 5.2, we can observe that they follow a similar trend. The first cavity has a slightly higher optical output power, with 0.3 mW versus 0.27 mW at a pump current of 200 mA, and 0.568 mW versus 0.575 mW at a pump current of 250 mA. However, in the case of the optimized cavity, the pump operates at 200 mA, while in the previous cavity, it operates at 250 mA, resulting in an effective power difference of 0.3 mW, or half as powerful. The reason for this is the operational stability. In the initial cavity, the best stability and ease of generating the dual-comb regime occurred at 250 mA, whereas in the optimized cavity, they were achieved at 200 mA. We can also compare the peak power difference of both cavities by multiplying the average power by the round trip time (25.5 ns and 83 ns) and dividing by the pulse width in ps (1.28 ps and 1.52 ps). These calculations indicate that the peak power of the new cavity pulse is 5.98 W compared to 31.77 W for the previous cavity pulse. Since the new optical coupler is a 25/75 coupler, redirecting 25% of the power out, and the initial cavity uses a 10/90 coupler, redirecting only 10% of the energy out, we can calculate the peak power inside the cavity as 24 W for the new cavity and 317 W for the old cavity. This tenfold decrease in pulse power inside the cavity likely contributes positively to the overall stability of the regime.

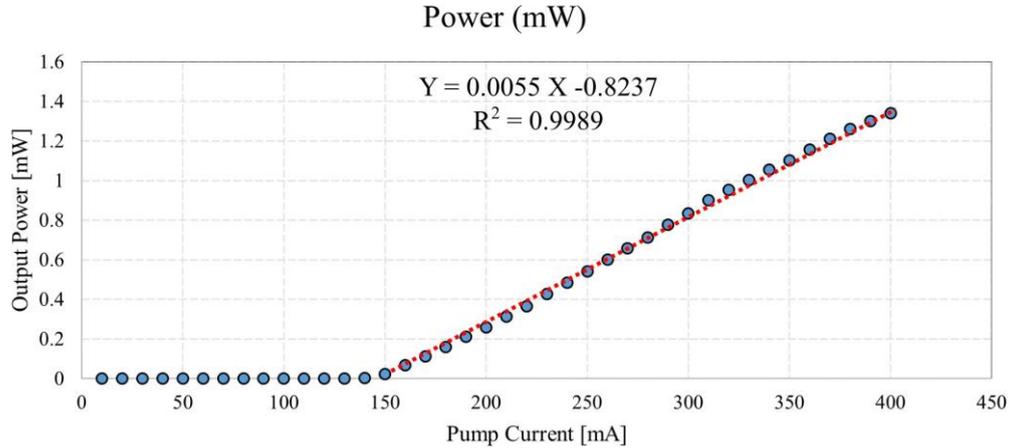

**Figure 5.2:** Pump current vs optical power output of the laser cavity of new laser

The setup used for ranging primarily employs the laser cavity discussed in Chapter 4 as the laser source. The output of the cavity is split using a combination of a Polarisation Controller (PC) and a Polarisation Beam Splitter (PBS). Both fast and slow combs are separated with an extinction ratio of 18.05 dBm for the fast axis and 17.9 dBm for the slow axis (see Figure 5.3 a)). This separation level leads to the OSA Spectrums shown in Figure 5.3 b). Out of these spectrums, we calculate a central wavelength, of 1559.78 nm for the fast axis, 1559.455 nm for the slow axis, and 1559.46 nm for the combined optical spectrum. At the same time, their FWHM broadband frequencies are $\Delta\lambda_{fast}$ = 1.836$nm$ for the fast axis, $\Delta\lambda_{slow}$ = 1.838$nm$ for the slow axis, and, $\Delta\lambda_{Original}$ = 1.838$nm$ for the original comb.

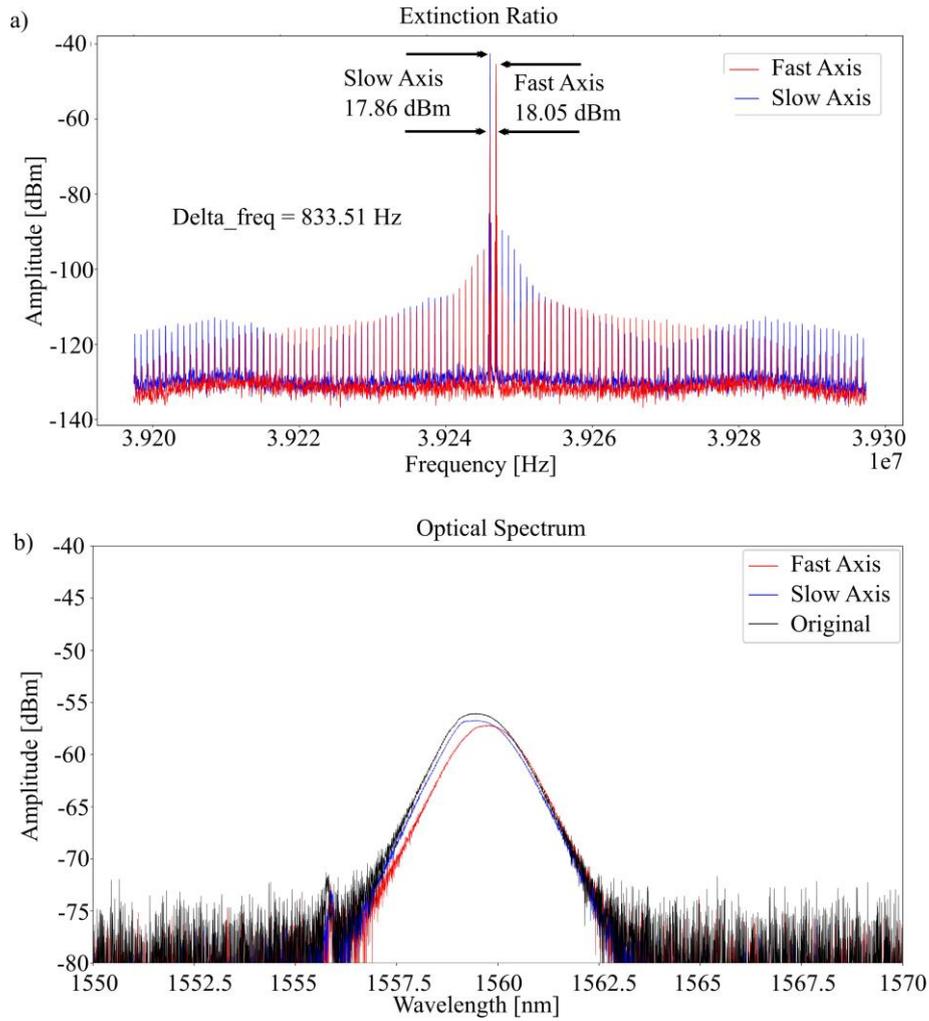

**Figure 5.3:** Comb Separation. b) RF spectrum of the fast (red) and slow (blue) axis with their respective extinction ratios after separation. a) OSA spectrum of the fast (red) and slow (blue) and the original signal (black).

Once both combs have been separated and their pulse trains have been divided into reference, target, and local oscillator. The fast axis is amplified and subdivided using a 50/50 Optical Coupler (OC). 50% of the signal is directed towards the target, while the other 50% is sent to the reference. Resulting in their electrical fields $E_{\text{ref}}$ (for the target) $E_{\text{ref}}$ (for the reference) and $E_{\text{ref}}$ (for the local oscillator) defined with the following equations:

$$E_{\text{ref}} = \sum_m A_m \cos\left[2\pi\left(mf_{\text{rep,sig}} + f_{\text{ceo,sig}}\right)t\right] \tag{5.5}$$

$$E_{\text{LO}} = \sum_n B_n \cos\left[2\pi\left(nf_{\text{rep,LO}} + f_{\text{ceo,LO}}\right)t\right] \tag{5.6}$$

$$E_{\text{tar}} = \sum_m A_m \cos\left[2\pi\left(mf_{\text{rep,sig}} + f_{\text{ceo,sig}}\right)(t - \tau)\right] \tag{5.7}$$

A telescope is aligned with the target, whereas the reference path consists of a segment of fibre of a known length that includes an attenuator. The attenuator is used to match the optical power of the target and reference signals. Without this, one signal might dominate the other, and the noise from the more powerful signal could obscure the weaker one. In some cases, instead of using a telescope to send and receive the signal from the target, a segment of fibre is used to avoid optical power losses. Finally, the slow axis, acting as a local oscillator, is passed through a PC and then combined in a 50/50 optical coupler. Both signals are mixed in a photodetector, and the resulting signal is measured using the same fast oscilloscope (OSC) discussed in Chapter 3, specifically the 32 GHz high-performance sampling oscilloscope (Agilent DSOX93204A Infiniium) and the RF spectrum analyser. The photodetector used for the OSC is a slow photodetector DC-125MHz and it works in combination with a low band pass filter capped at 15 MHz. This filter helps increasing the SNR of the signal in the heterodyne detection and therefore helps extract the values of distance with more precision.

**Figure 5.4:** Representation of the laser cavity & the lidar system: polarisation controller (PC), polarisation beam splitter (PBS), optical coupler 50/50 (OC), circulator, collimator telescope (T&C), a segment of referent fibre (Ref) and an attenuator. The amplifier is used just for the telescope and reference part. In combination with a photodiode (PD), an optical spectrum analyser (OSA), and a radio frequency spectrum analyser (RF).

## 5.3    Results.

The results presented in Figure 5.5 correspond to the dual-comb ranging obtained using the scheme shown in Figure 5.4. A 10 ms trace was captured with the fast oscilloscope (OSC) at a sampling rate of 80 million samples per second, as shown in Figure 5.5a. Another 10 ms trace was captured using the same setup and oscilloscope but with the EF526 Low-Pass Electrical Filter (≤ 15 MHz Passband) placed just before the oscilloscope. As seen in Figure 5.5b, the filter clarifies the signal burst. However, the signal is not as clean as reported in other publications [123], suggesting that further stabilisation of the setup may be required.

Despite this, the signal acquired in Figure 5.5b allows us to extract the bursts corresponding to the target and the reference. The setup is configured so that the reference signal is always slightly larger than the target signal. Moreover, we accurately know the repetition rates of both signals. The signal processing involves identifying the highest point in the oscilloscope trace, corresponding to the most intense beating in the signal. We then select the remaining points at multiples of $1/f_{rep}$ to identify all bursts generated by the beating of the reference signal. A window of 6000 values before and after the point is created, and these points and their positions are extracted. The same process is repeated to extract the burst signals derived from the beating of the target and local oscillator. The code used for the signal processing can be seen in Appendix F.

We average the distances between the maximum points of each target burst and their corresponding reference burst. Figure 5.5c shows the processed signal from Figure 5.5b. By doing that we are able to measure distances with a precision of 340 $\mu m$ for a 10ms averaging and an ambiguity range of 3.8 meters. We measured the distance of the reference segment of fiber first and the distance to a target using the telescope afterward. Figure 5.5 shows the data for the first measurement, when the segment of fiber is compared to a direct fiber connection without a telescope. In other measurements with free space optics the echoes from the target are weaker and have a worse SNR.

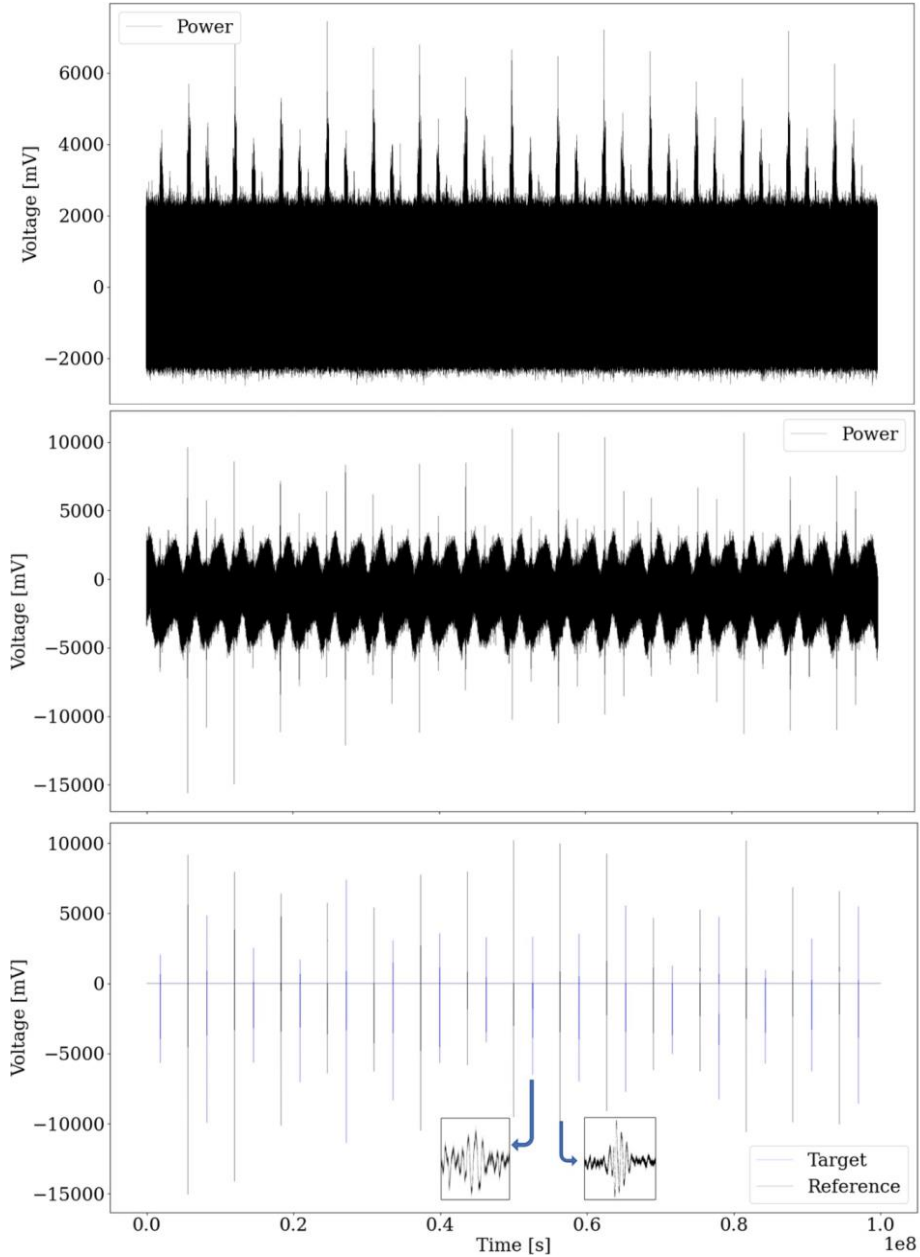

**Figure 5.5:** OSC trace obtained from a dual-comb LIDAR scheme. a) Original 10 ms OSC trace, b) 10 ms OSC trace with a Low Band Pass Filter, c) Post Processed OSC trace from figure b), in blue target echoes in black reference echoes.

One important aspect to be mentioned is that the pulse width was expanded using the DFT fiber of 11-km SMF-28 to a pulse width of 21.39 ps. Having this pulse width makes sure that the 10 $\Delta T \leq \Delta \tau$ condition is clearly satisfied. Nonetheless, under these conditions, the pulse power was not enough to get the signal back from the telescope and power was prioritized. Moreover, when the telescope was replaced by a segment of fiber of a known length to avoid losses, the result didn't show significant improvements.

Finally, an alternative ranging algorithm was tested to retrieve the distance from the phase difference. However, it was only tested using synthetic data, as it is not yet practical for real noisy data. This method can be found in Appendix E.

## 5.4  Discussion and summary.

The stability achieved in the previous chapter has enabled the use of this technology as a ranging tool. This chapter presented a proof of concept for its use, successfully measuring distances of objects with sub-millimetre accuracy. Given the demonstrated stability of this laser over 250 hours, this system has the potential to record highly detailed 3D images of surfaces with exceptional precision. However, realising this potential requires several additional conditions. A more advanced collimator system is needed to minimise the beam's focal point, coupled with a precise spatial controller to direct the laser beam accurately, thereby enabling surface mapping. Additionally, a system to reduce the data generated at each point is necessary. In our case, the OSC traces measured at each point require 0.5 to 4 GB of memory, with processing times ranging from 2 to 20 minutes. This data volume poses a significant challenge. However, innovative approaches are being developed to address the high data volume problem, including two-photon dual-comb techniques, which significantly reduce data volumes while maintaining the interferometric precision of the technique [123, 134].

The relatively low precision of our system, 0.34 mm, compared to other works [132, 123, 134, 36, 35, 133], can be attributed to the short averaging times of 10 ms and the higher short-term instability characteristic of non-stabilised lasers. Nonetheless, the reduction in precision can be largely compensated for by the decreased complexity and increased robustness of this laser system. This laser also has the capability to increase the ambiguity range by orders of magnitude using already published methodologies [36] or by simply adjusting the reference path to distances within the target range, provided the target distance can be estimated beforehand.

Finally, a future research direction for this system is its conversion into an ellipsometric LIDAR system, leveraging the studies in polarisation dynamics discussed in Chapter 2, which are similar to those observed in the optimised laser cavity. In fact, using the polar-isation analyser from Chapter 2 (Novoptel PM1000-XL-FA-N20 D), one could potentially obtain both distance measurements and the ellipsometric signature of targets simultaneously. In the concluding chapter, 7, I will outline the future steps planned for ranging using this system, particularly the steps towards employing it for polarimetric ranging.

# Chapter 6

# Effects and Mitigation of Harsh Environmental Conditions in LIDAR Systems Used in Autonomous Driving Vehicles

---

*This chapter is based on the work done at Aurrigo Ltd. during my secondment, the codes and some other pieces of work will not be disclosed as it is under IP protection.*

---

## 6.1   Aurrigo secondment

As dual-comb LIDAR technology advances towards commercialization [134], it must contend with the realities of outdoor operation that are not present in the lab. Harsh environmental conditions such as rain, fog, snow, and dust pose significant challenges to the performance and reliability of LIDAR systems.  This chapter delves into these challenges, providing a detailed analysis of how such conditions affect LIDAR performance and proposes ways to mitigate these effects. Thus, this chapter serves as a vital bridge connecting the theoretical advancements in dual-comb LIDAR technology with their practical deployment in real-world conditions.  It emphasizes the necessity of considering environmental factors in the design and optimization of commercial LIDAR systems, ensuring their reliability and effectiveness. Thus this chapter offers a more commercial and end-user enginnering approach than the previous chapters.

In order to gain practical insights and industrial experience, I undertook a secondment at Aurrigo International PLC in Coventry, UK, from Monday 5th of February to Friday 14th of July. As a Marie-Curie-funded scholarship collaboration, partnering with industry and gaining mobility experience is fundamental. This six-month secondment was planned between the second and third years of my PhD, allowing me to learn about industrial applications for the LIDAR technology I am developing. During this time, I also assisted Aurrigo International PLC in addressing some of its critical challenges with LIDAR systems.

Aurrigo was one of the industrial partners of MEFISTA and it has collaborated with Aston University in several different projects before. The origins of the company date back to 1993 when an automotive company called RDM (Richmond Design & Marketing) was created in the Coventry area. RDM specializes in automotive parts design and manufacture, and still to this day it provides ground-up vehicle design, development, and manufacturing services in a number of areas including electronics, software development, electric vehicle development, and wire harness design. Over the last 30 years, the Group has been supplying leading Tier 1 suppliers and vehicle manufacturers (OEMs) which include Aston Martin Lagonda, Bentley, Jaguar Land Rover, McLaren, and Rolls Royce. The Group's expertise and consistent delivery of high-quality products have built long-term customer relationships and continue to provide the Group with opportunities to expand its existing customer base. The transition from RDM to Aurrigo has gone hand in hand with the establishment of a new business model based on autonomous driving vehicles. Their driverless vehicles are focused on urban environment transportation and more importantly in the niche market sector of aviation, being Changi Airport in Singapore their main client. Within their list of vehicles, the three most successful commercially available vehicles to this day (Jan 2024) are Auto-Pod, Auto-Shuttle, and Auto-Dolly.

Auto-Pod was the first autonomous vehicle developed by the Group and is a four-seat product designed for the non-road going passenger transportation of small groups, for example to and from airports, city centres, sporting venues, university campuses, and age care communities. It is capable of operating on a fully autonomous basis, although current UK and most international safety standards require a safety supervisor to be on board whilst it is in an autonomous mode of operation.

The Auto-Shuttle is a ten-seat passenger vehicle with the ability to operate fully au-tonomously or driven manually as a conventional EV shuttle. It is the first road-legal vehicle to be manufactured by the Group. The vehicle has been designed to provide economical public transport in under-served or previously less cost-efficient parts of the country, as well as providing an excellent transport solution in airports to move passengers to and from terminals and to operate airside with applications such as VIP transport and crew movements.

Auto-Dolly is a unique and disruptive baggage transportation solution for airports. It has been specially designed to reduce baggage and cargo loading and unloading times, improve movement efficiencies, health & safety, drastically reduce the manpower required for an operation and reduce overall operation costs.

All of Aurrigo's autonomous driving vehicles operate using LIDAR as part of their localisation system. Initially, the area of operation is mapped using a sensor suite that includes a layer-dense LIDAR system such as a 128-layer LIDAR and once the information has been rendered and the 3D map of the area is established this map is loaded into the navigation & control system of the vehicle. Each vehicle uses its own sensor suite system to navigate through the pre-mapped environment. By obtaining 3D images of the area they are able to localize themselves in the map and evaluate the presence of obstacles that they need to avoid. Usually, their vehicles have two or more LIDAR systems to ensure reliability, avoid blind spots, and multiply the input data.

## 6.2   LIDAR devices used by Aurrigo's vehicles.

There are many commercial LIDAR systems available on the market, such as those from, RoboSense (Suteng Innovation Technology Co., Ltd.), Leishen Intelligent System Co., Ltd. (LSLiDAR), and Velodyne Lidar. Any LIDAR system considered for use must be evaluated to ensure they have appropriate specifications, can be integrated, are reliable, have an appropriate supply chain and cost point. Typical commercial LIDAR devices that are currently used or were previously tested have several features in common. Their horizontal field of view (FoV)is 360 degrees, and their maximum rotation speed is usually 1200 rpm (revolutions per minute). This rotation speed is equivalent to 20 Hz vision update range. All of them have an operating temperature range of -30 to +60 degrees and their operating voltage is usually between 9 and 32 V. In terms of the physical shape, their dimensions vary little from one device to another, being cylinders of 10 to 15 cm in height and roughly 10 cm in diameter with a total weight of 1 to 1.5 Kg. Moreover, they are all considered as class 1 in the eye-safe scale.

Table 6.1 displays some of the main characteristics of the commercial LIDAR devices used available on the market with suitable specifications for use on autonomous vehicles. It's necessary to highlight that all of these devices are suitable for localisation and navigation but not for mapping, where bigger and more information- and layer-dense LIDAR system is typically required.

| COMMERCIAL LIDAR SPECIFICATIONS | | | | | | |
|---|---|---|---|---|---|---|
| Product | RS-Helios 5515 | RS-Helios 1615 | RS-Helios-16P | C32 LSlidar | C16 LSlidar | Velodyne LiDAR PUCK |
| Laser Wavelength | 905nm | 905nm | 905nm | 905 nm | 905 nm | 905 nm |
| Layers | 32 | 32 | 16 | 32 | 16 | 16 |
| Range | 150m (110m@10% NIST) | 150m (110m@10% NIST) | 150m (110m@10% NIST) | 100 m @10% reflectivity; 150 m @70% reflectivity | 150 m (Reflectivity of 70%) | 100 meters |
| Range Accuracy (Typical) | ±2cm (1m to 100m) | ±2cm (1m to 100m) | ±2cm (1m to 100m) | ±3 cm | ±3 cm | ±3 cm |
| | ±3cm (0.1m to 1m) | ±3cm (0.1m to 1m) | ±3cm (0.1m to 1m) | ±3 cm | ±3 cm | ±3 cm |
| | ±3cm (100m to 150m) | ±3cm (100m to 150m) | ±3cm (100m to 150m) | ±3 cm | ±10 cm (50-70 m) | ±3 cm |
| Vertical FoV | 70° (-55°~ + 15°) | 31° (-16°~ + 15°) | 30° (-15°~ + 15°) | -16°~+15° | ±15° | ±15° |
| Horizontal Resolution | 0.2°/0.4° | 0.2°/0.4° | 0.2°/0.4° | 5 Hz: 0.09º / 10 Hz: 0.18º / 20 Hz: 0.36º | 5Hz: 0.09º 10Hz: 0.18º 20Hz: 0.36 | 0.1° - 0.4° |
| Vertical Resolution | Up to 1.33° | 1° | 2° | 1° (Uniform Distribution) | 2º | 2º |
| Points Per Second | ~576,000pt s/s (Single Return) | ~576,000pt s/s (Single Return) | ~288,000pt s/s (Single Return) | 640,000 pts/sec | ~320,000pt s/s (Single Return) | ~300,000pt s/s |
| Power Consumption | 12W | 12W | 11W | 12 W (Typical*); 25 W (Max) | 12 W (Typical) | 8 W |

**Table 6.1:** Commercial LIDAR devices used by Aurrigo ltd. for its autonomous navigation vehicles.

As it can be derived from the Table 6.1, these systems share many specifications in common. All of them operate in the same wavelength, 905 nm, and have similar ranges of around 150 m except for the case of Velodyne where the range is just 100m. Nonetheless, these ranges are conditioned by the reflectivity, and in the case of the target with lower reflectivity (less than 70%) the measuring range decreases to 100-110 meters. When it comes to range accuracy, they tend to have accuracies in the order of 3cm for targets at distances of 100 to 150m away. Except for the C16 LSlidar where the accuracy is 10 cm. Notably the accuracy increases with the proximity of the target [135]. The vertical FoV tends to be of 30° except for the RS-Helios 5515 where it is 70°. Those devices that use 32 layers have a vertical resolution of around 1° and generate between 640,000 and 576,000pts/s while those that operate with 16 layers have a vertical resolution of around 2° and generate between 320,000 and 288,000pts/s. Finally, the power consumption is similar for all of them with typical values of 12 W.

## 6.3 Effect of adverse atmospheric conditions in LIDAR.

When autonomous vehicles operate in harsh meteorological conditions such as rain the LIDAR systems can report the position of raindrops or snowdrops to the localisation and navigation system. Additional software processing is required else that might trigger some fake obstacle detection. The heavier the rain, the more raindrops or snowdrops are overlaid on the input to the navigation system. When the volume of rain or snow is so great that the baseline environment is totally obscured, then the system can detect that the specific sensor is compromised, so trigger a 'minimal risk maneuverer or trigger the safety stop mechanism, which is designed to avoid collision with people, animals, or other vehicles. The effects of harsh environmental conditions on LIDAR systems have been already studied in detail in the literature. In this section, I will provide a short summary of them.

The functionality of LIDAR systems can encounter significant impediments when subjected to adverse environmental conditions like rain, fog, snow, dust, and other atmospheric interferences. These formidable challenges pose a direct impact on the accuracy, range, and overall performance of LIDAR technologies. Understanding the nuanced effects of these environmental factors stands as a pivotal step toward optimising LIDAR systems, ensuring their reliability across diverse conditions, and advancing their applicability across various industries.

Inherent within current vision systems is their design tailored for optimal performance in clear weather conditions. However, the reality of outdoor applications dictates a necessity to contend with adverse weather scenarios. Incorporating mechanisms within vision systems becomes imperative to enable functionality amidst the presence of haze, fog, rain, hail, and snow.

The variability of weather conditions primarily resides in the types, sizes, and concentrations of particles within the atmosphere. Extensive efforts have been directed towards measuring these particle sizes and concentrations across a spectrum of conditions [136, 137].

The fundamental characteristics of light, encompassing aspects like intensity and colour, undergo alterations through interactions with the atmosphere. These interactions can be broadly categorized into three groups: scattering, absorption, and emission. Among these, scattering induced by suspended particles stands as the most detrimental factor for artificial vision reliant on LIDAR.

The susceptibility of LiDAR data to adverse weather conditions, notably rain and fog, has been extensively examined. Literature abounds with comparisons related to fog attenuations [138, 139, 140, 141, 142]. The attenuation caused by the fog is measured with the equation 6.1. It can vary from 0.2 dB/Km in exceptional clear sky conditions to more than 450 dB/Km in maritime fog.

$$\text{Attenuation} = 10 \log_{10}\left(\frac{P_{\text{in}}}{P_{\text{out}}}\right) \tag{6.1}$$

Fogs, comprised of fine suspended water droplets in the lower atmosphere, significantly reduce visibility by scattering light. These droplets play a crucial role in hindering visibility near the ground level. Different mechanisms explain the formation of fog. For example, radiation fog materializes due to radiative cooling during night-time, when the air becomes sufficiently cool and saturated. Particle diameters hover around 4 microns, with liquid water content between 0.01 and 0.1 g/m³. Advection fog, on the other hand, results from the movement of moist, warm air masses above colder surfaces, whether maritime or terrestrial. All types of fogs are characterised by parameters like liquid water content, particle size distribution, average particle size, and particle count per unit volume. Spherical in shape, fog particles range in radius from 0.01 to 15 microns, varying geographically.

Even fog is often referred to most of the weather conditions where water droplets are suspended in the atmosphere we can technically classify them in different forms according to the visibility. Fog for visibilities less than 500 meters, haze for visibilities exceeding 1000 meters, and a transitional phase termed mist for visibilities ranging between 500 and 1000 meters. Their composition also varies, haze primarily consists of microscopic dust, salt particles, or small droplets, ranging from a few microns to tenths of a micron. Fog emerges during high humidity, where water droplets form over haze particle nuclei. Mist, a transitional phase, arises as humidity approaches saturation, marked by an increase in droplet sizes that rapidly deteriorate visibility.

Consequently, exploration into different wavelengths has been a key area of research in mitigating adverse weather impacts on LIDAR systems, particularly focusing on the comparison between 905 nm and 1550 nm wavelengths [136]. While these studies have not revealed significant differences in target detection distances between the two wavelengths, they highlight a crucial factor: the variation in water absorption. Notably, water absorption is significantly higher at 1550 nm compared to 905 nm. This implies that under similar output power conditions, a LIDAR system operating at 1550 nm may be more susceptible to the effects of elevated humidity levels. However other studies [143] indicate that the 0.9 $\mu$m (905 nm) module can measure distances up to 60% longer than the 1.5 $\mu$m (1550 nm) module in fog, while still maintaining good visibility. There are older studies that have evaluated a wider range of wavelengths up to 10 $\mu$m and different sizes and concentrations of droplets [144]. Despite this, fog remains a challenging element for both wavelengths, as it severely limits rangefinder performance to just a few hundred meters. This observation underscores the complex relationship between wavelength selection and LIDAR performance in adverse weather, emphasizing that despite similar detection distances, the differential impact of humidity absorption is a critical factor in determining the efficacy of specific wavelengths in LIDAR technology.

Beyond the wavelength considerations, other atmospheric factors also play significant roles in affecting LIDAR performance in adverse weather conditions. Variables such as atmospheric pressure, temperature fluctuations, and the composition of particles in the air are pivotal in altering LIDAR readings and accuracy. The effect of atmospheric pressure is linked to changes in the density and composition of the atmosphere, which can alter the refractive index of the air. This, in turn, affects the speed of light, impacting the accuracy of distance measurements by LIDAR systems.

Temperature variations also play a crucial role. They influence the behaviour of atmospheric particles, affecting their movement, size, and density. These changes in particle behaviour directly impact the scattering and absorption of light, which are vital factors in the reliability and precision of LIDAR readings. Additionally, the specific composition of atmospheric particles, be it dust, fog, rain, or snow, significantly influences how light interacts with the environment. Each type of particle presents unique challenges for LI-DAR systems, necessitating adaptive mechanisms to ensure optimal performance in diverse weather conditions. In the case of Aurrigo, their main struggles with the weather is given in their fleet of Singapore where there are tropical rains in a daily basis during the wet season and the intensity of these rains is very heavy.

There are three main disruptive effects of rain in the LIDAR information which are the following. Number one - drops of water that interact with the LIDAR laser and reflect light from a shorter-than-expected distance. Number two - a 20 to 100-cm tall layer of fog that appears on the surface of the road or concrete due to the splits of the thick raindrops that hit the floor and divide into smaller drops and water aerosol. Number three – some water drops directly impact the LIDAR glass cover and remain there causing the system to detect a permanent obstacle next to the vehicle. Out of the three distortions this last one is without doubt the one that causes a larger disruptive impact in their autonomous driving operations.

## 6.4 Mitigation strategies - physical and neural network approaches.

Addressing adverse weather effects on LIDAR involves an array of mitigation strategies encompassing both physical interventions and sophisticated neural network methodologies.

Physical mitigation strategies encompass diverse options, ranging from mechanical components like shields or protective mechanisms designed to shield the system from water or humidity-induced alterations. For instance, in the realms of civil and military vehicle technology, spinning shields have been developed, notably integrated into autonomous vehicles, as seen in Waymo's initiatives [145]. However, when the impact of rain, fog, snow, or hail persists despite such mechanical approaches, signal processing techniques serve as a valuable resource.

Signal processing techniques constitute an arsenal of approaches to pre-process the optical signals received in LIDAR systems, offering methods like echo selection or threshold modification. Additionally, post-processing techniques such as moving averages, median filters, segmentation algorithms, or Kalman filters [146, 147] play a significant role in refining LIDAR data by eliminating or modifying incoherent data points.

Moreover, the realm of machine learning has emerged as a promising avenue for weather-resistant LIDAR systems. Numerous papers have delved into machine learning approaches tailored to enhance LIDAR performance amidst adverse weather conditions. Some machine learning strategies are grounded in generative adversarial networks (GAN), facilitating LIDAR translation between sunny and adverse weather conditions, crucial for autonomous driving and driving simulation [148]. Additionally, convolutional neural networks (CNN) are employed for denoising point clouds, enhancing data fidelity by removing noise and distortions [149].

In a dynamic and evolving landscape, the synergy between physical interventions, signal processing techniques, and advanced neural network methodologies stands poised to revolutionize the resilience of LIDAR systems in adverse weather conditions. These multifaceted approaches signify the ongoing quest to fortify LIDAR's efficacy across diverse environmental challenges.

## 6.5   Problem solution.

Having evaluated the nature of the problem, my main task within the company is to engineer, design and partially implement a test rig to mitigate the disruptive effects of rain with a special focus on the water drops that impact directly on the LIDAR glass cover. For such task, I designed and manufactured a test article for an implementation of a spinning shield Figure 6.1.

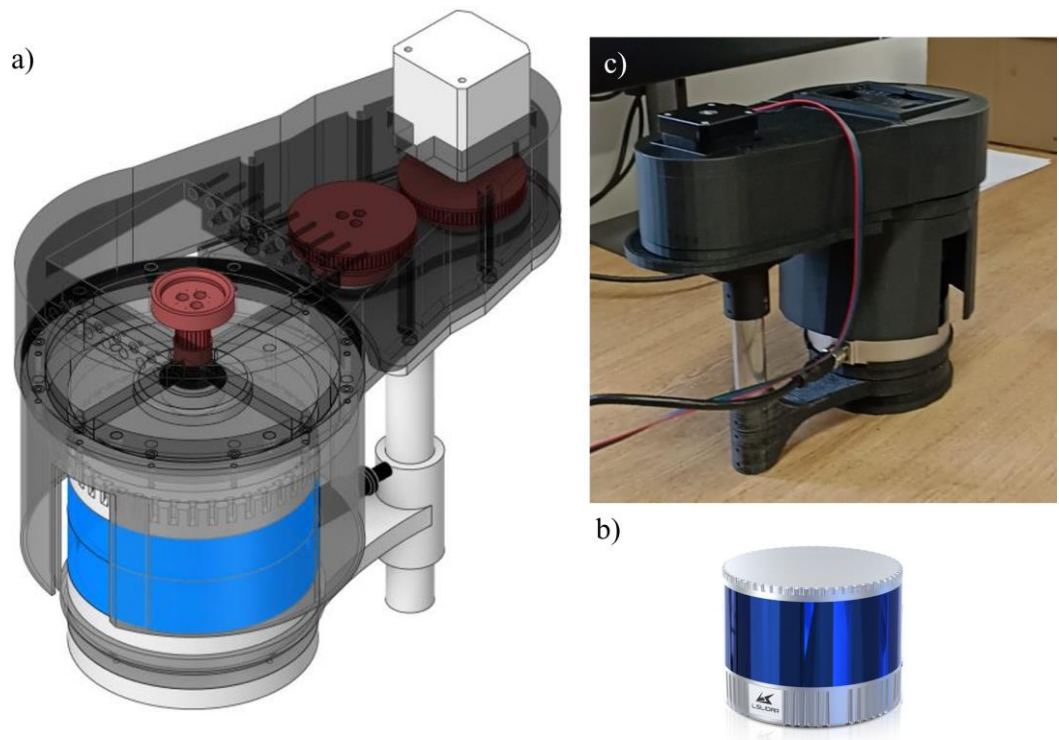

**Figure 6.1:** Spinning shield LIDAR. a) 3D design of the spinning shield designed to mitigate the effect of rain in the LIDAR system. b) Lslidar C16 was used as a model for the shield. c) 3D printed prototype of the spinning shield system.

The shield of Figure 6.1 is composed of 14 smaller pieces, and they require a total time of 114 hours to be printed while they use 468 grams of ABS filament. The shield consists of a static part that holds the LIDAR, the motor, and the internal gearbox and a dynamic part that is a circular cylinder with an aperture spinning around the LIDAR. The moving part is spun with a NEMA 17 motor connected to a gearbox to leverage the spinning speed of the shield and controlled with an Arduino. The aperture of the cylinder must spin exactly at the same time that the lasers within the LIDAR, that is 20 times per second. Furthermore, the aperture of the shield needs to be in perfect coordination with the Lasers, otherwise, the shield would block the light coming out of the LIDAR inputting wrong information to the navigation system. The LSlidar C16 used to implement the spinning shield operates at 905 nm and has a light filter that prevents any light outside of this wavelength from arriving at the photodetector. That filter, the blue part of the cylinder in Figure 6.1 b) blocks the light and one can not identify the actual position of the lasers within the system. Thus, to solve this problem I sniff the IP packages sent from the LIDAR to the controlling device, in this case a Jetson Nvidia computer. The LIDAR system sends continuously packages of information of 1248 Bytes with the data structure shown in Figure 6.2.

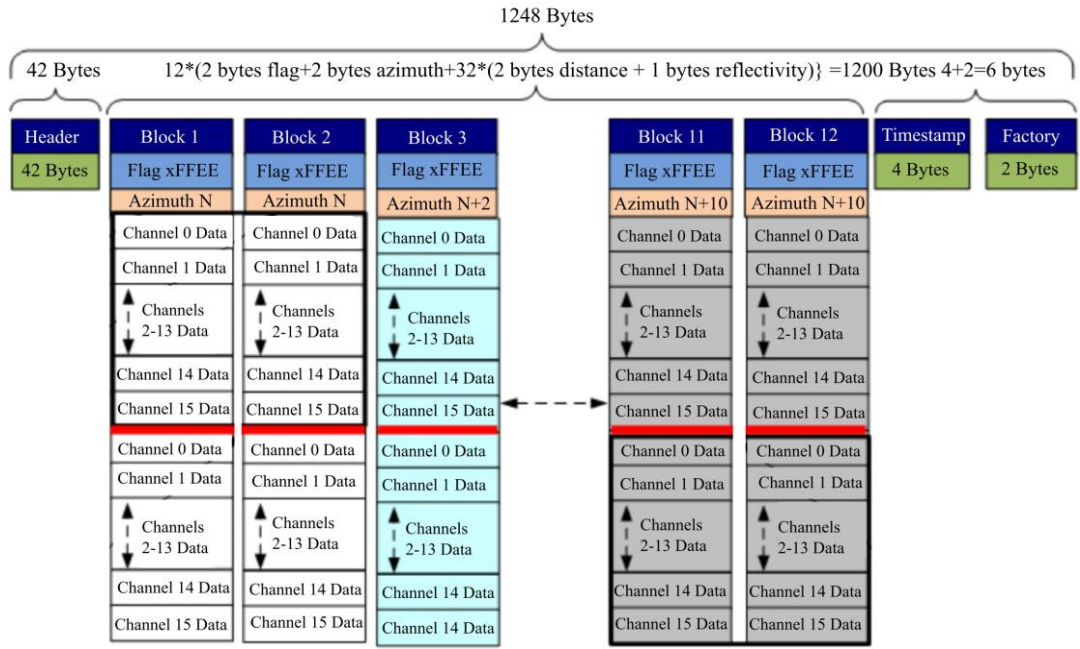

**Figure 6.2:** IP package structure. Echo packet data structure from the LSlidar C16

Each package is encoded in binary information and contains information on the vertical and horizontal angles, the distance the reflectivity of 384 data points. To decrypt all this binary information, I used the Python libraries of Scapy and Struct. In that way, I transformed binary information such as the one shower shown in Figure 6.3 to the 3D image shown in Figure 6.4.

A preliminary study of the aperture dimensions in the spinning shield was performed. The aim of this study is to determine the minimum possible vertical aperture so that the lens of the lidar is largely covered and no drop of water gets in contact with the lens. Thus, an initial estimation based on the horizontal resolution of the LIDAR of 0.35° suggests that an aperture of 2 mm in the shield cylinder is enough.

Nonetheless, after testing this aperture equivalent to 2 mm, there was no data whatsoever so the aperture was increased. With larger than 2mm apertures some points are detected but some others give wrong results probably due to an internal average of the LIDAR device. Nonetheless, when the aperture is around 35 mm or in other words an angle of 20° the measurements are always correct. Having a larger aperture is slightly detrimental for the purpose of the shield, protecting the LIDAR from the rain. Nonetheless, it makes the coordination and adjustment in the speed of rotation significantly simpler.

```
0000 48 B0 2D 2F 54 DD 50 3E 7C 20 03 CE 08 00 45 00 H.-/T.P>| ....E.
0010 04 D2 C4 A5 40 00 80 11 AC F6 C0 A8 01 C8 C0 A8 ....@...........
0020 01 66 09 41 09 40 04 BE 52 85 FF EE 18 19 08 00 .f.A.@..R.......
0030 0B 0A 00 13 0E 00 13 06 00 13 0A 00 13 0C 00 13 ................
0040 0A 00 0E 0D 00 19 08 00 14 0F 00 17 02 00 15 0E ................
0050 00 12 0C 00 19 11 00 15 0A 00 15 3C 00 16 0B 00 ................
0060 0C 0A 00 12 0F 00 11 07 00 15 09 00 11 0C 00 11 ................
0070 08 00 0F 0D 00 13 09 00 15 10 00 14 03 00 18 0E ................
0080 00 12 0C 00 15 11 00 15 0B 00 15 10 00 14 FF EE ................
0090 3C 19 09 00 09 09 00 14 10 00 13 07 00 0E 08 00 <...............
     ..............  ..............  ..............  .............
     ..............  ..............  ..............  .............
     ..............  ..............  ..............  .............
0480 12 00 11 0C 00 14 0A 00 0F 09 00 13 0C 00 13 10 ................
0490 00 10 09 00 13 0F 00 16 04 00 14 3C 00 16 10 00 ................
04a0 1A 10 00 12 0B 00 14 0E 00 14 06 00 0C 0E 00 14 ................
04b0 13 00 16 0C 00 12 0C 00 0F 09 00 11 0B 00 12 11 ................
04c0 00 13 0A 00 18 12 00 17 04 00 14 3B 00 16 0F 00 ................
04d0 13 11 00 17 0B 00 17 10 00 13 30 D4 32 41 37 10 ..........0.2A7.
```

**Figure 6.3:** IP package binary. Echo packet from the LSlidar C16

Figure 6.4 represents a sequence of 3D images reconstructed from the IP packages captured from the LIDAR device. These packages are captured using a sniffing library in Python and the information of the point cloud is extracted from the binary data. Every package has the structure displayed in Figure 6.2 and each image consists of around 84 ±1 packages and a total number of points of roughly 16000.

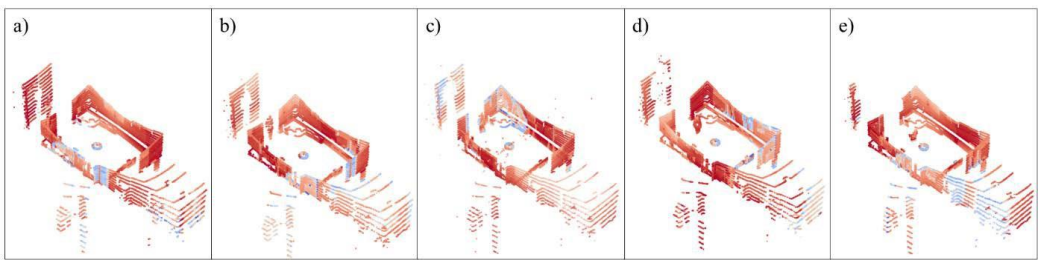

**Figure 6.4:** 3D images reconstructed from the IP packages captured in the LIDAR. The LIDAR is in the middle of the office and there is a person walking towards it at different moments in time. a) time =0 seconds, b) time = 0.75 seconds, c) time = 1.15 seconds, d) time = 1.75 seconds, e) time = 2.3 seconds.

Using this information, I was able to identify the angular position of the laser within the LIDAR at all times and set the basics for the coordination of the LIDAR and the Spinning shield. Not only that but also, by using the information contained in the data points of this system, I could filter out data points that have been distorted by rain in the first and second ways.

## 6.6   Summary and future works.

In this secondment, I studied the effect of rain, fog, and other harsh environmental conditions in commercial lidar operations for autonomous driving vehicles. Among them, rain is one of the conditions that needs to be mitigated to successfully localise an autonomous vehicle due to its prevalence and due to the significant distortion caused by its effects. To mitigate the effects of rain a software and hardware test rig was developed. The software intercepts the binary packages that are sent from the LIDAR device to the computer, it pre-processes the signal and extracts the azimuth angle which refers to the horizontal angle of the Laser inside the LIDAR. In addition to that, it extracts the information from the packages and represents the 3D images captured with the LIDAR. The Hardware is an initial prototype version of a spinning shield that protects the LIDAR system and prevents water drops from getting attached to the lens. This prototype has been manufactured using a 3D printer and commercially available motors compatible with Arduino as well as transmission belts. It consists of a static part that is fixed on the vehicle and holds the gearbox the motor and the lidar itself and a moving part which is basically a cylinder with an aperture to allow the light to be sent and received from the LIDAR.

This project means an important advance and a proof of principle for a potential physical implementation of a mitigation system that could be integrated with the existing mechanisms of software filtering and sensor fusion used on typical autonomous vehicle fleets. This proof of concept demonstrator has shown that mitigating water droplets landing on the surface of the LIDAR can be beneficial. Nonetheless, to fully mitigate this problem there are several steps yet to be taken. The spinning of the shield needs to be in coordination with the internal spinning of the laser in the LIDAR. To achieve that, the azimuth angle that is extracted from every IP package must be sent to the Arduino which on its own needs to keep track of the position of the spinning shield at every moment. Since there is some delay in the extraction of the information from the packages, in sending the order to the information to the Arduino, and in the calculation of the deviation and adjustment of the speed of the motor this is not a straightforward implementation. It would require from a precise estimation of the delays in processing the positional information and changing the RPM (revolutions per minute) of the NEMA 17 motor that is used. Moreover, although the system was tested with the motors and the spinning shield was spinning at high speed and achieving values of RPM (revolutions per minute) like the ones required, there was some friction of the components, therefore the design requires minor corrections to guarantee a smooth operation. On the other hand, the software needs to be programmed in such a way that the LIDAR status packages of information are neglected. Every few hundred packages the LIDAR system sends one with information related to the configuration parameters. Since it has a similar binary structure to the others the code developed in the proof of concept, processes this package and extracts information on the angular position of the laser (azimuth angle) nonetheless this information is incorrect and can lead to imprecisions in the case it is not removed. Since the LIDAR status packages are sent every X number of packages the software can be updated to neglect these packages before processing them.

Finally, this project is a proof of concept, so it would need to be commercialised, made from industrial components, ported to an embedded processing platform, and fully tested before the software and the hardware could be considered for integration in a commercially deployed vehicle and linked with the navigation and control environment used in such vehicles.

# Chapter 7

# Conclusion and Future Works

This thesis adopts a holistic approach to investigating the generation, intensity dynamics, and application of a dual-comb system produced in a single cavity using the polarization-multiplexing method. This PhD research combines a scientific evaluation of the fundamental physics of optical frequency combs (OFCs) with an engineering perspective to apply these principles. Additionally, this work serves as a proof of concept for future dual-comb LIDAR applications, linking fundamental research with commercial devices used in autonomous vehicles by Aurrigo Ltd. The key conclusions and future directions from this study are as follows.

### 7.0.1  Conclusions.

The thesis demonstrated the successful design and characterisation of a single-cavity polarisation-multiplexed fibre laser capable of generating two optical frequency combs (OFCs) with slightly different repetition rates. This system achieved a stable dual-comb regime with minimal drift, showcasing its potential for practical applications. The use of a polarisation beam splitter (PBS) and careful adjustment of the state of polarisation (SOP) were critical in achieving the desired extinction ratio for the two combs. Detailed studies on the build-up and propagation dynamics of the dual-comb system were conducted. It was observed that the initial energy spikes in the laser cavity played a crucial role in the formation of stable dual-comb pulses. The successful generation of dual combs was dependent on surpassing specific energy thresholds during the build-up phase. The energy dynamics were found to be similar to those observed in other dual-comb systems. Additionally, the collision dy-namics of solitons within the cavity did not affect the stability of the combs, indicating robust propagation characteristics essential for applications like dual-comb LIDAR. The research optimised the laser cavity design to enhance performance. By reducing the cavity length and adjusting the polarisation-maintaining (PM) fibre segment, the fundamental repetition rate was significantly increased, leading to improved stability and performance of the dual-comb system. The optimised setup demonstrated long-term stability and the ability to maintain dual-comb generation under varying environmental conditions. One of the practical applications of the dual-comb system explored in this thesis was its use in distance ranging (LIDAR). The system demonstrated sub-millimeter precision in measuring distances, highlighting its potential for high-accuracy ranging applications. The study showed that the stability achieved in the dual-comb regime could be effectively utilized in LIDAR systems to measure distances with precision and reliability. Another important outcome obtained not only from the results of this thesis but also from the wholistic experience of using science to develop technology and working for a company that uses that technology is that dual-comb ranging is not the right technology for navigation of autonomous driving vehicles since they require longer ambiguity range, faster update time and they do not need to exploit the best feature that it offers, micrometer to nanometer precision. When the resolution is lower than 1 cm they can operate at an optimal level and having a more precise system does not translate to better performance. Dual-comb ranging technology works better for those applications where high precision is key and fast mapping times are not so important. Applications such as industrial defect monitoring, reverse engineering, or structural building monitoring. Applications where the targets stay static for enough time so that a complete mapping can be done.

### 7.0.2   Future works.

Following the significant advancements and practical applications demonstrated in this thesis, several avenues for future research and development emerge. Firstly, while the dual-comb system has shown promising capabilities in distance ranging (LIDAR) with sub-millimeter precision, further experiments are necessary to refine and extend these capabilities. Increasing the power output of the system could improve its performance, particularly in long-range applications. Additionally, testing the system with a variety of materials as targets would provide a more comprehensive understanding of its ranging accuracy and versatility in different environments and conditions. Secondly, the accuracy of distance measurements could be further enhanced by employing phase shift ranging techniques. By capturing phase shifts, the system could achieve higher precision in distance measurements, potentially surpassing the accuracy of traditional time-of-flight methods. This approach would involve a detailed analysis of phase information to precisely determine distances. Moreover, there is the potential to expand the system's capabilities beyond distance measurements alone. By incorporating a polarisation analyser instead of solely relying on an OSC, the system could capture polarimetric signatures in addition to distances. This would involve using different input states of polarisation and analysing the corresponding output states. Employing the Mueller matrix could facilitate the evaluation of these polarimetric signatures, providing valuable insights into the material properties and structural characteristics of the targets. Testing the system outside the laboratory environment is another crucial step. The dual-comb system has been designed to be fully portable, requiring only a power supply, which makes it suitable for field experiments. Assessing the system's performance in real-world conditions would validate its robustness and practicality for various applications, including remote sensing and environmental monitoring. Finally, improving data analysis and incorporating automation for in vivo applications would significantly enhance the system's usability and efficiency. Developing advanced algorithms for data processing and real-time analysis would streamline the interpretation of results, making the system more user-friendly and applicable in dynamic settings. Automation would also facilitate continuous monitoring and data collection, which is particularly beneficial for applications requiring long-term observation and analysis. In summary, future research should focus on optimising power output, exploring phase shift ranging, integrating polarisation analysis, testing in real-world environments, and enhancing data analysis and automation. These efforts will further unlock the system's capabilities, paving the way for innovative applications and advancements in the field of optical frequency combs.

# List of References

# Appendix A

# Amplified Spontaneous Emission for Erbium Doped Fibre.

Here I present the amplified spontaneous emission for erbium doped measured as power output at the output of the erbium fiber vs pump current.

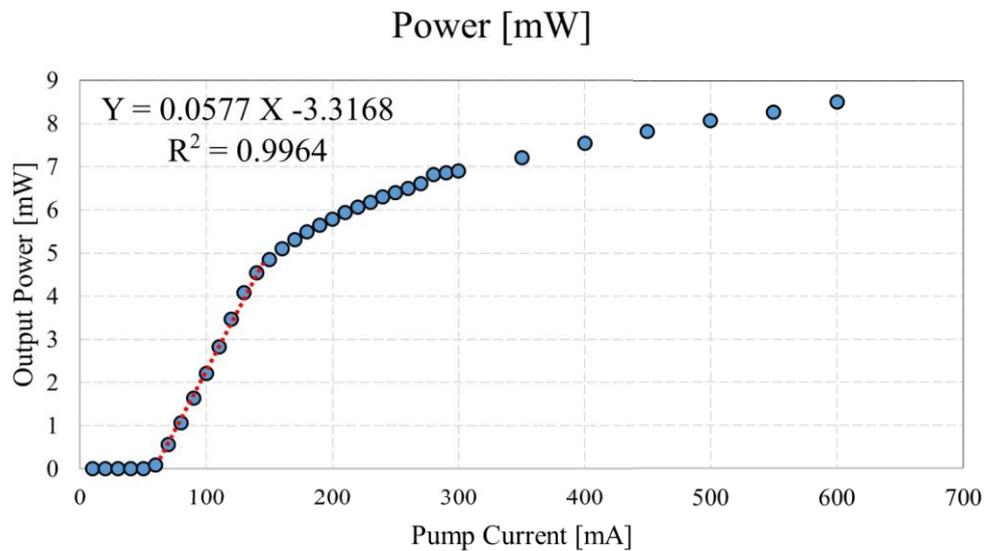

**Figure A.1:** Pump current vs output pump power inside the laser cavity

# Appendix B

# CNT Characterisation.

The non-linear absorbent used in our experiments consists of a carbon nanotube layer deposited on a thin film. This layer was manufactured by Mohammad Mubarak Mohammed Al Araim at the AiPT, Aston University, during his PhD research, 2015-2018. Since then the CNT film has been stored at the AiPT facilities and to evaluate the current properties as a non-linear absorber it was characterised using the laser described in Chapter 2, operating in the fundamental mode-locked laser regime. Since the average power output in this condition is around 0.5 mW an electronic attenuator is inserted at the laser output. Then a 50/50 beam splitter separated the light into two paths. One of them is directly connected to a power meter (Fibre Optic Power Meter, -50 to +26dBm - MP700122) while the other path is connected to the FC/UPC connectors where the CNT film is inserted. The optical power is evaluated using a power meter. The results of the CNT characterisation peak power vs transmission are presented in Figure B.1.

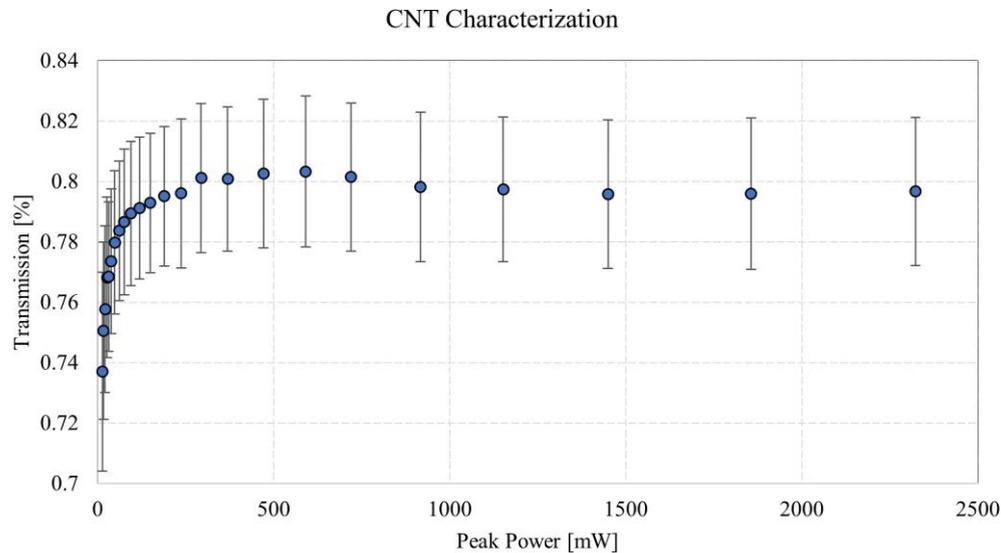

**Figure B.1:** CNT characterisation. Power out vs power in transmission

This characterisation is done by evaluating the level of transmission of light through the CNT compared to the path without CNT (Transmission) for different peak power levels according to the following equation B.1.

$$Transmission(\%) = 1 - \frac{PowerOutput}{PowerInput} \qquad (B.1)$$

The Peak Power can be calculated by taking into consideration the pulse length and dividing the average power measured with the power meter and the ratio of pulse length vs roundtrip time. The CNT gets saturated with values of transmission of 80% when the peak power passes 300 mW. Due to the limit of detection of the power meter (50 nW), the lowest measured transmission was 73.8% for a peak power of 8 mW. The margin of error in the results is rather similar in all the points measured roughly 2.5% positive and negative. This error is evaluated by replicating the characterisation of the CNT and is mainly due to the very low average powers being detected with the power meters and very high levels of attenuations which are sometimes slightly inaccurate even with the electronic attenuator ((OZ Optics Digital Variable Attenuator 62.5/125 MMF 1300nm SC With Power Supply).

# Appendix C

# Build-Up and Propagation Dynamics.

The build-up and propagation dynamics of the following Figures E.2 and C.2 were obtained from the original long OSC trace. Segmenting the trace into the single round trips that make up this trace and after that plotting a succession of those round trips where the build-up and propagation are clearly seen.

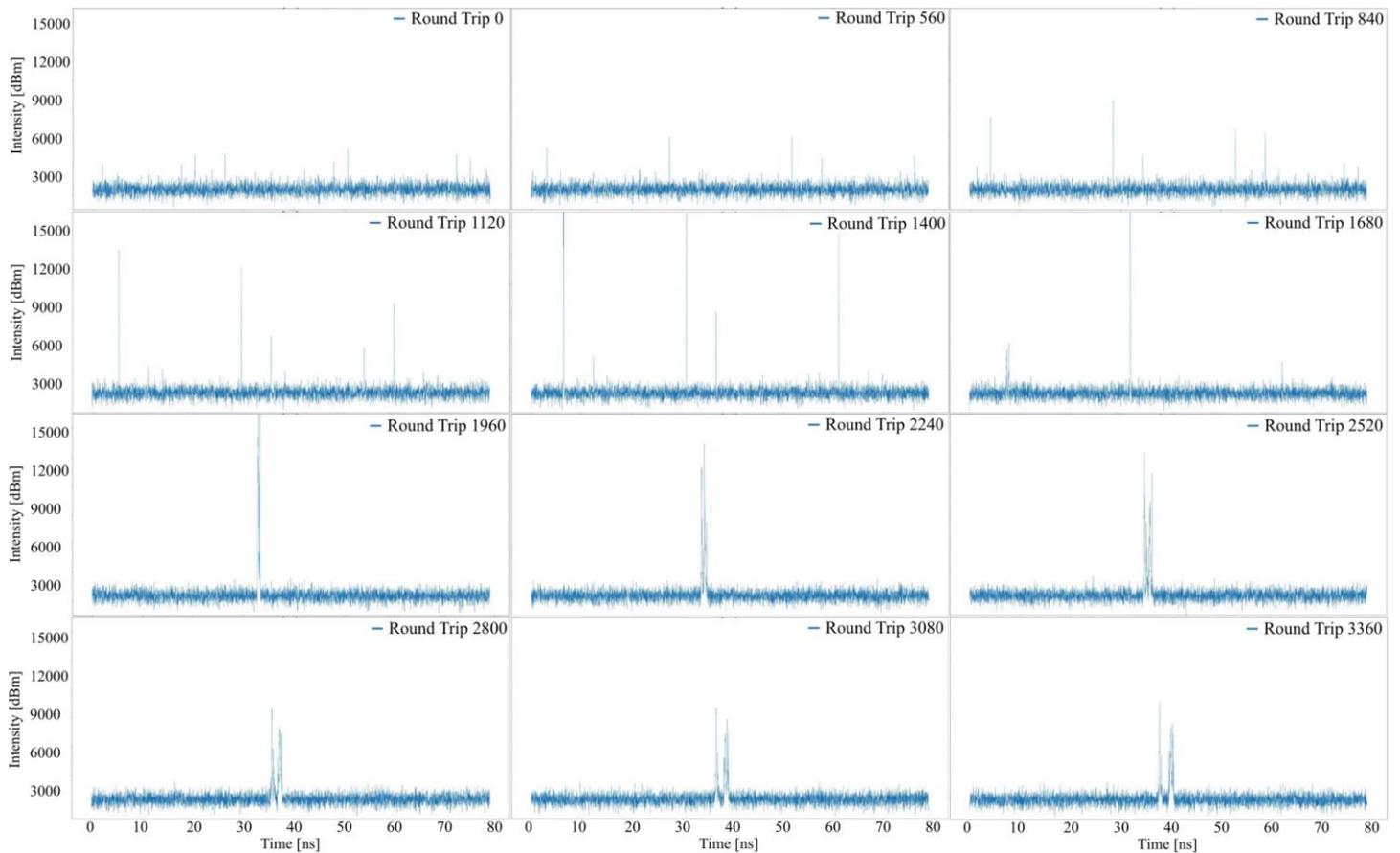

**Figure C.1:** Temporal evolution of the build-up dynamics of the dual-comb regime

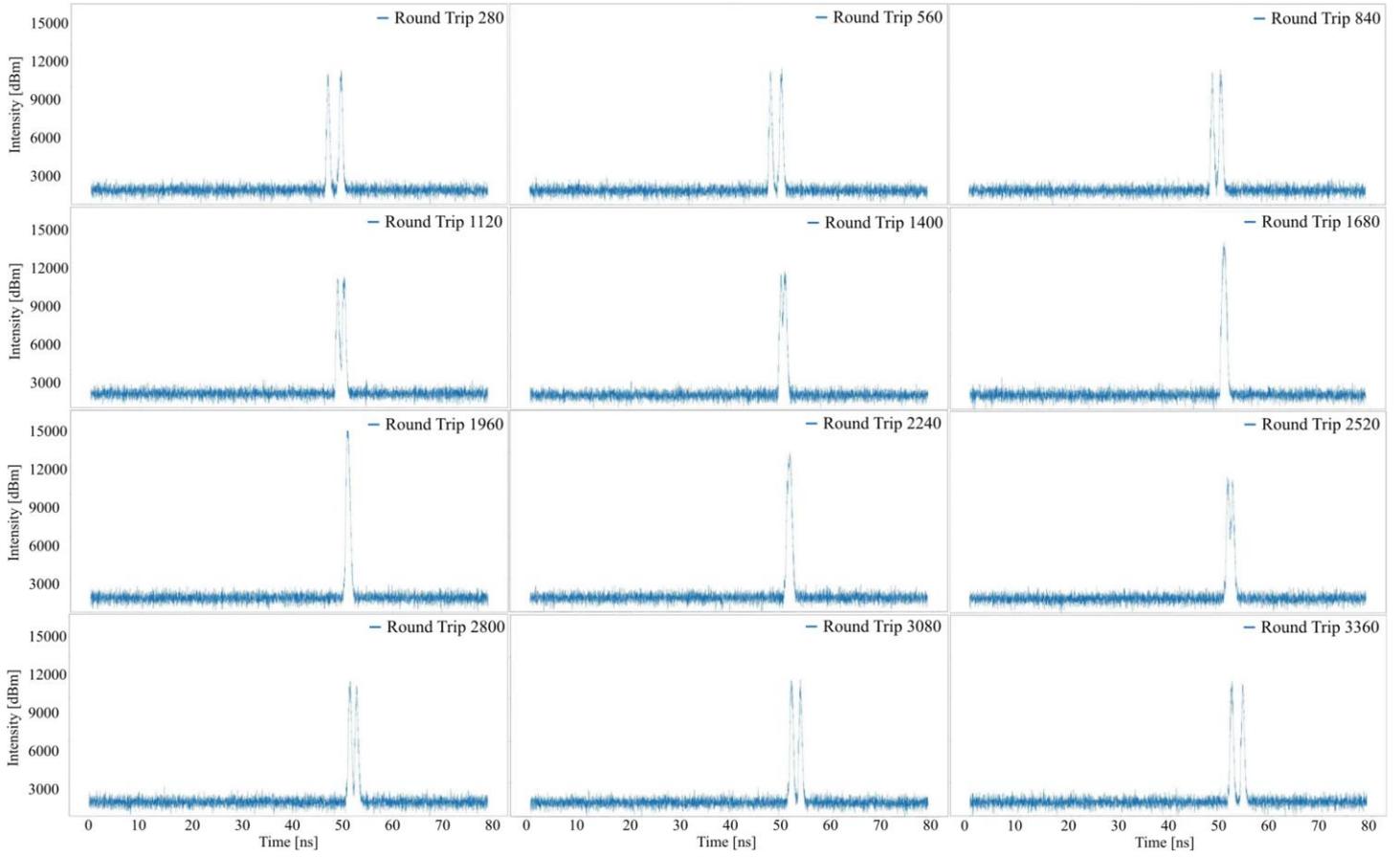

**Figure C.2:** Temporal evolution of the collision of the two combs

# Appendix D

# Housing Optimised Laser.

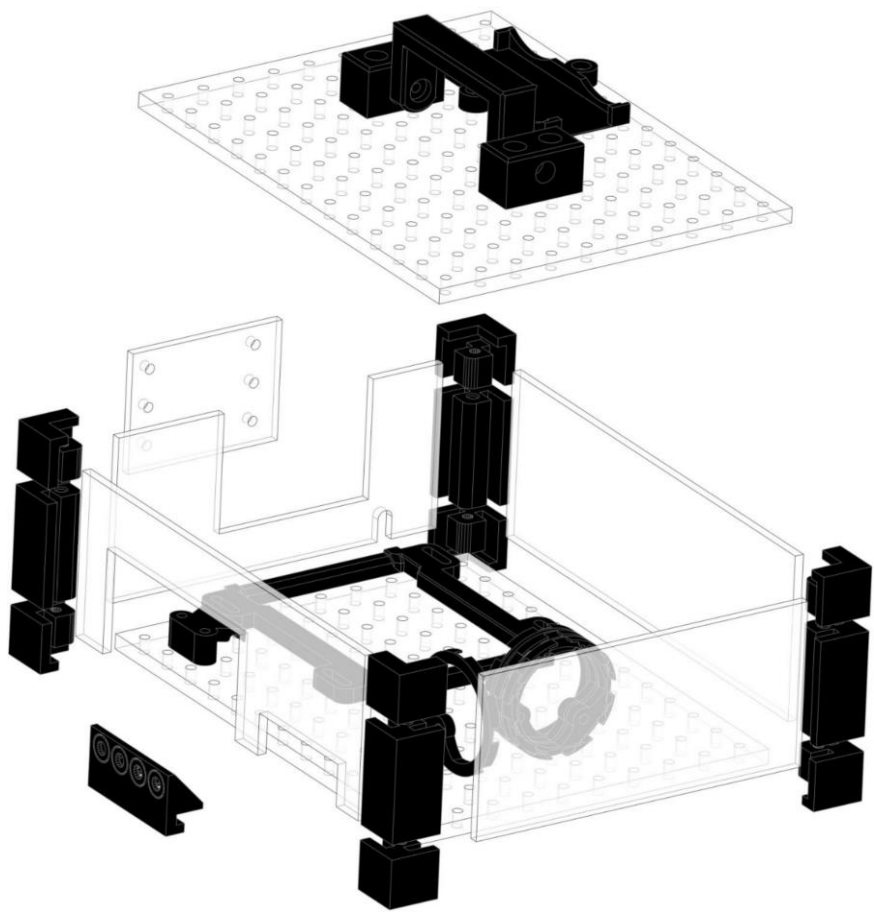

**Figure D.1:** Housing designed for the optimised laser cavity presented in the Chapter 4. In white, the laser-cut pieces in transparent polycarbonate. In black the 3D-printed pieces in black acrylonitrile butadiene styrene

# Appendix E

# Alternative Algorithm for Dual-Comb Distance Ranging

Signal from the target:

$$
\mathrm{FENB}(T, \tau, N) = \frac{1}{50} \sum_{k=1}^{N} \cos\left(\frac{2\pi k}{\tau \cdot 0.995} (T - 15)\right) + i \sin\left(\frac{2\pi k}{\tau \cdot 0.995} (T - 15)\right) \tag{E.1}
$$

Signal from the reference:

$$
\mathrm{FENC}(T, \tau, N) = \frac{1}{50} \sum_{k=1}^{N} \cos\left(\frac{2\pi k}{\tau \cdot 0.995} (\tau - 10)\right) + i \sin\left(\frac{2\pi k}{\tau \cdot 0.995} (T - 10)\right) \tag{E.2}
$$

Power of the Beating Target and Local Oscillator:

$$
F(T, \tau, N) = |(\mathrm{FENB}(T, \tau, N))|^2 + |(\mathrm{FENA}(T, \tau, N))|^2, \quad i = 0, 1, 2, \dots, 12 - i \tag{E.3}
$$

Power of the Beating Reference and Local Oscillator:

$$
F1(T, \tau, N) = |(\mathrm{FENC}(T, \tau, N))|^2 + |(\mathrm{FENA}(T, \tau, N))|^2 \tag{E.4}
$$

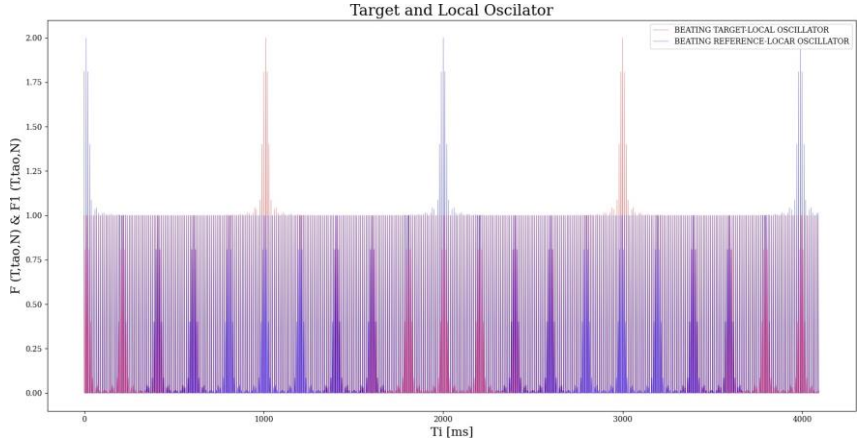

**Figure E.1:** Syntetic Dual Comb Data

Fourier transform:

$$\text{FTR} := \text{FFT}(\text{FF}) \tag{E.5}$$

$$\text{FTR1} := \text{FFT}(\text{FF1}) \tag{E.6}$$

$$\text{FTRL}_{mm} := \text{FTR}_{mm} \tag{E.7}$$

$$\Omega_{mm} = \frac{mm}{2^{12}} \tag{E.8}$$

$$\text{FTRL1}_{mm} := \text{FTR1}_{mm} \tag{E.9}$$

Local oscillator:

$$\text{FENA}(T, \tau, N) = \frac{1}{50} \sum_{k=1}^{N} \exp\left(\frac{i \cdot 2\pi \cdot k}{\tau}\right) \cdot T \tag{E.10}$$

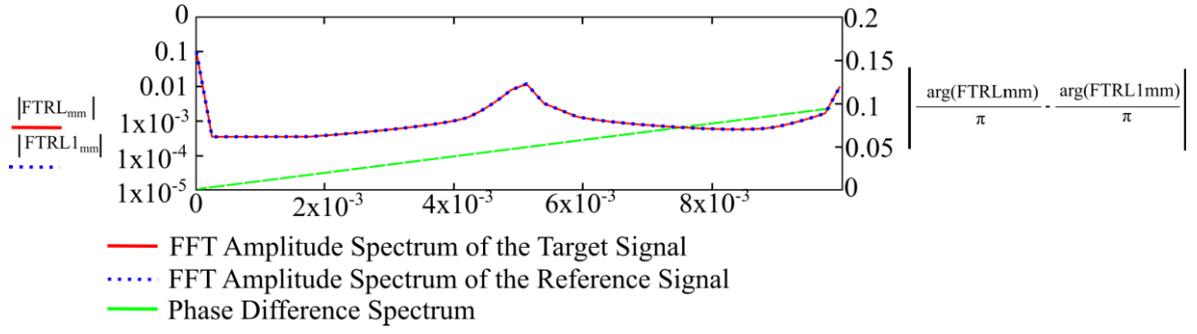

**Figure E.2:** Distance retrieved from the synthetic data tilt. Distance = Tf

$$\text{Tilt} := \frac{dy}{dx} \tag{E.11}$$

$$dy = 0.1 \tag{E.12}$$

$$dx = 0.01 \tag{E.13}$$

$$\text{Tf} := \frac{\text{Tilt}}{2} \tag{E.14}$$

$$\text{Tf} = 5 \tag{E.15}$$

# Appendix F

# Python Code for Signal Processing of the Dual-Comb Ranging OSC Trace.

```python
# -*- coding: utf-8 -*-
"""
Created on Fri Mar 15 15:28:52 2024

@author: rodriga3
"""
import h5py
import numpy as np
import matplotlib.pyplot as plt
# %matplotlib auto - ONLY USE THIS IN CONSOLE, NOT IN A SCRIPT
plt.rcParams.update({'font.size': 24, 'font.family': 'serif'})

Path = 'E:\Results_DC_LIDAR_01-06-2024/'
File = 'DC_LIDAR_5GSa.h5'

f = h5py.File(Path + File, 'r')
Variables0 = list(f.keys())
print(Variables0)

Frep = 39247491.7    # Hz
Delta_Frep = 843.36    # Hz
Sampling_rate = 5000000000
Speed_of_light = 299792458
Refractive_index = 1.45213

Overlapping = 1 / Delta_Frep    # Overlapping time in seconds.

with h5py.File(Path + File, 'r') as f:
    data = f.get(str((Variables0[0])))
    Variables00 = list(data.keys())
    string = str((Variables0[0]) + "/" + (Variables00[0]))
    data = f.get(string)
    print("length of ", string, data.shape)
```

```python
35      data = f.get(str((Variables0[1])))
36      Variables01 = list(data.keys())
37      string = str((Variables0[1]) + "/" + (Variables01[0]))
38      data = f.get(string)
39      print("length of ", string, data.shape)
40
41      data = f.get(str((Variables0[2])))
42      Variables02 = list(data.keys())
43      string = str((Variables0[2]) + "/" + (Variables02[0]))
44      data = f.get(string)
45      Variables020 = list(data.keys())
46      string = str((Variables0[2]) + "/" + (Variables02[0]) + "/" +
        ↪  (Variables020[0]))
47      data = f.get(string)
48      print("length of ", string, data.shape)
49      Values = np.array(data)
50
51  Valuesz = list(Values[:int(len(Values) / 50)])
52
53  # 80 GHz is equal to 0.0125 ns.
54  Time = np.linspace(0, len(Values), len(Values))
55  Time = Time / Sampling_rate
56  del f, data, File, Path, Valuesz, Variables0, Variables00, Variables01,
    ↪  Variables02, Variables020, string
57
58  #%% FIND THE HIGHEST VALUE IN THE SERIES
59
60  from Moving_average import moving_average
61  Test2 = moving_average(Values, 20)
62  Test3 = moving_average(Test2, 20)
63  Test4 = moving_average(Test3, 2000)
64
65  Overlapping_2 = int(Overlapping * Sampling_rate)
66
67  Reference_x = np.argmax(Test3)
68  Reference_y = np.max(Test3)
69
70  #%% GENERATE A VECTOR OF WINDOWS WITH ones_length VALUES USING THE PREVIOUS
    ↪  REFERENCE
71  step = Overlapping_2
72  ones_length = 12000
73
74  vector1 = np.zeros(len(Values))
75
76  for i in range(0, len(vector1), step):
77      end = min(i + ones_length, len(vector1))
78      vector1[i:end] = 1
79
80  vector1 = vector1 * int(Reference_y)
81
82  Remainder = Reference_x % step
83  Remainder = Remainder - 6000
84  aux = np.zeros(Remainder)
85
```

```python
86   vector1 = np.insert(vector1, 0, aux)
87   vector1 = vector1[:-int(len(aux))]
88
89   vector1 = vector1 / int(Reference_y)
90   N_Vector = [vector1[i] * Test2[i] for i in range(len(vector1))]
91
92   #%% FIND THE SECOND BATCH OF VALUES
93   vector2 = np.zeros(len(Values))
94   for i in range(0, len(vector2), step):
95       end = min(i + ones_length * 10, len(vector2))
96       vector2[i:end] = 1
97
98   Remainder = Reference_x % step
99   Remainder = Remainder - 6000
100  aux = np.zeros(Remainder)
101
102  vector2 = np.insert(vector2, 0, aux)
103  vector2 = vector2[:-int(len(aux))]
104
105  vector2 = abs(vector2 - 1)
106
107  Test5 = [Test3[i] * vector2[i] for i in range(len(Test3))]
108
109  Reference_x2 = np.argmax(Test5)
110  Reference_y2 = np.max(Test5)
111  del Test4, Test3, Test5
112
113  #%% GENERATE A VECTOR OF WINDOWS WITH ones_length VALUES USING THE SECOND
     ↪  REFERENCE
114  vector3 = np.zeros(len(Values))
115
116  for i in range(0, len(vector3), step):
117      end = min(i + ones_length, len(vector3))
118      vector3[i:end] = 1
119
120  vector3 = vector3 * int(Reference_y2)
121
122  Remainder = Reference_x2 % step
123  Remainder = Remainder - 6000
124  aux = np.zeros(Remainder)
125
126  vector3 = np.insert(vector3, 0, aux)
127  vector3 = vector3[:-int(len(aux))]
128
129  vector3 = vector3 / int(Reference_y2)
130  N2_Vector = [vector3[i] * Test2[i] for i in range(len(vector3))]
131
132  #%% SMOOTH THE VECTORS USING MOVING AVERAGE
133  NN_Vector = np.abs(N_Vector)
134  NN_Vector = moving_average(NN_Vector, 1000)
135  NN_Vector = moving_average(NN_Vector, 1000)
136
137  NN2_Vector = np.abs(N2_Vector)
138  NN2_Vector = moving_average(NN2_Vector, 1000)
```

```python
139     NN2_Vector = moving_average(NN2_Vector, 1000)
140
141     del vector1, vector2, vector3
142
143     #%% IDENTIFY INTERVALS AND FIND MAXIMUMS
144
145     def find_high_points(vector):
146         max_per_interval = []
147         interval_start = None
148         max_value = 0
149         max_index = 0
150
151         for i in range(len(vector)):
152             if vector[i] != 0:
153                 if interval_start is None:  # Start of a new interval
154                     interval_start = i
155                     max_value = vector[i]
156                     max_index = i
157                 else:  # Within an interval
158                     if vector[i] > max_value:
159                         max_value = vector[i]
160                         max_index = i
161             else:
162                 if interval_start is not None:  # End of an interval
163                     max_per_interval.append((max_index, max_value))
164                     interval_start = None
165
166         # Ensure the last interval is captured if vector doesn't end in zero
167         if interval_start is not None:
168             max_per_interval.append((max_index, max_value))
169
170         return max_per_interval
171
172     #%% CALCULATE DISTANCE BETWEEN TARGETS AND REFERENCES
173     References = find_high_points(NN_Vector)
174     Targets = find_high_points(NN2_Vector)
175
176     Distances = []
177     count = 0
178
179     try:
180         for i in range(len(Targets)):
181             Distances.append(abs(References[i][0] - Targets[i][0]))     # Use abs() for
                 ↪  absolute distance
182             count += 1
183
184     except Exception as e:
185         print(f"An error occurred: {e}")
186
187     Average_distance = sum(Distances) / count
188     Average_distance = 1000 * (Average_distance * (Delta_Frep / Frep) / Sampling_rate
         ↪  * (Speed_of_light / Refractive_index))
189     Deviation_distance = np.std(Distances)
```

```python
190   Deviation_distance = 1000 * (Deviation_distance * (Delta_Frep / Frep) /
      ↪   Sampling_rate * (Speed_of_light / Refractive_index))
191
192   print(f"Average distance: {Average_distance}mm ({Deviation_distance} std)")
193
194   del NN2_Vector, NN_Vector, aux
195   del count, end, i, ones_length, Overlapping, References, Overlapping_2
196   del Reference_x, Reference_x2, Reference_y, Reference_y2
197   del Refractive_index, Remainder, step, Deviation_distance, Distances
198
199   #%% PLOT RESULTS
200
201   plt.figure(figsize=(10, 8))   # You can adjust the figure size to fit your data
      ↪   better.
202   plt.title(f"Average  distance: {Average_distance}  mm")
203   # First subplot
204   plt.subplot(2, 1, 1)
205   plt.plot(Time, Values, color="k", linewidth=0.2, label="Original")
206   plt.xlabel("Time  [s]")
207   plt.ylabel("Voltage  [mV]")
208   plt.legend()
209
210   # Second subplot
211   plt.subplot(2, 1, 2)
212   plt.plot(Time, N2_Vector, color="b", linewidth=0.2, label="Target")
213   plt.plot(Time, N_Vector, color="k", linewidth=0.2, label="Reference")
214   plt.xlabel("Time  [s]")
215   plt.ylabel("Amplitude  [dBm]")
216   plt.legend()
217
218   plt.tight_layout()   # Adjusts subplot params so that subplots are nicely fit in
      ↪   the figure.
219   plt.show()
220
```